\documentclass[11pt,letterpaper]{article}
\pdfoutput=1
\usepackage{jheppub}
\usepackage{color}
\usepackage{graphicx}
\usepackage{tabularx}
\usepackage{xspace}
\usepackage{empheq}
\usepackage{verbatim}
\usepackage{amsmath}
\usepackage{amssymb}
\usepackage[caption=false]{subfig}
\usepackage{url}
\usepackage{bbold}
\usepackage{slashed}
\usepackage{array}

\usepackage{multirow}
\usepackage{threeparttable}
\usepackage{paralist}
\usepackage{xspace}
\usepackage{upgreek}
\usepackage{lipsum}
\usepackage[T1]{fontenc}

\usepackage{tcolorbox}
\usepackage[framemethod=TikZ]{mdframed}
\usepackage[framemethod=TikZ]{mdframed}

\usepackage{setspace}

\usepackage{mathtools}

\usepackage[shortlabels]{enumitem}

\newcommand\colvec[3][]{\begin{pmatrix}\ifx\relax#1\relax\else#1\\\fi#2\\#3\end{pmatrix}}

\newcommand{\beq}{\begin{equation}}
\newcommand{\beqn}{\begin{eqnarray}}
\newcommand{\eeq}{\end{equation}}
\newcommand{\eeqn}{\end{eqnarray}}

\DeclareRobustCommand{\Eq}[1]{Eq.~(\ref{#1})}
\DeclareRobustCommand{\Eqs}[2]{Eqs.~(\ref{#1}) and (\ref{#2})}

\DeclareRobustCommand{\Sec}[1]{Sec.~\ref{#1}}
\DeclareRobustCommand{\App}[1]{Appendix~\ref{#1}}
\DeclareRobustCommand{\Fig}[1]{Fig.~\ref{#1}}

\def\nn{{\nonumber}}

\newcommand{\fd}[2]{\parbox{#1}{\includegraphics[width=#1]{#2}}}

\newcommand{\hard}{\mathrm{hard}}
\newcommand{\dyn}{\mathrm{dyn}}
\newcommand{\tree}{\mathrm{tree}}

\def\cB{\mathcal{B}}

\def\cL{\mathcal{L}}
\def\cM{\mathcal{M}}

\def\cO{\mathcal{O}}

\def\cP{\mathcal{P}}

\newcommand{\zcut}{z_\mathrm{cut}}

\newcommand{\SCETi}{\mbox{${\rm SCET}_{\rm I}$}\xspace}
\newcommand{\SCETii}{\mbox{${\rm SCET}_{\rm II}$}\xspace}

\newcommand{\sdt}{\!\cdot\!}

\newcommand\bn{{\bar n}}

\newcommand{\la}{\lambda}

\newcommand{\lp}{\tilde p}        

\newcommand{\bea}{\begin{eqnarray}}
\newcommand{\eea}{\end{eqnarray}}

\newcommand{\aW}{\alpha_{\scriptscriptstyle W}}
\newcommand{\TaW}{\tilde{\alpha}_{\scriptscriptstyle W}}
\newcommand{\mW}{m_{\scriptscriptstyle W}}
\newcommand{\gW}{g_{\scriptscriptstyle W}}
\newcommand{\thetaW}{\theta_{\scriptscriptstyle W}}
\newcommand{\ralpha}{{\tilde \alpha_{\scriptscriptstyle W}}}

\newcommand{\id}{\mathbf{1}}

\begin{document}

\preprint{\vbox{\hbox{CALT-TH-2017-066}\hbox{LA-UR-17-31169}\hbox{MIT-CTP 4959}}}

\title{
Resummed Photon Spectra for WIMP Annihilation
}

\author[1,2]{Matthew Baumgart,}
\author[3]{Timothy Cohen,}
\author[4,5]{Ian Moult,}
\author[6]{Nicholas L. Rodd,}
\author[6]{\\[3pt]Tracy R. Slatyer,}
\author[7]{Mikhail P. Solon,}
\author[6]{Iain W. Stewart,}
\author[8]{and Varun Vaidya}

\affiliation[1]{\footnotesize Department of Physics, Arizona State University, Tempe, AZ 85287}
\affiliation[2]{\footnotesize New High Energy Theory Center, Rutgers University, Piscataway, NJ 08854}
\affiliation[3]{\footnotesize Institute of Theoretical Science, University of Oregon, Eugene, OR 97403}
\affiliation[4]{\footnotesize Berkeley Center for Theoretical Physics, University of California, Berkeley, CA 94720}
\affiliation[5]{\footnotesize Theoretical Physics Group, Lawrence Berkeley National Laboratory, Berkeley, CA 94720}
\affiliation[6]{\footnotesize Center for Theoretical Physics, Massachusetts Institute of Technology, Cambridge, MA 02139}
\affiliation[7]{\footnotesize Walter Burke Institute for Theoretical Physics, California Institute of Technology, Pasadena, CA 91125}
\affiliation[8]{\footnotesize Theoretical Division, MS B283, Los Alamos National Laboratory, Los Alamos, NM 87545}

\date{\today}

\abstract{We construct an effective field theory (EFT) description of the hard photon spectrum for heavy WIMP annihilation.  This facilitates precision predictions relevant for line searches, and allows the incorporation of non-trivial energy resolution effects.  Our framework combines techniques from non-relativistic EFTs and soft-collinear effective theory (SCET), as well as its multi-scale extensions that have been recently introduced for studying jet substructure. We find a number of interesting features, including the simultaneous presence of SCET$_{\text{I}}$ and SCET$_{\text{II}}$ modes, as well as collinear-soft modes at the electroweak scale.  We derive a factorization formula that enables both the resummation of the leading large Sudakov double logarithms that appear in the perturbative spectrum, and the inclusion of Sommerfeld enhancement effects. Consistency of this factorization is demonstrated to leading logarithmic order through explicit calculation.  Our final result contains both the exclusive and the inclusive limits, thereby providing a unifying description of these two previously-considered approximations. We estimate the impact on experimental sensitivity, focusing for concreteness on an SU$(2)_{W}$ triplet fermion dark matter -- the pure wino -- where the strongest constraints are due to a search for gamma-ray lines from the Galactic Center.  We find numerically significant corrections compared to previous results, thereby highlighting the importance of accounting for the photon spectrum when interpreting data from current and future indirect detection experiments.}

\maketitle
\setcounter{page}{2}

\pagebreak

\section{Introduction}
The discovery of the dark matter (DM) particle(s) is one of the central goals of the high energy physics program. While the Weakly Interacting Massive Particle (WIMP) paradigm with DM masses of order the electroweak scale $\sim100$ GeV has received the most attention, it is also a reasonable possibility that the WIMP could be much heavier.  The canonical example is the neutral component of a new Majorana SU$(2)_W$ triplet fermion -- this \emph{wino} DM will be the concrete example studied here, although many of the results presented below will hold for a wide class of heavy WIMPs.  Assuming no other new states are present, the wino mass is the only free parameter in this model.  The wino is a prototypical heavy WIMP: a calculation of the relic density for winos annihilating to electroweak gauge bosons (including the impact of the charged wino states via the Sommerfeld enhancement~\cite{Hisano:2002fk, Hisano:2003ec, Hisano:2004ds, Hisano:2006nn, Cirelli:2007xd}) yields a mass of around 3~TeV.  The wino as DM is motivated both from a ``complete'' theory perspective in the context of split supersymmetry \cite{ArkaniHamed:2004fb, ArkaniHamed:2004yi, Giudice:2004tc, Wells:2004di, Pierce:2004mk, Arvanitaki:2012ps, ArkaniHamed:2012gw, Hall:2012zp}, but it is also interesting due to its economy, \emph{i.e.}, minimal DM~\cite{Cirelli:2005uq, Cirelli:2007xd, Cirelli:2008id, Cirelli:2009uv, Cirelli:2015bda}.   

Multi-TeV WIMPs are unobservable at the LHC: 14 TeV projected limits on winos are in the few hundred GeV range, and they will even be challenging to find at a future collider~\cite{Low:2014cba, Cirelli:2014dsa}.  Furthermore, the cross section at direct detection experiments suffers an accidental cancellation between the spin-0 and spin-2 contributions, yielding a rate that is near the neutrino floor~\cite{Hill:2011be, Hill:2013hoa,Hisano:2015rsa}. The one known channel that holds promise for detecting multi-TeV winos is via astrophysical searches for their annihilation products. Annihilation to photons could provide a very clean signal visible to ground-based air Cherenkov array telescopes~\cite{Cirelli:2007xd, Cohen:2013ama, Fan:2013faa}, and constraints from the observed flux of antiproton cosmic rays can also be relevant, but require modeling of cosmic-ray propagation and backgrounds~\cite{Cuoco:2017aa}. In particular, a search for line photons by the HESS~experiment~\cite{Abramowski:2011hc} provides a powerful constraint for thermal winos with mass near 3 TeV, although this is subject to large uncertainties from the unknown shape of the DM density profile in the inner Galaxy~\cite{Cohen:2013ama}.  Furthermore, there are many upcoming experimental searches which could discover heavy WIMPs via indirect detection of gamma rays, including new data from HESS~\cite{Hinton:2004eu,Abramowski:2013ax}, HAWC~\cite{Sinnis:2004je,Harding:2015bua,Pretz:2015zja}, CTA~\cite{Consortium:2010bc}, VERITAS~\cite{Weekes:2001pd,Holder:2006gi,Geringer-Sameth:2013cxy}, and MAGIC~\cite{FlixMolina:2005hv,Ahnen:2016qkx}.  We would therefore like to have reliable theoretical predictions for the particle physics contribution to the cross section over a wide range of DM masses.   One key feature of these ground-based experiments is that their resolution for line searches is not particularly sharp, implying that finite bin effects should be accounted for when making a precise prediction of the annihilation cross section. A main goal of the present work is to address this.

It is by now well understood that the calculation of the annihilation rate is complicated by the presence of multiple hierarchical scales, namely $\mW$ and $M_\chi$.  For models with $M_\chi \gg \mW$, this separation of scales invalidates the standard perturbative expansion, introducing a number of effects that must be treated to all orders, in particular Sommerfeld enhancement, which resums terms of the form $\left( \aW M_\chi/\mW \right )^k$ \cite{Hisano:2003ec,Hisano:2004ds,Cirelli:2007xd,ArkaniHamed:2008qn,Blum:2016nrz}, and Sudakov double logarithms $\aW\log^2(M_\chi/\mW)$~\cite{Hryczuk:2011vi,Baumgart:2014vma,Bauer:2014ula,Ovanesyan:2014fwa,Baumgart:2014saa,Baumgart:2015bpa,Ovanesyan:2016vkk}. These can be conveniently treated using effective field theory (EFT) techniques, which allow for a systematic expansion in $\mW/M_\chi \ll 1$, and the identification of universal behavior in this limit. This has attracted recent attention, resulting in calculations from different groups, with differing assumptions. Two groups~\cite{Bauer:2014ula,Ovanesyan:2014fwa,Ovanesyan:2016vkk} resummed the logarithms that appear assuming the final state was specified as $\gamma\,\gamma$ or $\gamma \,Z$ (referred to here as exclusive), while \cite{Baumgart:2014vma,Baumgart:2014saa,Baumgart:2015bpa} calculated a resummed cross section using the operator product expansion (OPE) and assuming a $\gamma+X$ final state (referred to here as inclusive). Due to these differing assumptions, distinct conclusions were reached on the importance of the logarithmically enhanced terms.

In reality, the situation is more subtle and lies somewhere in between these two extremes. Due to the finite energy resolution of the detector, the state $X$ recoiling against the detected photon, which we take to have energy $E_\gamma$, is not forced to be a single electroweak boson.  However, $X$ is constrained to lie near the light cone, namely it is a \emph{jet}.  In this region it is well known that the standard OPE breaks down, and a more complicated factorization, describing the dynamics of the radiation within the jet, is required.  Explicitly, this introduces another small parameter $(1-z)\ll 1$, where 
\begin{align}
z = \frac{E_\gamma}{M_\chi} \in [0,1]\,,
\end{align}
controls the distance from the endpoint, thereby further complicating the perturbative structure. In particular, large logarithms of $(1-z)$ appear. We will refer to these as endpoint logarithms since they become important as $z\to1$. The importance of these endpoint logarithms in the DM case was noticed in \cite{Baumgart:2015bpa} where an attempt was made to extend the OPE based expansion beyond its region of validity into the endpoint region.\footnote{Similar effects have also been seen in fixed order calculations of $\chi\, \chi\to W^+ W^- \gamma$ in the WIMP DM literature~\cite{Beacom:2004pe, Birkedal:2005ep, Bergstrom:2004cy, Bergstrom:2005ss, Bringmann:2007nk}.} However, this framework did not provide a way to exponentiate these logarithms. Their resummation is one of the goals of this paper. 

In this paper we develop a comprehensive EFT framework to compute the photon spectrum for annihilating (or decaying) DM. We use the soft-collinear effective theory (SCET) \cite{Bauer:2000yr, Bauer:2001ct, Bauer:2001yt}, and its recent extensions developed for treating similar multi-scale problems in jet substructure, to factorize the dynamics at the scales $\mW$ (electroweak breaking scale), $M_\chi (1-z)$ (soft scale), $M_\chi \sqrt{1-z}$ (jet scale), and $M_\chi$ (hard scale).  In order to perform the resummation, we will need to refactorize the cross section using techniques for multi-modal field theories~\cite{Bauer:2011uc,Larkoski:2014tva,Procura:2014cba,Larkoski:2015zka,Pietrulewicz:2016nwo}.  All large logarithms present in the cross section are then captured by renormalization group evolution  between the relevant scales. The end result is a completely factorized description that allows for systematically improvable calculations of the photon spectrum. In this paper we will use this framework to compute the resummed spectrum for pure wino annihilation.  The extension to Higgsinos and more general representations will be left for future work.

\begin{figure}
\begin{center}
\includegraphics[width=0.75\columnwidth]{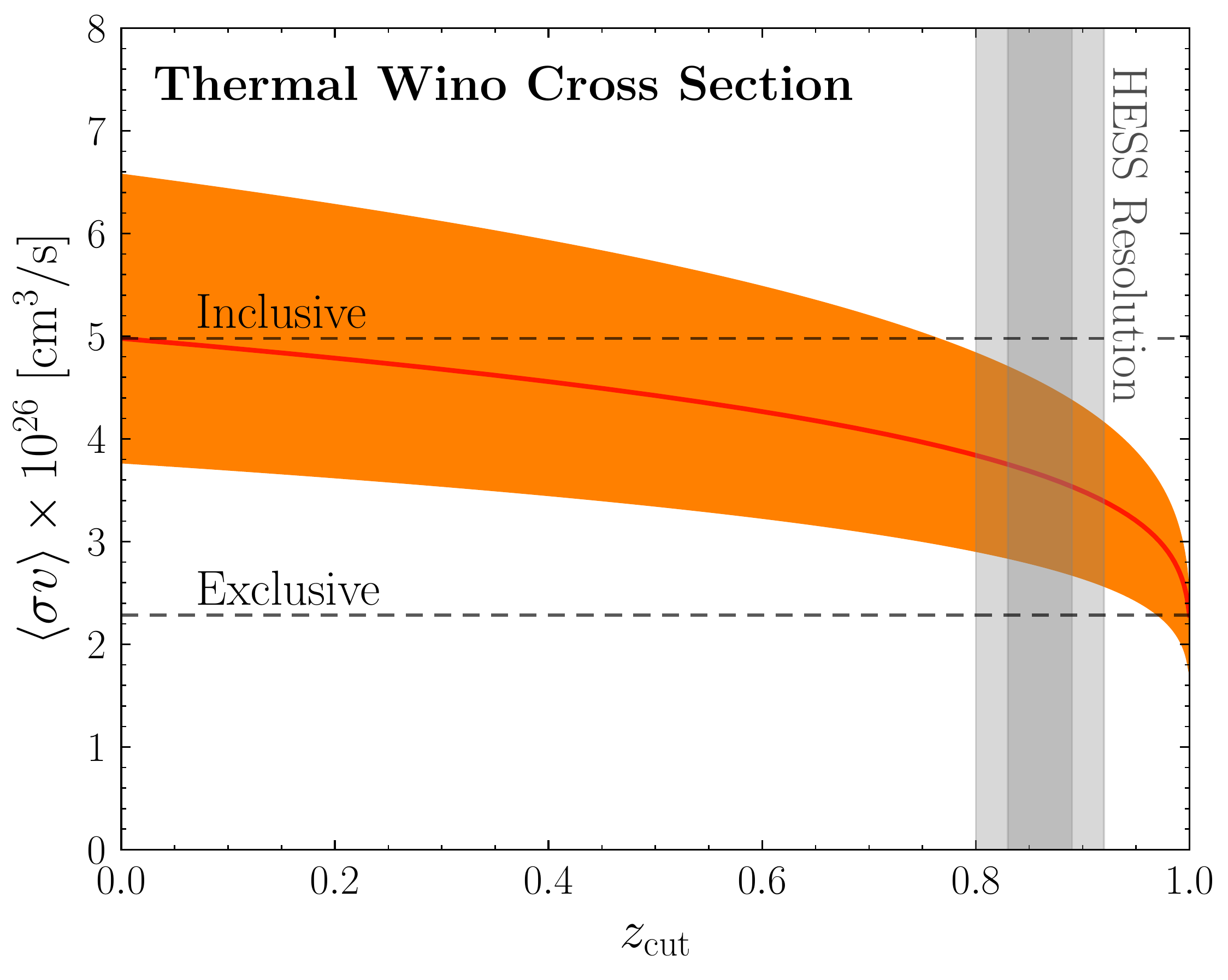}
\end{center}
\caption{The resummed cross section as a function of the experimental resolution parameter $\zcut$ for a $3$ TeV wino, showing the transition between the fully inclusive ($\zcut=0$) and the fully exclusive ($\zcut=1$) cases. For $\zcut \sim 0.8$-$0.9$, as relevant for the HESS experiment, the prediction is half way between the two limiting cases, emphasizing the importance of properly treating $\zcut$.}
\label{fig:cumulative_zcut_intro}
\end{figure}

An example of the result from our calculation is shown in Fig.~\ref{fig:cumulative_zcut_intro}.   Here we have plotted the cumulative spectrum,
\begin{align}\label{eq:cumulative_def_intro}
\sigma(\zcut) = \int\limits_{\zcut}^{1}\text{d}z \frac{\text{d} \sigma}{\text{d}z}\,.
\end{align}
A value of $\zcut=0$ corresponds to the fully inclusive case, and $\zcut=1$ to the fully exclusive case.  As a benchmark, we have taken the wino mass to be 3 TeV -- a wider range of masses are presented below in Sec.~\ref{sec:num_transition}.  Here we see the impact of resumming the endpoint logarithms: there is the known factor of $2.2$ difference between the exclusive and inclusive calculations, and when we take $z_\text{cut} \sim 0.8-0.9$ (which is motivated by the HESS~energy resolution), we find that the prediction falls almost half way between the inclusive and exclusive limits.  This demonstrates the importance of the study presented below.

An outline of this paper is as follows. In \Sec{sec:kinematics} we carefully review the kinematics of indirect detection, highlighting the different regions of the photon spectrum, the appropriate field theoretic techniques that are required for their description, and the differing approximations made in previous presentations. In \Sec{sec:review} we review the different effective theories that we will make use of in our analysis, namely non-relativistic DM effective theory (NRDM) and SCET. In \Sec{sec:factorization} we present our factorization formula for the region $\mW \ll M_\chi (1-z) \ll M_\chi$. We describe in detail the multi-step matching procedure used in its derivation, and the physical role of the different functions appearing in the factorization. In \Sec{sec:LL_resum} we perform the LL resummation, and derive a compact analytic expression for the resummed spectrum.  In \Sec{sec:limits} we show that our EFT reproduces the resummation in both the OPE region, and the exclusive endpoint by taking appropriate limits, hence tying together different results in the literature. In \Sec{sec:num_transition} we present numerical results for the case of wino DM, comparing with previous results obtained using the exclusive and inclusive calculations, allowing us to demonstrate that properly accounting for the finite resolution has a numerically significant effect. In \Sec{sec:results} we estimate the impact of our newly derived predictions on indirect detection constraints using a simplified mock analysis of the HESS data. We conclude in \Sec{sec:conc}.  Two appendices are provided:  in Appendix~\ref{sec:one_loop_app}, we provide many technical aspects of the one-loop calculations presented in the text, and Appendix~\ref{sec:continuum} demonstrates the minimal impact of photons from cascade decays (\emph{e.g.}~$\chi\,\chi\rightarrow W^+ W^- \rightarrow \text{many }\gamma$s) on our mock reanalysis of the HESS data.

\vspace{0.4cm}
\noindent {\textbf{Guide for the Reader}}  
\addcontentsline{toc}{section}{\protect\numberline{}Guide for the Reader}%
\vspace{0.4cm}

\noindent We anticipate that our audience's interests span from the technical aspects of the EFT-based calculation to an interest in the implications for the indirect detection experimental predictions.  We therefore provide two road maps for navigating this paper, depending on the expertise of the reader.  For the EFT enthusiasts, the main technical details of the factorization are presented in Secs.~\ref{sec:review}-\ref{sec:LL_resum}.  While we have attempted to make the presentation as self contained as possible, in particular by reviewing the relevant technology, these sections necessarily assume a higher level of familiarity with EFT techniques, and are as such more mathematically intensive.  These sections provide the details which yield the final prediction, but can be skipped without affecting one's big picture understanding of this work.

For the reader interested primarily in the results, and the resolution of previous differing approximations and conclusions in the literature, we recommend \Sec{sec:inc_exc} and Secs.~\ref{sec:num_transition}-\ref{sec:results}. \Sec{sec:inc_exc} emphasizes the physical differences between the different approximations previously made in the literature, and explains the necessity of pursuing our approach to derive a complete understanding for the range of parameters of interest to current and future experiments. The main results of our study are shown in graphical form in \Sec{sec:num_transition}, where we highlight the numerical impact of the resummation of logarithms of $\zcut$, and compare with numerical results from previous approximations. This clearly illustrates the importance of properly including the finite resolution of the experiments. Finally, the impact of our updated numerical results on DM exclusions are given in \Sec{sec:results}.

\section{Kinematics for Heavy WIMP Annihilation}\label{sec:kinematics}

In this section, we discuss in detail the kinematics of heavy DM decay or annihilation to photons as relevant for indirect detection. We carefully analyze all relevant scales, identifying regions where large ratios of scales exist, which will give rise to logarithms that need to be resummed. This analysis will also make clear the differences between the previous studies in the literature. We will also highlight how collinear-soft modes appear in the broken theory, highlighting the distinction with the case of the naively similar $B\to X_s \gamma$ that has been thoroughly treated in the literature (see \emph{e.g.}~\cite{Neubert:1993um,Ligeti:1999ea,Bauer:2000ew,Neubert:2004dd,Becher:2006pu}). The discussion of this section is completely independent of the details of the DM, allowing us to simultaneously consider decay and annihilation, and depends only on the kinematics of indirect detection.

\subsection{Three Effective Field Theory Regimes}\label{sec:inc_exc}

\begin{figure}
\begin{center}
\subfloat[]{\label{fig:exclusive}
\includegraphics[width=4.5cm]{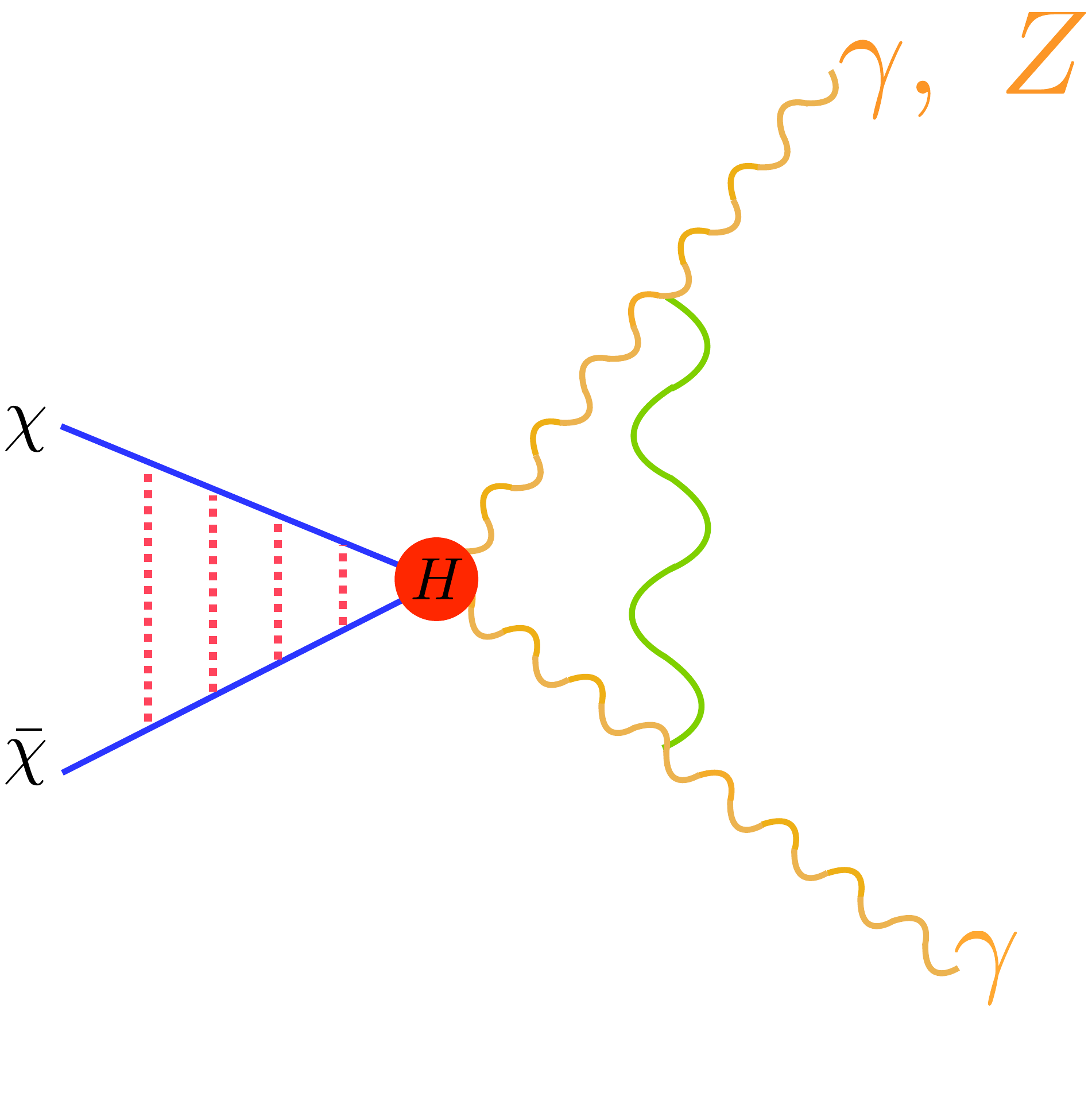}    
}\qquad
\subfloat[]{\label{fig:OPE}
\includegraphics[width=4.05cm]{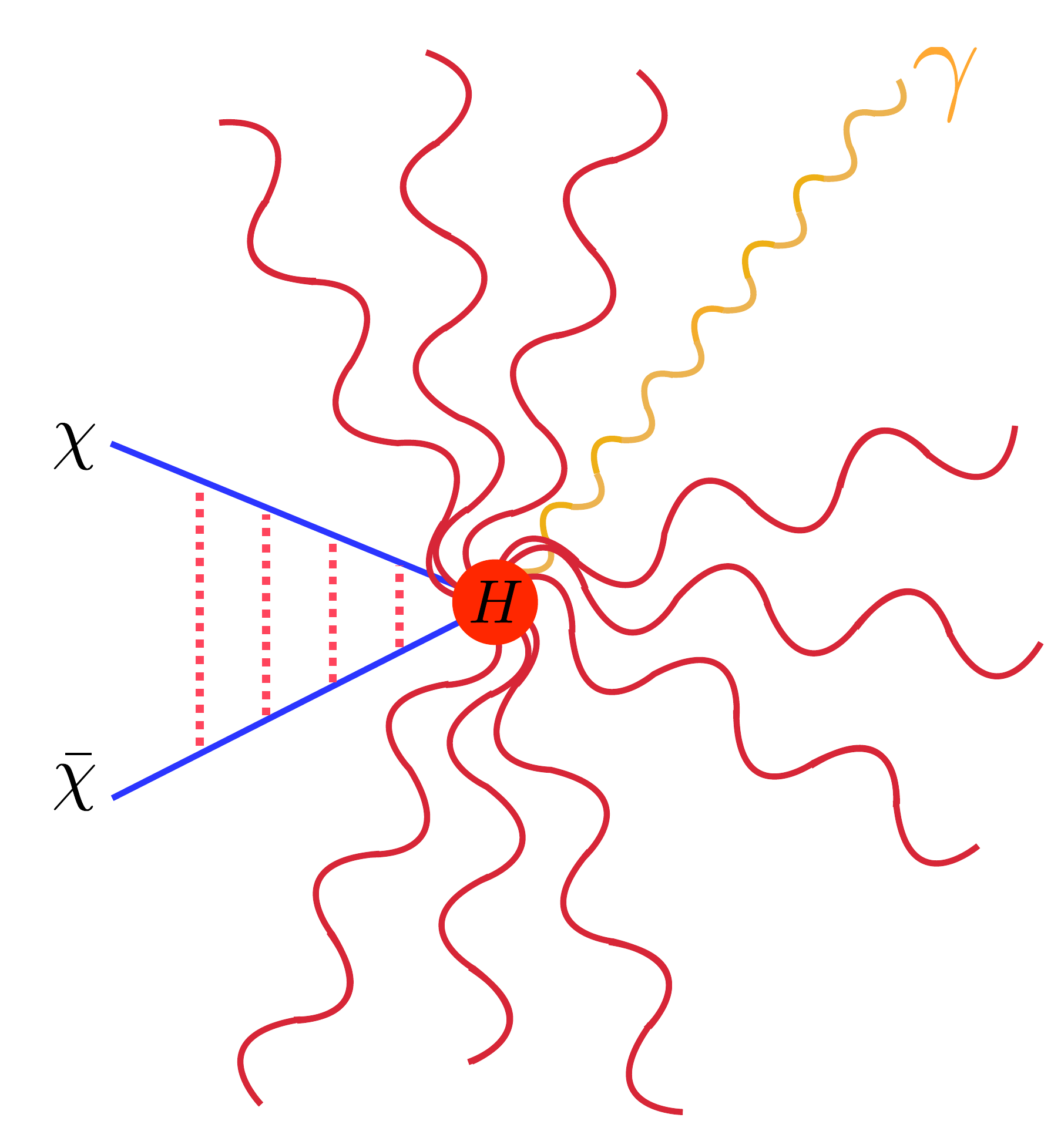}
}\qquad
\subfloat[]{\label{fig:endpoint}
\includegraphics[width=4.5cm]{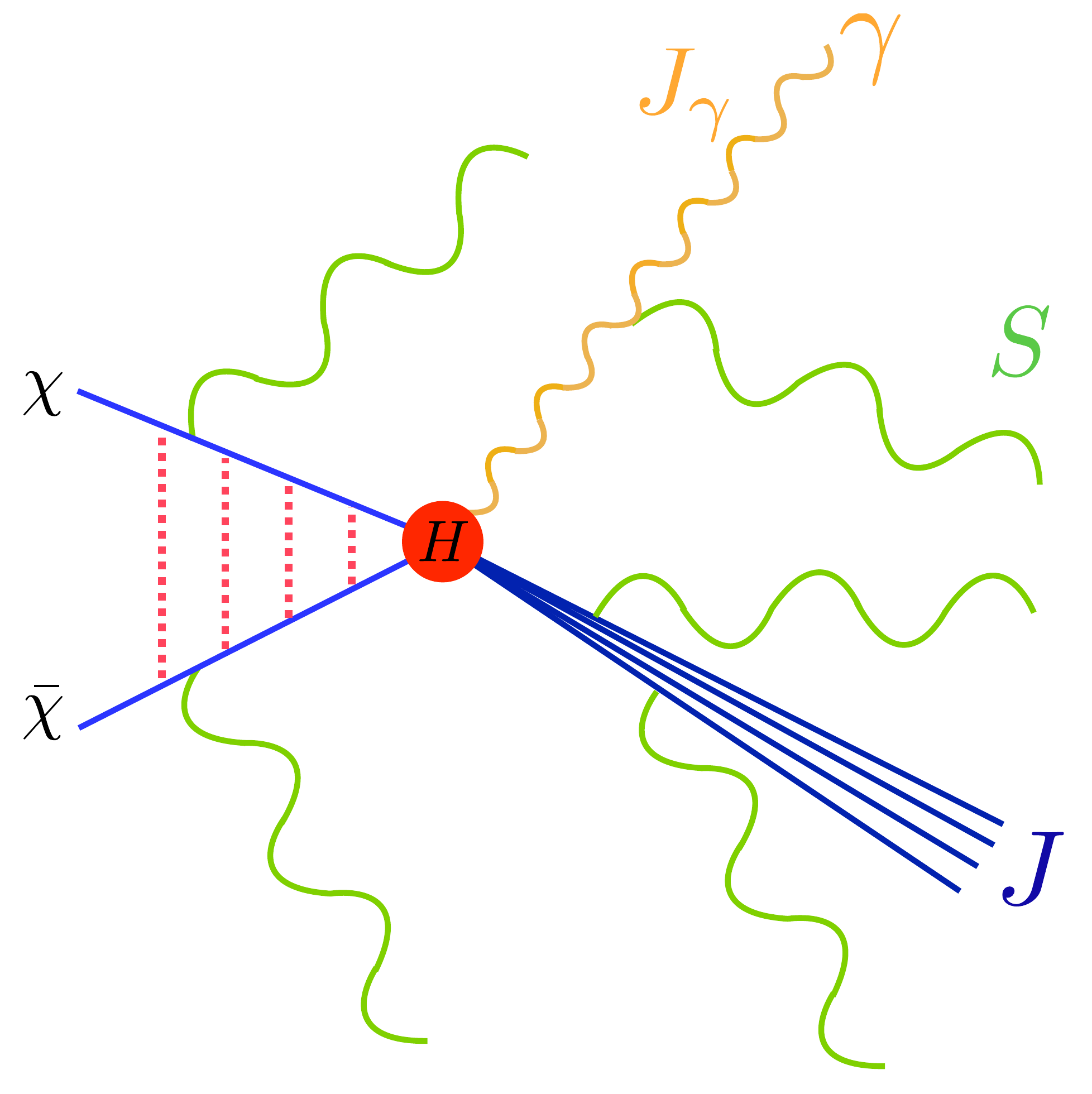}
}
\end{center}
\caption{(a) Fully exclusive production, which contributes only at the endpoint where $z=1$. Only virtual corrections are present. (b) Operator Product Expansion for $\gamma+X$ with $m_X \sim M_\chi$. Here the state $X$ has a large invariant mass and can be integrated out. (c) The endpoint region, $m_X \ll M_\chi$. Here the measurement on the final state $X$ constrains it to have a small invariant mass.  This implies that $X$ cannot be integrated out and must be treated as a dynamical object in the EFT.  In all cases, the dashed lines dressing the annihilating DM represent the Sommerfeld enhancement. 
}
\label{fig:factorization_regions}
\end{figure}

We consider for concreteness the annihilation of  two nearly stationary DM particles of mass $M_\chi$ decaying to $\gamma+X$, where the $\gamma$ is assumed to be detected by the experiment. Here $X$ denotes all final state radiation apart from the photon. The case of DM decay for a particle of mass $2M_\chi$ is identical. We use a dimensionless variable $z$ to characterize the energy fraction of the photon
\begin{align}
E_\gamma= M_\chi\, z\,,
\label{eq:EGamma}
\end{align}
or equivalently, 
\begin{align}\label{eq:mass}
 m_X^2=4\, M_\chi^2(1-z)\,,
\end{align}
where $m_X$ is the invariant mass of the final state $X$.  The result of the calculation will be a differential cross section as a function of $z$, which will be integrated from $z=\zcut \rightarrow 1$.  Depending on the value of $\zcut$, a number of different field theoretic descriptions are required:\footnote{At this level of discussion, namely the description of kinematics, the different regions are identical to those for $B\to X_s \gamma$ and related processes. In the $B$-physics literature, the endpoint region, which will be the focus of this paper, is also referred to as the shape function region \cite{Bigi:1993ex,Neubert:1993um,Mannel:1994pm}.}   
\begin{itemize}
\item {\bf Exclusive Final State} $((1-\zcut) =0)$~\cite{Bauer:2014ula,Ovanesyan:2014fwa,Ovanesyan:2016vkk}: Here the final state is exactly specified, either $\gamma\,\gamma$ or $\gamma\,Z$, and we have $z_\text{cut}=1$. Electroweak Sudakov double logarithms, $\log^2 (2\,M_\chi/\mW)$, appear in the perturbative expansion.  See~\Fig{fig:exclusive}.
\item {\bf Inclusive Final State} $((1-\zcut) \sim1)$~\cite{Baumgart:2014vma, Baumgart:2014saa, Baumgart:2015bpa}: Here the final state is $\gamma + X$, and the final state $X$ is fully inclusive.  This implies that $m_X$ is large, such that the state $X$ can be integrated out using a local OPE~\cite{Wilson:1969zs}. See~\Fig{fig:OPE}.
\item {\bf Endpoint Region} $(0<(1-\zcut) \ll 1$):  In this region, the invariant mass of the final state $m_X\to 0$ and as such it cannot be integrated out using a local OPE.  The photon of interest is taken to lie along one lightcone.  Then $X$ consists of collimated high energy radiation along an orthogonal light cone, with transverse spread $p_T \sim M_\chi \sqrt{1-z}$, as well as isotropic soft radiation with $E \sim M_\chi (1-z)$.  The standard OPE approach is not sufficient, and a more complicated factorization theorem describing the dynamics of the soft and collinear radiation is required~\cite{Korchemsky:1994jb}. Deriving an analogous factorization for the case of WIMP annihilation is one of the main results of this paper.  In this region, Sudakov double logarithms, $\log^2 (1-z)$ appear in addition to electroweak Sudakov double logarithms $\log^2(2\,M_\chi/\mW)$.  See~\Fig{fig:endpoint}. 
\end{itemize}

We can now determine which of the above regions are most relevant to model the input photon spectrum for a search for DM lines. In principle, if the energy resolution of the detector is sufficiently precise, the appropriate cross section would only include the exclusive final state consisting of a photon and a single recoiling electroweak boson. In this case, the kinematics dictate that this condition is equivalent to requiring $z \gtrsim 0.99$ ($0.9999$) for $M_\chi \sim 500$ GeV (10 TeV).  The corresponding energy resolution is well beyond the capabilities of existing detectors.  For example, translating the Gaussian width of the resolution quoted in the HESS line search~\cite{Abramowski:2013ax} to a hard cut, would naively imply that $\zcut$ varies from 0.83 to 0.89 as $M_\chi$ goes from 500 GeV to 10 TeV.  This range additionally implies that we are outside the inclusive region, such that factors of $\log^2(1-z)$ are potentially large and resummation should be performed. We conclude then that the theory which best describes the line observations made by air Cherenkov telescopes has a state $X$ that is recoiling against the photon with $m_X \ll M_\chi$, \emph{i.e.}, the endpoint region EFT.  The theoretical descriptions of the matching to the exclusive region, as well as the OPE region, are also important for a complete description of the spectrum.  We will see that these limits arise naturally from our endpoint EFT. 

\subsection{Kinematics of the Endpoint Region}\label{sec:csoft_intro}

\begin{figure}
\begin{center}
\subfloat[]{\label{fig:csoft_intro_a}
\includegraphics[width=5.5cm]{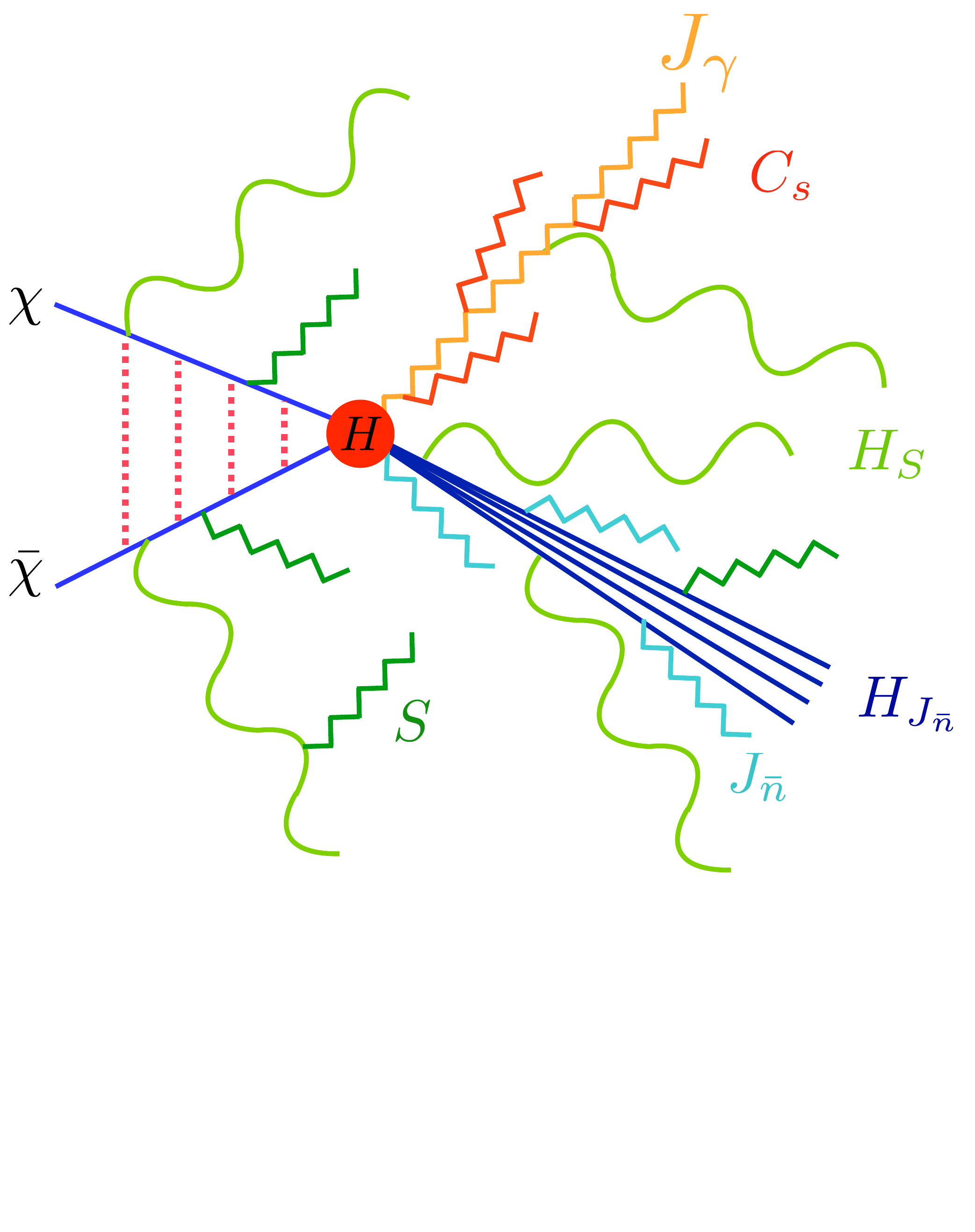}    
}\hspace{40pt}
\subfloat[]{\label{fig:csoft_intro_b}
\includegraphics[width=7.5cm]{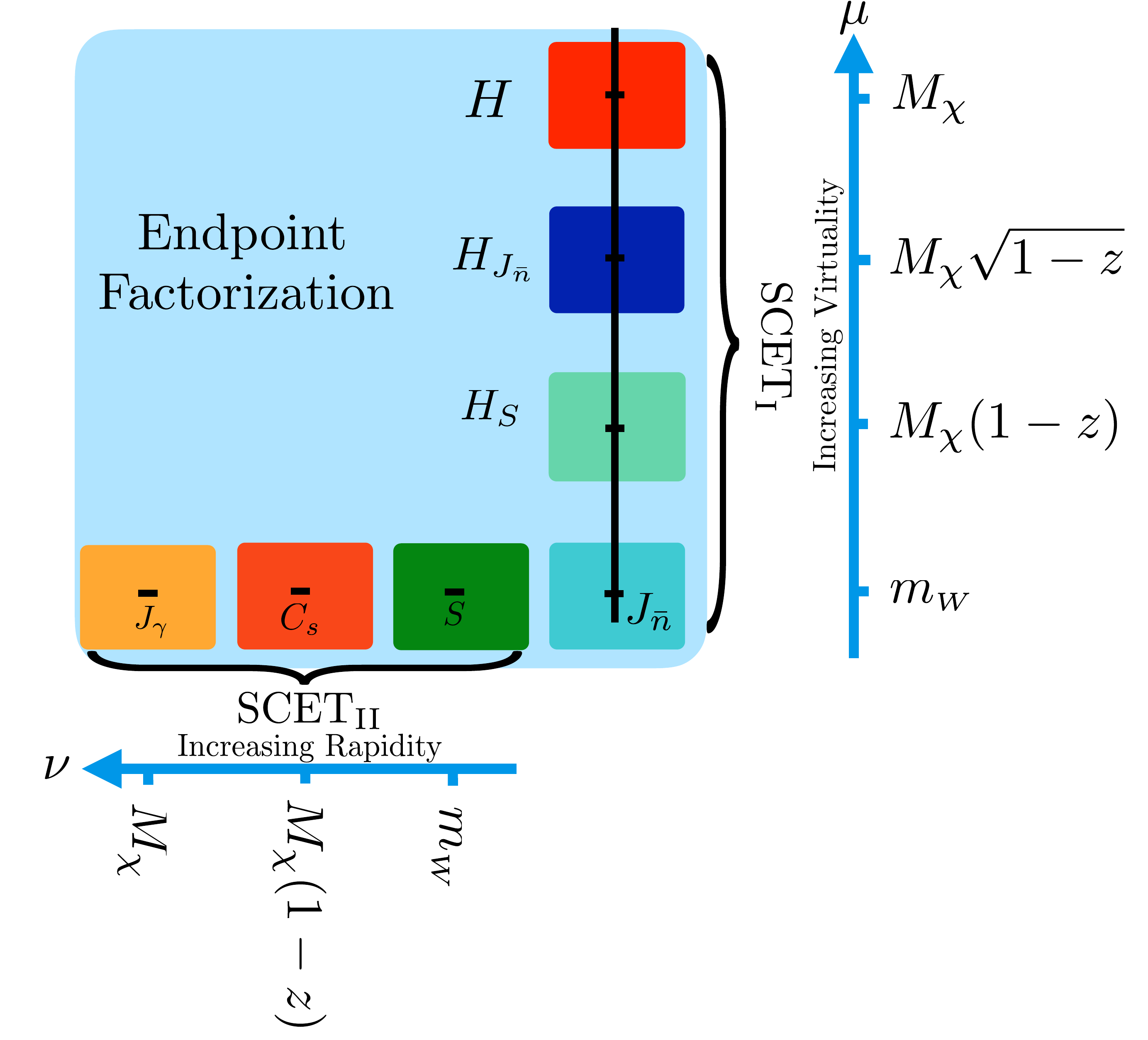}
}
\end{center}
\vspace{-0.2cm}
\caption{(a) A schematic depiction of the relevant modes in the effective theory for DM annihilation near the endpoint. Modes which are sensitive to the mass of the electroweak bosons (broken theory) are in zig-zag, while those that behave as effectively massless (unbroken theory) are curvy. (b) Rapidities and virtualities of the modes describing the final state. The complicated modal structure of the EFT is driven by the simultaneous presence of the scales $M_\chi$, and $\mW$, as well as the constraint on the mass of the final state.
}
\label{fig:csoft_intro}
\end{figure}

Having determined that experimental considerations drive us to focus on the endpoint region, next we describe the relevant kinematics.  This will expose the corresponding modes that will be required to construct the EFT description.  These modes are shown schematically in \Fig{fig:csoft_intro}, along with their virtualities and rapidities.  Our goal in this section is twofold. First, this discussion will motivate the EFTs introduced in \Sec{sec:review}.   Second, it will allow us to provide context and highlight the new features of the factorization needed here in a physical manner, motivating the technical discussion of \Sec{sec:factorization}.  The later sections will then provide a comprehensive mathematical treatment, to complement the simple picture that follows from kinematic arguments.

We begin with the kinematics of the initial state, namely the annihilating DM.  The DM in the halo has a typical velocity $v\sim 10^{-3}$, so a non-relativistic description is appropriate.  The DM will be modeled as heavy sources (in analogy with heavy quark EFT or non-relativistic QCD) emitting ultra-soft radiation, as shown in \Fig{fig:csoft_intro_a}.  There is one well known complication in the heavy mass limit.  Winos carry electroweak charge such that the Sommerfeld enhancement due to the exchange of electroweak gauge bosons must be included.  This can be appropriately accounted for in the non-relativistic DM (NRDM) EFT by including the relevant potentials, see \Sec{sec:NRDM}.  A feature of the NRDM EFT is that it allows a factorization of the Sudakov corrections from the Sommerfeld effects.  

The final state is more complicated, and a full characterization will require a multi-modal EFT.  Recapping the discussion above, as $\zcut\to 1$ the final state consists of both a jet of collimated energetic particles and wide angle low energy radiation.  As is well known, this can be captured by SCET.  However, due to the multi-scale nature of the problem, we will show that additional modes, illustrated in \Fig{fig:csoft_intro}, will be required to fully factorize all the logarithms. The origin of the multi-modal structure, and its complexity compared to that seen in previous approaches to heavy WIMP annihilation, can be understood from kinematic arguments.  Specifically, logarithms appear due to two types of phase space restrictions:
\begin{itemize}
\item {\bf Kinematic Restrictions on Final States of Massless Particles:}\footnote{Here we mean massless in perturbation theory, as relevant for scales appearing in logarithms in the weak coupling expansion. Other mass scales can appear non-perturbatively, for example, hadron mass effects in QCD event shapes have been studied in \cite{Salam:2001bd,Mateu:2012nk}.} These include kinematic restrictions via event shape observables, such as thrust, or restrictions from kinematics that force one into an endpoint region, as in $B\to X_s \gamma$, and have been discussed above. EFT descriptions in these cases typically involve three scales: the hard scale, which in our case will be $M_\chi$; the scale of the transverse momenta of particles in the jet (whose modes are called collinear), namely $M_\chi \sqrt{1-z}$; and the energy scale of soft radiation, namely $M_\chi (1-z)$. This class of problems is well understood and can be treated using $\SCETi$, discussed in \Sec{sec:scet}. The radiation in the final state is factorized into energetic modes, referred to as collinear ($c$), which comprise the dynamics of the jet, and wide angle low energy radiation, referred to as ultrasoft ($us$). Decomposed into light cone coordinates $(n\cdot p, \bar n \cdot p, p_\perp)$ (see \Eq{eq:lightcone_dec}), along the direction of the jet, these modes have momentum scaling as\footnote{Note that here and throughout the text, when we describe the scaling of modes we indicate only the parametric scaling as a function of the relevant scales in the problem, namely $M_\chi$, $\mW$, and $1-z$. Any $\cO(1)$ numerical factors do not modify this scaling, and are therefore neglected.}
\begin{align}
 p_c \sim M_\chi \big(1, \lambda^2, \lambda\big), \qquad p_{us} \sim M_\chi \big(\lambda^2, \lambda^2, \lambda^2\big)\,; \qquad \lambda=\sqrt{1-z}\,.
\end{align}
\item {\bf Exclusive Final States of Massive Particles:} These include the classic massive Sudakov form factor \cite{Collins:1989bt}, and more recently, exclusive electroweak production \cite{Chiu:2007dg,Chiu:2007yn,Chiu:2008vv,Chiu:2009mg,Chiu:2009ft,Fuhrer:2010eu}, and the exclusive approximation for DM annihilation discussed above \cite{Bauer:2014ula,Ovanesyan:2014fwa,Ovanesyan:2016vkk}. Here there are two relevant mass scales, namely the hard ($h$) scale $M_\chi$, and the scale of the massive boson, $\mW$. Problems of this type can be treated using an $\SCETii$ theory, discussed in \Sec{sec:scet}. The relevant modes in the effective theory are collinear ($c$) and soft ($s$) modes.
Decomposed into light cone coordinates (see \Eq{eq:lightcone_dec}), along the direction of the jet, these modes have momentum scaling as
\begin{align}
p_c \sim M_\chi \big(1, \lambda^2, \lambda\big), \qquad p_s \sim M_\chi \big(\lambda, \lambda, \lambda\big)\,; \qquad \lambda=\frac{\mW}{M_\chi}\,.
\end{align}
Note the distinction in scaling between the ultrasoft and soft modes. While in this case the collinear and soft modes are at the same virtuality $p^2 \sim M_\chi^2 \lambda^2$, they are separated in rapidity.\footnote{We will typically use a dimensionful rapidity, $\nu$, as in \Fig{fig:csoft_intro_b}. This should be thought of in analogy with the dimensional regularization scale, $\mu$, and is introduced in \Sec{sec:scet} where we discuss the regularization of rapidity singularities.} This explains the appearance of the rapidity axis in \Fig{fig:csoft_intro_b}.
\end{itemize}

The annihilation of WIMP DM in the endpoint region is a more complicated problem, since it involves the physics of both types of restrictions.  There is both a constraint on the final state radiation, as well as the presence of the mass scale of the electroweak bosons and the measurement of just the photon state from among the SU$(2)$ $\times$ U$(1)$ gauge bosons.  Indeed, we will find that all the scales (in both rapidity and virtuality) present in both individual cases will appear.  This is illustrated in \Fig{fig:csoft_intro_b}, which shows the modes that live at each of these mass and rapidity scales.  We will show how to factorize the dynamics at each of these scales when large hierarchies are present, thereby facilitating resummation.  The final form involves a component where the gauge boson can be treated as massless, so that the scale is set by the final state kinematic restriction, and a component where the relevant scale is $\mW$.  For example, the description of the final state jet will be split into a massless jet function, described using standard techniques in $\SCETi$, as well as a function describing the dynamics at the scale $\mW$, using $\SCETii$. 

In addition to these $\SCETi$ and $\SCETii$ ingredients, we will show that an extra mode is required to achieve the fully factorized result.  This mode has a virtuality $\mu^2 \sim \mW^2$, but it has a large momentum component along the direction of the recoiling photon of size $M_\chi (1-z)$ (the momentum scale of the soft function):
\begin{align}
p_{cs}\sim M_\chi (1-z) \big(\lambda^2, 1, \lambda\big)\,, \qquad \lambda = \frac{\mW}{ M_\chi (1-z)}\,.
\end{align}
In the case that both $M_\chi (1-z) \ll M_\chi$ and $\mW/(M_\chi (1-z))\ll 1$, these modes are neither (ultra)soft, or collinear, \emph{i.e.}, they do not appear in either $\SCETi$ or $\SCETii$ EFTs, but are instead an example of \emph{collinear-soft} modes, see \Sec{sec:scet}. Our factorization formula allows for the separate treatment of these collinear-soft modes, which allows us to resum all large logarithms, but also ensures continuity of the cross section as we move away from the endpoint region, where these modes are no longer distinguishable from the standard soft modes. It is the simultaneous presence of the scales $M_\chi (1-z)$ and $\mW$ that gives rise to the presence of these collinear-soft modes -- they would not appear if only a subset of the scales were present.\footnote{Here we have argued for the existence of collinear-soft modes based only on kinematics. The fact that these modes are actually required is also related to the fact that there are external states with electroweak charges, as will be discussed in \Sec{sec:match}.}  The structure of the results presented below shares similarities with the factorization formulae for jet substructure observables, where a measurement in addition to the mass has been performed~\cite{Bauer:2011uc,Larkoski:2014tva,Procura:2014cba,Larkoski:2015zka,Larkoski:2015kga,Pietrulewicz:2016nwo,Larkoski:2017iuy,Larkoski:2017cqq}. 

The complete description of the final state therefore combines the $\SCETi$ collinear and ultrasoft modes with the $\SCETii$ soft and collinear modes in the direction of the jet, along with the collinear-soft modes describing additional radiation along the direction of the photon.  Each of these will yield distinct functions in our factorization formula \Eq{eq:factorization}, implying that each of these functions has a clear physical origin in terms of the scales of the problem.  This seemingly complicated description is in fact a significant simplification, since the description of the dynamics at any one of these scales has been reduced to its elemental form.  In the next section, we will introduce the EFT ingredients, and in \Sec{sec:factorization} we give the technical details of the factorization.

\section{Review of Relevant Effective Field Theories}\label{sec:review}

In this section we briefly review the different EFTs that we will use, primarily to establish our notation. Our use of non-relativistic (NR) field theories will be standard in the context of QCD~\cite{Caswell:1985ui,Bodwin:1994jh,Luke:1999kz} (for reviews, see \cite{Rothstein:1999vz,Rothstein:2003mp,Hoang:2002ae}), and will focus on aspects relevant for annihilating DM (for applications of NRDM EFT to the scattering of DM with nucleon targets, see~\cite{Fan:2010gt,Fitzpatrick:2012ix,Hill:2011be,Hill:2013hoa}). As we review SCET, we will highlight necessary extensions that are perhaps less familiar.

\subsection{Non-Relativistic Dark Matter Effective Theory}\label{sec:NRDM}

In the NRDM EFT, large fluctuations of the DM field $\chi$ about a particular velocity $v$ are integrated out.  The non-relativistic DM is described by a field $\chi_v$ with a label velocity $v$, just as in heavy quark EFT~\cite{Neubert:1993mb,Manohar:2000dt}. Here $v$ is a dimensionless four vector describing the velocity of the DM, which for concreteness we will take to be $v=(1,0,0,0)$. The freedom in the choice of $v$ is represented in the EFT as a symmetry known as reparametrization invariance~\cite{Luke:1992cs,Heinonen:2012km}. The dynamics of $\chi_v$ describe the residual fluctuations of the heavy state, as in non-relativistic QCD. The EFT captures the interactions of the non-relativistic particles whose momenta $p_\mu = (E,\vec{p}\,)$ scale as soft $(M_\chi v, M_\chi v)$, ultrasoft $(M_\chi v^2, M_\chi v^2)$, and potential $(M_\chi v^2, M_\chi v)$. The ultrasoft modes describe radiation, while the soft modes give rise to the running of potentials. 

The leading power interactions of the heavy DM particle(s) with the ultrasoft radiation can be eliminated using a field redefinition $\chi_v^{(r)} \rightarrow S_v^{(r)} \chi_v^{(r)}$ \cite{Bauer:2014ula,Ovanesyan:2014fwa,Ovanesyan:2016vkk,Baumgart:2014vma, Baumgart:2014saa, Baumgart:2015bpa}, where
\begin{align}
S^{(r)}_v(x)=\bold{P} \exp \left [ ig \int\limits_{-\infty}^0 \text{d}s\, v\cdot A^a_{us}(x+sv)  T_{(r)}^{a}\right]\,,
\label{eq:SWilsonLine}
\end{align}
where $\bold{P}$ denotes path ordering, $g$ is the relevant gauge coupling, and $T^a_{(r)}$ is the generator for the DM representation $r$.  Furthermore, soft radiation is not required at the order to which we work. This implies that all dynamical radiation in NRDM is completely captured by Wilson lines along the directions of the heavy particles, greatly simplifying the field theory treatment.  

After decoupling the soft radiation, the leading power Lagrangian is given by
\begin{align}\label{eq:NRDM_Lagrangian}
{\cal L}_{\rm NRDM}^{(0)} =  \chi_v^\dagger \left(i\, v\cdot \partial + \frac{\vec{\nabla}^2}{2\,M_\chi} \right) \chi_v + \hat V\Big[\,\chi_v, \chi_v^\dagger\,\Big](m_{W,Z}),
\end{align}
which describes the interactions of the heavy particles as the sum of a kinetic and potential term. The potential $\hat V$ describes potential exchanges of the $W, Z, \gamma$, and its explicit form can be found in Ref.~\cite{Hisano:2006nn}. Note that going to higher orders and powers is well understood in the context of NRQCD (see e.g. Refs.~\cite{Manohar:1999xd,Manohar:2000hj}).  The dynamics of the heavy particles are governed by low energy matrix elements evaluated with the above Lagrangian.  Since this is a non-relativistic description, the number of heavy particles is fixed, and there exists an associated Schr\"odinger equation. These low energy matrix elements give rise to the Sommerfeld enhancement, which must be included when computing the DM cross section.  We will therefore briefly review the structure of the low energy matrix elements and the Sommerfeld factors.

\subsubsection{Sommerfeld Factors}\label{sec:sommerfeld}

Since we have chosen to work with pure wino DM, the model includes a Majorana fermion DM candidate $\chi^0$,  and an electrically charged fermion $\chi^\pm$. For the calculation of the Sommerfeld factors, we include a mass splitting, that is neglected when performing the Sudakov resummation.  Including this splitting is important as it plays a role in determining the positions of the Sommerfeld resonances. For winos, electroweak corrections yield a mass splitting $\delta \equiv M_{\chi^\pm} - M_{\chi^0} \simeq 164.4 \text{ MeV}$~\cite{Ibe:2012sx}.

In our formalism, the Sommerfeld enhancement will be captured by low energy matrix elements of the heavy annihilating particles.  As discussed  in \Sec{sec:factorization} where we derive the factorization formula, the following matrix elements appear
\begin{align}\label{eq:Lfunction}
F^{a'b'ab} = \Big\langle \big(\chi^0 \chi^0\big)_S \Big| \big( \chi_v^{a' T}\, i \sigma_2\, \chi^{b'}_v\big)^\dagger \Big|0 \Big\rangle \Big\langle 0 \Big| \big(\chi_v^{aT}\, i \sigma_2\, \chi_v^b \big) \Big|\big(\chi^0 \chi^0\big)_S \Big\rangle \, ,
\end{align}
where $T$ denotes transpose, $\sigma_2$ is the second Pauli matrix, and the external state is given by the $S$-wave combination $(\chi^0\chi^0)_S$. Here the color indices $a,b,a',b' = 1,2,3$, and we have the usual relations $\chi^0 = \chi^3$ and $\chi^\pm = (\chi^1 \mp i \chi^2)/\sqrt{2}$.
In terms of the charge eigenstates, we will find that the relevant components of $F^{a'b'ab}$ are
\begin{align} \label{eq:wavefunction}
 &  \Big\langle 0 \Big|\, \chi_v^{3\,T}\, i\sigma_2 \,\chi_v^{3\,\,} \,\Big| \big(\chi^0 \chi^0\big)_S \Big\rangle = 4 \sqrt{2} \, M_\chi\, s_{00} \,, \\[5pt]
 & \Big\langle 0 \Big|\, \chi_v^{+T}\, i\sigma_2 \,\chi_v^-\, \Big| \big(\chi^0 \chi^0\big)_S \Big\rangle= 4\, M_\chi \,s_{0\pm} \,,\nn
\label{eq:sdefn}
\end{align}
where the Sommerfeld enhancement is captured by the factors $s_{00}$ and $s_{0\pm}$, which must be evaluated non-perturbatively. In practice we do this by numerically solving the associated Schr\"odinger equation.  We summarize some of the most important aspects here; a detailed discussion can be found in Appendix A of \cite{Cohen:2013ama}. For other detailed studies of both phenomenological and formal aspects of Sommerfeld enhancement, we refer the reader to Refs.~\cite{Beneke:2012tg,Beneke:2014gja,Beneke:2016ync,Braaten:2017kci,Braaten:2017gpq}.

The first step in solving for the Sommerfeld factors is to compute a wavefunction $\big(\psi^i\big)_j$, where the index $i$ labels the asymptotic state and $j$ is the component index for the resulting solution, and the indices $i,j =1, 2$ refer to the $(00), (+-)$ states respectively. A discussion of the relevant boundary conditions can be found in Ref.~\cite{Cohen:2013ama}. Once the solutions $\psi$ have been obtained, the Sommerfeld enhancement matrix is 
\begin{align}
s_{ij} = \big(\psi^i(\infty)\big)_j\,.
\end{align}
In practice, one must choose a velocity when computing $s_{ij}$.  As is well known, the Sommerfeld enhancement saturates at low velocities, and we have checked that this occurs for the range relevant for DM annihilations, \emph{i.e.}, $v\lesssim 10^{-3}$, for the wino mass range of interest.  Therefore, we can neglect any velocity profile dependence, and treat all velocity dependence as constant for the parameter range of interest.

Once we know $s_{ij}$, using \Eq{eq:wavefunction} we can then determine the relevant components of $F^{a'b'ab}$ given in \Eq{eq:Lfunction}. From this point, the annihilation cross section can be computed as
\begin{align}\label{eq:cross1}
\sigma = \sum_{a'b'ab} F^{a'b'ab}\,\hat{\sigma}^{a'b'ab}(\zcut)\,,
\end{align}
where $\hat{\sigma}^{a'b'ab}(\zcut)$ denotes the resummed perturbative cross section as a function of $\zcut$, whose computation is the subject of this paper (see \Eq{eq:factorization} below).

As a final comment, we note that we have glossed over the fact that we will be working in a theory with a spontaneously broken gauge symmetry, as opposed to standard NRQCD.  There will several manifestations of this fact. First, and most trivially, it impacts the Sommerfeld enhancement calculation, as well as the color algebra, due to the identification of a color index for the external photon. More non-trivially, a significant portion of this paper (see in particular \Sec{sec:factorization}) will relate to the refactorization of the function describing wide angle soft radiation, including that from the incoming DM particles. This is required, since $\mW$ introduces another scale for the soft radiation in addition to that imposed by the final state measurement.

\subsection{Soft-Collinear Effective Theory}\label{sec:scet}

Soft-Collinear Effective Theory (SCET)~\cite{Bauer:2000yr, Bauer:2001ct, Bauer:2001yt} will provide the framework for describing radiation in the final state. SCET describes the dynamics of soft and collinear radiation in the presence of a hard scattering.  While originally developed for applications to QCD with massless gauge bosons, the formalism was extended to the electroweak sector with massive gauge bosons in~\cite{Chiu:2007yn,Chiu:2008vv,Chiu:2007dg}. In what follows, we will provide a brief review of the features of SCET that will be used for our heavy DM annihilation process (along with a few more general comments).

\subsubsection{Modes, Fields, and Wilson Lines}\label{sec:scet_a}

SCET is a theory of both soft and collinear particles. Collinear particles have a large momentum along a particular light-like direction, while soft particles have a small momentum, and no preferred direction.  For each relevant light-like direction, we define two reference vectors $n^\mu$ and $\bn^\mu$ such that $n^2 = \bn^2 = 0$ and $n\sdt\bn = 2$.  The typical choice of $n^\mu=(1,0,0,1)$ and $\bar n^\mu=(1,0,0,-1)$ will be used below. The freedom in the choice of $n$, as in the case of $v$ for non-relativistic EFTs, is represented in the EFT through a reparameterization invariance~\cite{Manohar:2002fd,Chay:2002vy}.   Any four-momentum $p$ can be decomposed with respect to $n^\mu$ as
\begin{equation} \label{eq:lightcone_dec}
p^\mu = \bn\sdt p\,\frac{n^\mu}{2} + n\sdt p\,\frac{\bn^\mu}{2} + p^\mu_{\perp}\
\,.\end{equation}

The SCET expansion is defined by a formal power counting parameter  $\la \ll 1$, which is determined by the measurements or kinematic restrictions imposed on the radiation.  Then the momenta for the different particles in the EFT scale as
\begin{align}
\text{Collinear}&: ~\big(n\!\cdot\! p, \bn \!\cdot\! p, p_{\perp}\big) \sim Q \,\big(\la^2,1,\la\big)\,, \nn \\[5pt]
\text{Soft}&: ~\big(n\!\cdot\! p, \bn \!\cdot\! p, p_{\perp}\big) \sim Q \,\big(\la,\la,\la\big)\,,  \\[5pt]
\text{Ultrasoft}&: ~\big(n\!\cdot\! p, \bn \!\cdot\! p, p_{\perp}\big) \sim Q \,\big(\la^2,\la^2,\la^2\big)\,,\nn
\end{align}
 where $Q$ is a typical scale of the hard interaction.  A theory with collinear and ultrasoft modes is typically referred to as $\SCETi$, while that with collinear and soft modes is referred to as $\SCETii$ \cite{Bauer:2002aj}.\footnote{In the presence of Glauber modes, soft modes are always required to run the Glauber potentials \cite{Rothstein:2016bsq,Moult:2017xpp}. Whether or not ultrasoft modes are required depends on the physical observable in question.} 

In order to expand the full theory fields around a particular direction, the momenta are decomposed into label $\tilde{p}^\mu$ and residual $k^\mu$ components
\begin{equation} \label{eq:label_dec}
p^\mu = \lp^\mu + k^\mu = \bn \sdt\lp\, \frac{n^\mu}{2} + \lp_{\perp}^\mu + k^\mu\,.
\,\end{equation}
Then for a collinear particle, $\bn \cdot\lp \sim Q$ and $\lp_{\perp} \sim \la Q$, while $k\sim \la^2 Q$ describes small fluctuations about the label momentum.  EFT modes with momenta of definite scaling are obtained by performing a multipole expansion of the full theory fields.  SCET involves independent gauge bosons\footnote{The standard formalism also incorporates collinear scalars and fermions as well.  These are not required for the calculation presented here, so we will not discuss them.} for each collinear direction $A_{n,\lp}(x)$, which are labeled by their collinear direction $n$ and their large label momentum $\lp$, as well as (ultra)soft gauge boson fields $A_{(u)s}(x)$.  Independent gauge symmetries are enforced for each set of fields. Overlap between different regions is removed by the zero-bin procedure \cite{Manohar:2006nz}. This ensures that there is no double counting of momentum regions. 

The leading power SCET Lagrangian takes the form
\begin{align} \label{eq:SCETLagExpand}
\cL_{\text{SCET}}=\cL_\hard+\cL_\dyn= \cL_\hard^{(0)}+ \cL^{(0)} + 
 {\cal L}_G^{(0)}  \,.
\end{align}
Here  $ \cL_\hard^{(0)}$  contains the hard scattering operators and is determined by an explicit matching calculation. The Lagrangian $\cL^{(0)}$ describes the universal leading power dynamics of the soft and collinear modes and can be found in Refs.~\cite{Bauer:2000yr, Bauer:2001ct, Bauer:2001yt}.  Finally, ${\cal L}_G^{(0)} $ is the leading power Glauber Lagrangian \cite{Rothstein:2016bsq}, which describes the leading power coupling of soft and collinear degrees of freedom through potential operators. We will not need to consider it in this paper.

Hard scattering operators involving collinear fields are constructed out of products of collinear gauge invariant fields~\cite{Bauer:2000yr,Bauer:2001ct}.  The gauge invariant gauge boson operator is given by
\begin{align} \label{eq:chiB}
\cB_{n\perp}^\mu(x)
= \frac{1}{g}\Bigl  [W_{n}^\dagger(x)\,i  D_{{n}\perp}^\mu W_{n}(x)\Bigr]
 \,.
\end{align}
Here $D_{n\perp}$ is the collinear gauge covariant derivative, and $W_n$ is a collinear Wilson line\footnote{Note that when the label momentum is large compared to the virtuality of the EFT modes, it is convenient to use a mixed position/momentum space representation space Wilson line, where the label is in momentum space and the residual fluctuations are in position space.  Otherwise, Wilson lines will be written in position space, \emph{e.g.} \Eq{eq:SWilsonLine}. It is also possible to formulate SCET entirely in position space, see  \emph{e.g.} Refs.~\cite{Beneke:2002ni,Beneke:2002ph}, although we will not use the position space formalism here.}
\begin{align}
W_n(x) =\left[ \, \sum\limits_{\text{perms}} \exp \left(  -\frac{g}{\bar{n}\cdot \cP}\, \bar n \cdot A_n(x)  \right) \right]\,,
\end{align}
where $\cP^\mu$ is an operator that returns the label momentum.  The collinear Wilson line, $W_{n}(x)$, is localized with respect to the residual position $x$ so that $\cB_{{n\perp}}^\mu(x)$ can be treated as local gauge boson fields from the perspective of the ultrasoft degrees of freedom.  For the leading power calculation presented here, ultrasoft and soft fields will not appear explicitly in our hard scattering operators, other than through Wilson lines via the field redefinition
\begin{align} \label{eq:BPSfieldredefinition}
\cB^{a\mu}_{n\perp}\to Y_n^{ab} \cB^{b\mu}_{n\perp}\,,
\end{align}
which is performed in each collinear sector. For a general representation, $r$, the ultrasoft Wilson line is defined by\footnote{Here we give the explicit result for an incoming Wilson line. Depending on whether particles are incoming our outgoing, different Wilson lines must be used. When done correctly, the BPS field redefinition accounts for the full path of the particles \cite{Chay:2004zn,Arnesen:2005nk}.}
\begin{align}
Y^{(r)}_n(x)=\bold{P} \exp \left [ ig \int\limits_{-\infty}^0 \text{d}s\, n\cdot A^a_{us}(x+sn)  T_{(r)}^{a}\right]\,,
\label{eq:YWilsonLine}
\end{align}
where as before $\bold P$ denotes path ordering.  This so-called BPS field redefinition has the effect of decoupling ultrasoft and collinear degrees of freedom at leading power \cite{Bauer:2002nz}.  We will also need soft Wilson lines,
\begin{align}
S^{(r)}_n(x)=\bold{P} \exp \left [ ig \int\limits_{-\infty}^0 \text{d}s\, n\cdot A^a_{s}(x+sn)  T_{(r)}^{a}\right]\,.
\end{align}

Finally, the refactorization of the soft sector (see Sec.~\ref{sec:soft_refact} below) will require the inclusion of collinear-soft modes from SCET$_+$ \cite{Bauer:2011uc,Larkoski:2014tva,Procura:2014cba,Larkoski:2015zka,Pietrulewicz:2016nwo}. Collinear-soft modes have both a collinear and soft scaling
\begin{align}
p_{cs} \sim Q\, \tilde\la\, \big(\la^2, 1, \la\big)\,,
\end{align}
where $\la$ and $\tilde{\la}$ are distinct power counting parameters. Such modes first appeared in calculations of jet substructure when multiple simultaneous measurements are made on a jet \cite{Bauer:2011uc,Larkoski:2014tva,Procura:2014cba,Larkoski:2015zka,Pietrulewicz:2016nwo}.  This introduces additional scales, implying the need for both $\la$ and $\tilde{\la}$. For contrast, the measurement of a single observable, such as the mass of a jet, only introduces a single scale; the mass can either fix the angular spread of the mode, resulting in a  collinear mode, or it can fix the energy of the mode, resulting in soft or ultrasoft modes, but it cannot fix both, as required for collinear-soft modes. In our case, the collinear-soft modes will arise due to the presence of both the mass scale of the final state $m_X$, and the mass scale of electroweak symmetry breaking $\mW$.  Our study provides a new application of collinear-soft modes.

Since the collinear-soft modes arise from a refactorization of the soft sector, they couple eikonally and their interactions can be absorbed using additional Wilson lines defined as
\begin{align}
X^{(r)}_n(x)=\bold{P} \exp \left [ ig \int\limits_{-\infty}^0 \text{d}s\, n\cdot A^a_{cs}(x+sn)  T_{(r)}^{a}\right]\,,
\end{align}
and
\begin{align}
V^{(r)}_n(x)=\bold{P} \exp \left [ ig \int\limits_{-\infty}^0 \text{d}s\, \bar n\cdot A^a_{cs}(x+s\bar n)  T_{(r)}^{a}\right]\,.
\end{align}
This notation is chosen to reflect that the $X$ Wilson lines will arise from a BPS field redefinition, similar to the $Y$ Wilson lines in $\SCETi$ (and $X$ precedes $Y$ in the alphabet), and the $V$ Wilson lines are generated by integrating out interactions with particles in the $\bar n$ direction, similar to the $W$ Wilson lines that accompany the collinear fields (and $V$ precedes $W$ in the alphabet).  As with (ultra) soft fields, at the order to which we work, collinear-soft fields will appear only in Wilson lines. For example, they will arise from the BPS field redefinition, which allows the all orders decoupling of interactions between collinear-soft and collinear particles. This is identical to the transformation in \Eq{eq:BPSfieldredefinition} but with a collinear-soft Wilson line.
For a more detailed discussion of the BPS field redefinition for collinear-soft fields, see \cite{Bauer:2011uc}.

\subsubsection{Renormalization Group Evolution}\label{sec:scet_b}

SCET allows for the resummation of large logarithms through the renormalization group (RG) evolution of matrix elements of collinear, (ultra)soft, collinear-soft fields. Since we will use both $\SCETi$ and $\SCETii$, this RG evolution can be either in virtuality, $\mu$, or rapidity, $\nu$ \cite{Becher:2011dz, Chiu:2011qc,Chiu:2012ir}. We use the regulator of~\cite{Chiu:2011qc,Chiu:2012ir}, modifying the Wilson lines as
\begin{align}\label{eq:reg_soft}
S_n(x)= \left [ \, \sum\limits_{\text{perms}} \exp \left(  -\frac{g}{n\cdot \cP} \frac{\omega\, |2\,\cP^z|^{-\eta/2}}{\nu^{-\eta/2}} n \cdot A_s(x)  \right) \right] \,,
\end{align}
\begin{align}\label{eq:reg_coll}
W_{n}(x) = \left[\, \sum\limits_{\text{perms}} \exp \left(  -\frac{g}{ \bar n\cdot \cP} \frac{\omega^2 \,|\bar n\cdot \cP|^{-\eta/2}}{\nu^{-\eta/2}}  \bar n \cdot A_{n}(x)  \right) \right] \,,
\end{align}
Here $\nu$ is a rapidity scale, analogous to $\mu$ in dimensional regularization, $\eta$ is the regulating parameter, and $\mathcal{P}^z$ returns the $z$-component of the label momentum.  This allows us to define a dimensional regularization-like RG in terms of $\nu$. Here $\omega$ is a formal bookkeeping parameter which satisfies
\begin{align}
\nu \frac{\partial}{\partial \nu} \omega^2(\nu) =-\eta~ \omega^2(\nu) \,, \qquad \lim_{\eta \to 0}\, \omega(\nu)=1\,.
\end{align}
For convenience, we set $\omega=1$ throughout our calculations since it can be trivially restored.  Rapidity divergences for the collinear-soft modes will also be regulated with the appropriately modified versions of \Eqs{eq:reg_soft}{eq:reg_coll}.

In our factorization, we will encounter functions that satisfy both multiplicative and convolutional renormalization group equations. For a function $F(\mu,\nu)$ which is renormalized by a multiplicative factor $Z_F(\mu,\nu)$, we have
\begin{align}
F^{\text{bare}}=Z_F(\mu, \nu) F(\mu, \nu)\,,
\end{align}
from which we derive the RG equations
\begin{align}
\frac{\text{d}}{\text{d} \log\mu} F(\mu,\nu) =\gamma^\mu_F(\mu,\nu) F(\mu,\nu)\,, \qquad    \frac{\text{d}}{\text{d} \log\nu} F(\mu,\nu) =\gamma^\nu_F(\mu,\nu) F(\mu,\nu)\,,
\end{align}
with
\begin{align}
\gamma^\mu_F(\mu, \nu) =-\frac{1}{Z_F(\mu, \nu)}  \frac{\text{d}}{\text{d}\log\mu}Z_F(\mu, \nu)\,, \qquad   \gamma^\nu_F(\mu, \nu) =-\frac{1}{Z_F(\mu, \nu)} \frac{\text{d}}{\text{d}\log\nu}Z_F(\mu, \nu)\,.
\end{align}
Convolutional renormalization in a variable $\tau$ takes the form
\begin{align}
F^{\text{bare}}(\tau) =\int \text{d}\tau' Z_F(\tau-\tau';\mu,\nu) F(\tau'; \mu, \nu)\,,
\end{align}
giving rise to the RG equations
\begin{align}
\frac{\text{d}}{\text{d}\log\mu} F(\tau;\mu,\nu) &=\int \text{d}\tau'\, \gamma^\mu_F(\tau-\tau';\mu, \nu) F(\tau';\mu, \nu)\,,
\end{align}
\begin{align}
\frac{\text{d}}{\text{d}\log\nu} F(\tau;\mu,\nu) &=\int \text{d}\tau'\, \gamma^\nu_F(\tau-\tau';\mu, \nu) F(\tau';\mu, \nu)\,,
\end{align}
where the anomalous dimensions are given by
\begin{align}
\gamma_F^\mu(\tau; \mu, \nu)&=-\int \text{d}\tau' \,Z^{-1}_F(\tau-\tau';\mu, \nu)  \frac{\text{d}}{\text{d}\log \mu} Z_F(\tau';\mu, \nu)\,, \\[5 pt]
\gamma_F^\nu(\tau; \mu, \nu)&=-\int \text{d}\tau' \,Z^{-1}_F(\tau-\tau';\mu, \nu)  \frac{\text{d}}{\text{d}\log\nu} Z_F(\tau';\mu, \nu)\,.
\end{align}
Convolutional RG equations are most easily treated in a conjugate space (we will use Laplace space below), in which they are multiplicative.

The RG evolution can be used to run functions from their natural scale, where all large logarithms are minimized, to an arbitrary scale.
The independence of the RG path is guaranteed by the fact that the anomalous dimensions sum to zero, schematically
\begin{align}
\sum\limits_F \gamma^F_\mu=0\,, \qquad \sum\limits_F \gamma^F_\nu=0\,,
\end{align}
where the sum is over the functions $F$ that appear in the factorization formula,
along with the fact that evolution in $\mu$ and $\nu$ commutes:
\begin{align}
\left [ \frac{\text{d}}{\text{d}\log \mu},    \frac{\text{d}}{\text{d}\log \nu} \right ]=0\,.
\end{align}
The consistency of the anomalous dimensions will provide a strong check on our calculation. We will use the path independence to choose a particularly simple path to resum all large logarithms in the EFT, see Fig.~\ref{fig:RG_path} below.

\section{Factorization Formula for the Endpoint Region}\label{sec:factorization}

In this section, we present the factorization formula for the endpoint region of heavy WIMP annihilation -- this is one of the main results of this paper. We focus here on the short-distance component of the cross section, denoted  $\hat{\sigma}(z_\text{cut})$ in \Eq{eq:cross1}. As discussed below, the long-distance contributions, \emph{i.e.}, the Sommerfeld enhancement, also arise naturally from the factorization of the matrix elements presented in this section; we refer the reader to \Sec{sec:sommerfeld} for the details of how these factors are (numerically) computed.

In \Sec{sec:factorization_formula}, we present the factorization formula, and discuss each of its components in turn.  This section is aimed at readers without a technical EFT background, and as such emphasizes the physical content of each ingredient.  In \Sec{sec:match}, we provide the technical discussion of the multi-stage matching used to derive the factorization formula, emphasizing the operator definitions for the functions and key aspects of the refactorization. Tree level and one-loop results for all functions in both the intermediate and final EFT, as well as details of the calculations can be found in \App{sec:one_loop_app}.

\subsection{Factorization Overview}\label{sec:factorization_formula}

The main result of this section is a factorization formula for the photon spectrum in the endpoint region. We find that
the differential cross section for the heavy WIMP annihilation $\chi\, \chi \to \gamma + X$ factorizes in the limit that $z\rightarrow 1$ as
\begin{mdframed}[linewidth=1.5pt, roundcorner=10pt]
\vspace{-5pt}
\begin{align}
\label{eq:factorization}
\frac{\text{d} \hat{\sigma}^{\text{LL}}}{\text{d}z}=~&H(M_\chi, \mu)\, J_\gamma(\mW,\mu, \nu)\, J_\bn(\mW,\mu, \nu)\, S (\mW,\mu, \nu) \nn \\[3pt]
&\times H_{J_\bn}(M_\chi, 1-z,\mu) \otimes H_{S}(M_\chi, 1-z,\mu) \otimes C_S(M_\chi, 1-z,\mW,\mu, \nu)  \,,\\[-10pt] \nn
\end{align}
\end{mdframed}\vspace{0.0cm}
where $z$ is defined in \Eq{eq:EGamma}, and we use $\otimes$ to denote a convolution between the functions in the second line, as explained in detail below. Here $\hat{\sigma}$ denotes the short-distance component of the cross section in \Eq{eq:cross1} with suppressed initial/final state indices. The indices are to be contracted with the matrix element $F^{a'b'ab}$ in \Eq{eq:Lfunction}. This function also arises naturally when considering the factorization of the cross section, but to keep our discussion focused on the Sudakov factors, we will not consider $F^{a'b'ab}$ in this section. When we present the final cross section results in Sec.~\ref{sec:resum}, $F^{a'b'ab}$ will be included. The LL superscript indicates that this factorization as written is only true for the leading logarithmic contributions. Beyond this order additional functions are required, as will be described in this section.

\begin{figure}
\begin{center}
\includegraphics[width=0.75\columnwidth]{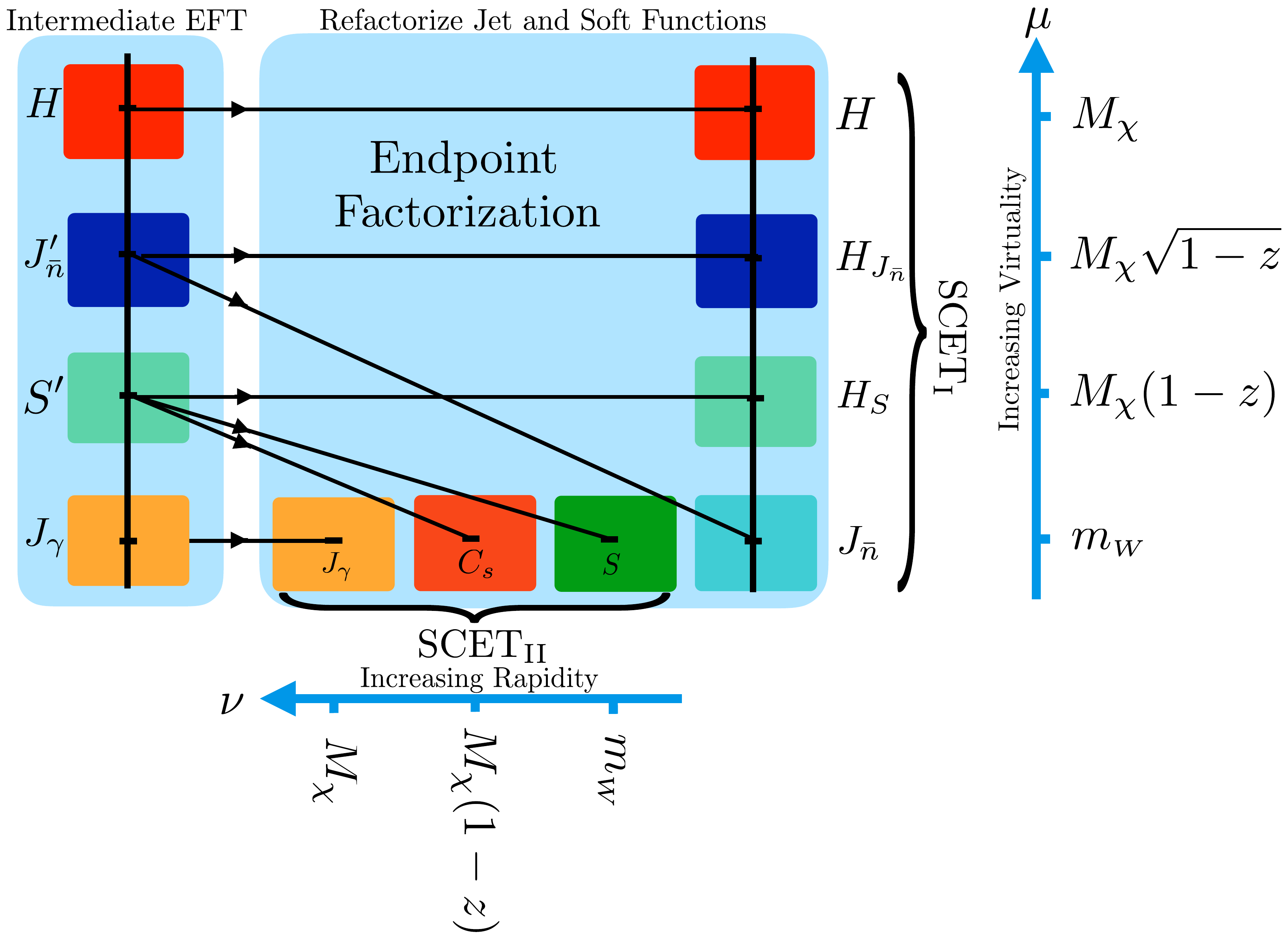}
\end{center}
\caption{A schematic of the multistage matching procedure used to derive the factorization formula for heavy WIMP annihilation in the endpoint region. The jet and soft functions appearing in the first stage of matching are refactorized into components that depend either on $\mW$, or on the phase space restriction implemented by $z$. }
\label{fig:matching}
\end{figure}

The iterative matching procedure used to derive this result is shown schematically in \Fig{fig:matching}.  In the first stage, we match onto a standard SCET theory, leading to the standard factorization into functions that describe the underlying hard scattering ($H$), the collinear radiation along the jet ($J_\bn^\prime$) and photon ($J_\gamma$) directions, and soft radiation ($S^\prime$).  In the second stage, we match onto a (electroweak symmetry breaking) theory with massive soft and collinear modes.  In particular, this manifests as a refactorization of the soft function $S^\prime$ into the functions $H_S$, $S$ and $C_S$, and of the jet function $J_\bn^\prime$ into the functions $H_{J_\bn}$ and $J_\bn$ -- these additional functions are described below.

The final EFT description consists of a collection of independent sectors, each corresponding to the functions appearing in the factorization formula \Eq{eq:factorization}. The procedure for factorizing the full cross section into these functions is illustrated in \Fig{fig:matching}.  The interpretation of each of the functions is discussed in the following, which is organized by the characteristic scale $\mu$ for these sectors. In particular, we separate it into two classes of functions, namely those that depend on $\mW$, and those that do not.

\vspace{10pt}
\noindent The first class of functions depend on scales far above the electroweak scale, $\mu \gg \mW$, and are thus independent of electroweak symmetry breaking effects.  
\begin{itemize}
\item[$\bullet$] $H(M_\chi, \mu)$ describes the underlying hard scattering process of $\chi\, \chi \to \gamma\, \gamma, \gamma\,Z$, and includes contributions from modes with virtuality $\mu \sim M_\chi$.   
\item[$\bullet$] $H_{J_\bn}(M_\chi, 1-z,\mu)$ describes collinear radiation along the jet direction with virtuality $\mu \sim M_\chi \sqrt{1-z}$ such that it contributes to the final state mass.
\item[$\bullet$] $H_{S}(M_\chi, 1-z,\mu)$ describes soft wide-angle radiation with virtuality $\mu \sim M_\chi (1-z)$ such that it contributes to the final state mass.  \end{itemize}
The second class of functions encode electroweak symmetry breaking effects, and have $\mu \sim \mW$, so that the gauge fields are treated as massive.  Additionally, these functions all depend on a rapidity renormalization scale $\nu$.  
\begin{itemize}
\item[$\bullet$] $J_\gamma(\mW,\mu, \nu)$ describes the final state photon, and results purely from modes with energy $E_\gamma$ and virtuality $\mu \sim \mW$.  This function receives only virtual corrections, since the final state is exactly specified.
\item[$\bullet$] $S(\mW,\mu, \nu)$ describes homogenous soft radiation with virtuality $\mu \sim \mW$ such that it does not contribute to the final state mass. \item[$\bullet$] $C_S(M_\chi, 1-z,\mW,\mu, \nu)$ describes radiation that is simultaneously soft and collinear to the photon direction. The momentum for this radiation has  collinear scaling, virtuality $\mu \sim \mW$, and contributes to the final state mass. 
\item[$\bullet$] $J_\bn(\mW,\mu, \nu)$ describes collinear radiation along the jet direction with virtuality $\mu \sim \mW$ such that it does not contribute to the final state mass.  
\end{itemize}
This full factorization simultaneously involves functions from NRDM, $\SCETi$, $\SCETii$, and SCET$_+$, and resummation requires RG evolution in both virtuality and rapidity.

For the analysis here, we will be interested in resumming only the leading logs (LL).  Our approach to the factorization persists at higher logarithmic order. However, as written, the refactorization of the soft function $S^\prime$ is only valid at LL order.  The origin of this effect, as well as the mechanism for disentangling these scales, is akin to the case of non-global logarithms (NGLs), and is discussed in \Sec{sec:soft_refact}. 

While we will present the factorization formula using the concrete example of an SU$(2)_W$ triplet of Majorana fermions, this choice merely affects the particular spin and charge structure of the operators involved, and as such the main features of the factorization and the relevant modes are universal.  The same factorization will also apply, \emph{e.g.} to the annihilation of heavy SU$(2)_W$ doublets or the decay of a heavy dark bound state~\cite{Mitridate:2017oky}.  Furthermore, some of the structure is generic to situations where event shape observables are measured on jets of massive radiation, and thus variants of \Eq{eq:factorization} may find applications for future high energy colliders~\cite{Bauer:2016kkv, Chen:2016wkt}.

\subsection{Multi-Stage Matching}\label{sec:match}

In this section, we discuss the derivation of the factorization formula given in \Eq{eq:factorization}. In \Sec{sec:sudakov} we present the first stage of matching, including the structure of the hard scattering operators, the factorization of the Hilbert space and measurement function for soft and collinear modes, and the matrix element definitions of the functions. In \Sec{sec:jet_refact} and \Sec{sec:soft_refact} we present the details for the second stage of matching, namely the refactorization of the collinear and soft sectors. For the soft sector, we give a detailed discussion of the relevant soft and colinear-soft modes.

\subsubsection{Soft-Collinear Factorization}\label{sec:sudakov}

We begin by determining the hard scattering Lagrangian in SCET, denoted by $\cL_\text{hard}$ in \Eq{eq:SCETLagExpand}. This is done through matching the full theory consisting of the Standard Model and an SU$(2)_W$ triplet of Majorana fermions onto SCET, and is identical to the fully exclusive case~\cite{Bauer:2014ula, Ovanesyan:2014fwa, Ovanesyan:2016vkk}. The Lagrangian describing the hard scattering is
\begin{align}\label{eq:hardscattering}
 {\cal L}^{(0)}_\text{hard} &=\sum_{r=1}^2 C_r(M_\chi,\mu)\, \cO_r \nn \\
 &= \sum_{r=1}^2 \: C_r(M_\chi,\mu)\: \left( \chi_v^{aT} \,i\sigma_2\, \chi_v^b  \right) \left( Y_r^{abcd} \,\cB_{n\perp}^{ic}\, \cB_{\bar n \perp}^{jd}    \right) i\, \epsilon^{ijk} (n-\bar n)^k  \,,
\end{align}
with the Wilson line structures
\begin{align}\label{eq:softwilsonY}
Y_1^{abcd}=\delta^{ab} \Big(Y_n^{ce} Y_{\bar n}^{de}\Big)\,, \qquad Y_2^{abcd} =\Big(Y_v^{ae}Y_n^{ce}\Big)\Big(Y_v^{bf} Y_{\bar n}^{df}\Big)\,,
\end{align}
obtained through the BPS field redefinition.
The Wilson coefficients $C_r$ are IR finite, and independent of the scale $\mW$. Performing a tree-level matching at the scale $\mu \sim M_\chi$, we find
\begin{align}
C_1(\mu)=-C_2(\mu)=-\pi\, \frac{\aW(\mu)}{M_\chi}\,.
\end{align}
The $C_r(\mu)$ encode the underlying hard scattering process and determine the hard function $H(M_\chi, \mu)$ appearing in our factorization formula, as will be defined in \Eq{eq:hardfunction}. Together with $\cL_\text{dyn}$ in \Eq{eq:SCETLagExpand}, the hard scattering operators in \Eq{eq:hardscattering} describe the annihilation at scales $\mu \lesssim M_\chi$. 

The factorization formula for the cross section for $\chi\, \chi \to \gamma + X$ depends on the squared matrix elements of these hard scattering operators.  For contrast, in the exclusive case there are only virtual contributions, and thus the factorization can be done at the level of the amplitude~\cite{Bauer:2014ula, Ovanesyan:2014fwa, Ovanesyan:2016vkk}. In the present analysis, there are both real and virtual contributions that are sensitive to $\mW$ as well as the scales imposed by the endpoint restrictions though $z$. These low-energy dynamics are not yet factorized at this stage.

First, we consider the factorization of the Hilbert space for the final state $|X\rangle$.  Since the soft and collinear modes are decoupled, the final state can be written as
\begin{align}
\big|X\big\rangle=\big|X_s\big\rangle\,\big |X_c \big\rangle.
\end{align}
Next, we expand out the contributions to the final state mass $m_X^2$, 
\begin{align}\label{eq:measure}
(1-z)&=\frac{1}{4\,M_\chi^2} \,m_X^2 \nn=\frac{1}{4\,M_\chi^2} \left( \sum\limits_{\,\,i\in X_s} p_i^\mu + \sum\limits_{\,\,i\in X_c}p_i^\mu \right)^2 \nn \\
&=\frac{2}{4\,M_\chi^2} \left( \sum\limits_{\,\,i\in X_s} p_i^\mu \right )\cdot \left( \sum\limits_{\,\,i\in X_c}p_i^\mu \right)  +\frac{1}{4\,M_\chi^2} \left( \sum\limits_{\,\,i\in X_c}p_i^\mu \right)^2 +\cO(\lambda^4) \nn \\
&=\frac{2}{4\,M_\chi}  \sum\limits_{\,\,i\in X_s} \bar{n}\cdot p_i +\frac{1}{4\,M_\chi^2} \left( \sum\limits_{\,\,i\in X_c}p_i^\mu \right)^2 +\cO(\lambda^4) \nn \\
&\equiv (1-z_s) +(1-z_c)+\cO(\lambda^4)\,,
\end{align}
which shows that contributions to the final state radiation from soft and collinear modes can be separated to leading power.
The last line in \Eq{eq:measure} defines the contributions from the soft and collinear modes as $(1-z_s)$ and $(1-z_c)$, respectively, and demonstrates the factorization of the final state restriction.  This allows us to define soft and collinear measurement operators, $\widehat\cM_s$ and $\widehat\cM_c$, as
\begin{align}
\widehat\cM_s\,\big  |X_s\big\rangle&=\frac{1}{2\,M_\chi}  \sum\limits_{\,\,i\in X_s} \bar{n}\cdot p_i\, \big |X_s\big\rangle\,, \qquad
\widehat\cM_c\, |X_c\rangle= \frac{1}{4\,M_\chi^2} \left( \sum\limits_{\,\,i\in X_c}p_i^\mu \right)^2 \big|X_c\big\rangle\,.
\end{align}
These measurement operators can be written in terms of the energy momentum tensor of either the full or effective theories~\cite{Sveshnikov:1995vi,Korchemsky:1997sy,Lee:2006nr}.  Here their role will simply be to return the value of the observable for a particular perturbative state in momentum space.

With the above ingredients, we can algebraically manipulate the cross section into a factorized form involving matrix elements of either soft or collinear fields. These matrix elements will be coupled together both through color indices and the convolutions that are present as a result of enforcing the measurements. This procedure is standard (see, \emph{e.g.} the review~\cite{iain_notes}) and we simply give the final result.  At the first stage of matching, the differential cross section with factorized dynamics in SCET is given in terms of the hard function $H$, the jet functions $J_\bn^\prime$ and $J_\gamma$ for $X$ and the photon respectively, and the soft function $S^\prime$ as
\begin{align}\label{eq:matching1}
\frac{\text{d}{\hat \sigma}}{\text{d}z}&=\int \!\text{d}z_s \,\text{d}z_c \, \delta (1+z-z_c-z_s)\,H_{ij}( M_\chi)\, J_\bn^\prime(M_\chi, 1-z_c, \mW)\,  J_\gamma(\mW)\, S_{ij}^\prime(1-z_s, \mW) \nn \\[5pt]
&\equiv H_{ij}( M_\chi)\, J_\gamma(\mW) \, J_\bn^\prime(M_\chi, 1-z, \mW) \otimes S_{ij}^\prime(1-z, \mW) \,,
\end{align}
where we have suppressed the color indices and the dependence on the RG scales $\mu$ and $\nu$ for simplicity. As in \Eq{eq:factorization}, we have used $\otimes$ to denote the convolution in $z$. The convolution arises due to the fact that the total invariant mass of the final state is a sum over the soft and collinear sectors, see \Eq{eq:measure}. 

The functions labeled with a superscript prime are those that require further factorization.  Note that the $J_\bn^\prime$ and $S^\prime$ functions still depend on both the $\mW$ and $(1-z)$ scales. This complication did not occur for the fully exclusive case, where the above factorization was sufficient since there is no intermediate scale $(1-z)$. The refactorization of the jet and soft functions  will be discussed in \Sec{sec:jet_refact} and \Sec{sec:soft_refact}.

Next, we provide field-theoretic definitions for the functions appearing in \Eq{eq:matching1}. The hard function is defined in terms of the Wilson coefficients of the hard scattering operators in \Eq{eq:hardscattering} as 
\begin{align}\label{eq:hardfunction}
H_{ij}= C_i^* C_j\,.
\end{align}
The soft function is a vacuum matrix element of the soft Wilson lines $Y_r$ in \Eq{eq:hardscattering},
\begin{align}\label{eq:softfunction_def}
S^\prime_{ij}(1-z_s, \mW,\mu, \nu)= \Big\langle 0 \Big | \, {\rm\bar T}\, Y_i^{ \dagger}(0) \ \delta\Big( (1-z_s) -\widehat{\cM}_s \Big) {\rm T}\, Y_j(0)\,  \Big |  0 \Big \rangle ,
\end{align}
where the color indices are suppressed, T and ${\rm \bar T}$ denote time ordering and anti-time ordering respectively, and the $Y_r$ factors are the products of Wilson lines defined in \Eq{eq:softwilsonY}.  The components of the soft function with explicit color indices are
\begin{align}
 S_{11}^{\prime \ a'b'ab} &= \bigg \langle 0 \, \bigg | \left(Y_n^{3k}Y_{\bar n}^{dk}\right)^{\!\dagger}\!\!\,\delta\Big( (1-z_s)-\widehat{\cM}_s \Big)  \left(Y_n^{3j}Y_{\bar n}^{dj}\right)\! \bigg| \, 0 \bigg \rangle\, \delta^{a'b'} \delta^{ab} \,, \nn \displaybreak[4]\\[5pt]
 S_{22}^{\prime \ a'b'ab} &= \bigg \langle 0 \, \bigg |  \left(Y_n^{3f'}Y_{\bar n}^{dg'}Y_v^{a'f'}Y_v^{b'g'}\right)^{\!\dagger}\!\!\,\delta\Big( (1-z_s) -\widehat{\cM}_s\Big)  \left(Y_n^{3f}Y_{\bar n}^{dg}Y_v^{af}Y_v^{bg}\right)\! \bigg| \, 0 \bigg \rangle \,, \nn \\[5pt]
S_{12}^{\prime \ a'b'ab} &= \bigg \langle 0 \, \bigg |  \left(Y_n^{3k}Y_{\bar n}^{dk}\right)^{\!\dagger}\!\!\,\delta\Big( (1-z_s)-\widehat{\cM}_s\Big)  \left(Y_n^{3g}Y_{\bar n}^{df}Y_v^{ag}Y_v^{bf}\right)\!  \bigg| \, 0 \bigg \rangle\, \delta^{a'b'}\,, \nn\\[5pt]
S_{21}^{\prime \ a'b'ab}&=  \bigg \langle 0 \, \bigg |  \left(Y_n^{3f'}Y_{\bar n}^{dg'}Y_v^{a'f'}Y_v^{b'g'}\right)^{\!\dagger}\!\!\,\delta\Big( (1-z_s) -\widehat{\cM}_s\Big)  \left(Y_n^{3k}Y_{\bar n}^{dk}\right)\!  \bigg| \, 0 \bigg \rangle\,  \delta^{ab}\,,
\label{eq:SprimeDefs}
\end{align}
where the color indices are explicit, but we have dropped the arguments and scale dependence of the functions for simplicity. Here, as well as in the expressions below, we keep the time ordering convention and the dependence on $x=0$ implicit. Note that the color index $3$ corresponds to the photon final state.  

The indices $i,j$ in the hard and soft functions span the space of the operators given in \Eq{eq:hardscattering} and are contracted with each other as $H_{ij} S_{ij}^\prime$. To reduce the number of indices appearing in later formulas, we introduce the following notation:
\begin{align}\label{eq:1indexnotation}
\begin{array}{ccc}
H_{1} \equiv H_{11} \,, \qquad \qquad & H_{2} \equiv H_{22} \,, \qquad \qquad & H_{3} \equiv H_{12} = H_{21} \,, \\[7pt]
S_{1}^\prime \equiv S_{11}^\prime \,, \qquad \qquad  & S_{2}^\prime \equiv  S_{22}^\prime \,,  \qquad \qquad  &S_{3}^\prime \equiv  S_{12}^\prime + S_{21}^\prime \,,
\end{array}
\end{align}
such that $H_{ij} S_{ij}^\prime = H_i S_i^\prime$.

The jet functions for the recoiling jet $X$ and the photon are color-singlet matrix elements of collinear fields. Explicitly, we have 
\begin{align}
 J_{\bar n}^{\prime \ dd'}\big(M_\chi, 1-z_c,\mW,\mu \big) &=  \Big\langle 0 \Big| B_{\bar n\perp}^{d'}
\, \delta\Big( (1-z_c) - \widehat{\cM}_c\Big) \delta\big(2\, M_{\chi} - \bar n\cdot\mathcal{P}\big)\, \delta^2\big( \vec{\mathcal{P}}_{\perp} \big) \, 
 B_{\bar n\perp}^d \Big| 0 \Big\rangle \,, \nn \\[8pt]
 J_{\gamma} \big(\mW,\mu,\nu\big) &=  \Big\langle 0 \Big| B_{n \perp}^c \Big| \gamma \Big\rangle \Big\langle \gamma \Big| B_{n \perp}^c    \Big| 0 \Big\rangle \, ,
 \label{eq:JDefs}
\end{align}
where $\vec{\mathcal{P}}_\perp$ returns the perpendicular component of the label momentum.   As discussed above, this is the final form for $J_\gamma$, but the jet function for $X$ will require further factorization -- we turn to this in the next section.

\subsubsection{Refactorization of the Jet Sector}\label{sec:jet_refact}

As currently formulated, the jet function $J_\bn^\prime$ in \Eq{eq:matching1} results from dynamics at both the scale $M_\chi \sqrt{1-z}$ and the scale $\mW$. To be able to resum logarithms of $\mW/(M_\chi \sqrt{1-z}\,)$, which can become large as we move towards the endpoint, we must factorize these two scales. This factorization is similar to that performed in the fully exclusive case, where one is separating $M_\chi$ from $\mW$ using a hard matching coefficient that is independent of the IR scale $\mW$, along with jet and soft functions which describe the dynamics at the scale $\mW$. Here we will write the jet function $J_\bn^\prime(M_\chi,1-z, \mW,\mu, \nu) $ as a hard matching coefficient $H_{J_{\bar{n}}}(M_\chi, 1-z, \mu)$, and a jet function $J_{\bar{n}}(\mW, \mu,\nu)$. 

The collinear state, $X_c$, factorizes into two types of collinear modes as 
\begin{align}\label{eq:jetrefactH}
\big|X_c\big\rangle= \big|X_{c_z}\big\rangle\, \big|X_{c_{\scriptscriptstyle W}}\big\rangle\,,
\end{align}
where $c_z$ is in the Hilbert space containing the collinear modes that are sensitive to the measurement enforced as a function of $z$, while $c_{\scriptscriptstyle W}$ is in the Hilbert space that contains the modes with mass $\mW$.  This follows from the same logic as the standard hard-collinear factorization.
Here the $c_z$ modes which contribute to the jet mass measurement have the standard scaling for an $\SCETi$ collinear mode associated with the mass measurement,
\begin{align}
p_{c_z}\sim M_\chi \big(\la^2, 1, \la\big)\,, \qquad \la= \sqrt{1-z}\,.
\end{align}
The modes sensitive to the $\mW$ scale are standard $\SCETii$ collinear modes at the scale $\mW$, with scaling
\begin{align}
p_{c_{\scriptscriptstyle W}}\sim M_\chi \big(\la^2, 1, \la\big)\,, \qquad \la={\mW \over M_\chi}\,,
\end{align}
and do not contribute to the mass of the final state at leading power.

The factorization of the measurement function is trivial since, at leading power, the low-energy collinear modes have an invariant mass $p_{c_{\scriptscriptstyle W}}^2 \sim \mW^2 \ll M_\chi^2 (1-z)$, and do not contribute to the mass of the final state. We therefore only have
\begin{align}\label{eq:jetrefactM}
\widehat\cM_{c_z} \big|X_{c_z}\big\rangle&= \frac{1}{4\,M_\chi^2} \left( \sum\limits_{\,\,i\in X_{c_z}}p_i^\mu \right)^2   \big|X_{c_z}\big\rangle\,.
\end{align}
 
The separation of collinear modes through Eqs.~(\ref{eq:jetrefactH}) and~(\ref{eq:jetrefactM}) allows us to fully factorize the jet function as 
\begin{align}\label{eq:jet_refact}
J'_\bn\big(M_\chi, 1-z, \mW, \mu,\nu\big) = H_{J_\bn}\big(M_\chi, 1-z, \mu\big)\, J_\bn\big(\mW, \mu, \nu\big)+\cO\left( \frac{\mW}{M_\chi \sqrt{1-z}} \right)\,.
\end{align}
This factorization is a power expansion in $\mW/(M_\chi \sqrt{1-z}\,)$. The matching coefficient $H_{J_\bn}$ can be evaluated in the unbroken theory with massless electroweak bosons, and is IR finite due to the mass measurement. The dependence on the electroweak scale is completely captured by the function $J_\bn(\mW, \mu, \nu)$. 

\subsubsection{Refactorization of the Soft Sector}\label{sec:soft_refact}

Next, we turn to the refactorization of the soft function $S^\prime$.  The goal is to have separate EFTs for the dynamics at scales $\mu \sim M_\chi (1-z)$ and $\mu\sim \mW$.  Comparing to the discussion of the jet refactorization in the previous section, the physics of the soft sector is more interesting, as logarithms due to collinear-soft modes appear.

Consider the possible classes of soft modes with virtuality $\mu^2 \sim \mW^2$.  The virtuality of the soft modes with scaling $p_{S^\prime} \sim M_\chi (1-z) (1,1,1)$ can be lowered uniformly to yield modes with $p_{S} \sim (\mW, \mW, \mW)$.  When acting on these states, the measurement function in \Eq{eq:softfunction_def} can be expanded as
\begin{align}
\delta \Big( (1-z_s)  - \widehat{\cM}_s \Big)= \delta \left( 1-z_s \right) +\cO\left( \frac{\mW}{M_\chi (1-z)^2}\right)\,.
\end{align}
We conclude that these soft modes do not contribute to the measurement, which allows a simplification of the operator structure. As an explicit example, the soft functions $S^\prime_{1}$ and $S^\prime_{2}$ become
\begin{align}\label{eq:S11collapse}
S_{1}^{\prime \ a'b'ab}  \ \to \ S_{1}^{a'b'ab} 
&= \delta^{a'b'} \delta^{ab} \,\delta\!\left( 1-z_s \right)\,,\\[5pt]
S_{2}^{\prime \ a'b'ab}  \ \to \ S_{2}^{a'b'ab} 
&=\delta(1-z_s) \bigg( \delta^{a'b'}  \Big\langle 0 \Big | \, \big(Y_{\bar{n}}^{\dagger}\big)^{e3}\, Y_v^{ae} Y_v^{bf} Y_{\bar{n}}^{3f} \,\Big | 0 \Big\rangle \nn \\
&\qquad\qquad\quad + \delta^{ab}  \Big\langle 0 \Big | \, \big(Y_{\bar{n}}^{\dagger}\big)^{e3}\, Y_v^{a'e} Y_v^{b'f} Y_{\bar{n}}^{3f} \,\Big | 0 \Big\rangle \bigg) \,,
\end{align}
where we have used the unitarity of the Wilson lines.  These new functions $S_{i}$ are now independent of $\mW$.  Physically, the simplification (collapse) of the Wilson lines occurs because the measurement operator has been expanded away, implying that the refactorized soft functions are now inclusive.  However, we are still specifying the photon as the final state, and therefore violate the assumptions of the Bloch-Nordsieck~\cite{Bloch:1937pw} or KLN~\cite{Kinoshita:1962ur,Lee:1964is} theorems, as originally pointed out in \cite{Ciafaloni:1998xg,Ciafaloni:1999ub,Ciafaloni:2000df}.  This explains why the Wilson lines in $S_{2}$ do not completely simplify, as compared to $S_{1}$ where the Wilson line dependence has collapsed to the unit operator leaving behind only color and kinematic factors.

\begin{figure}
\begin{center}
\subfloat[]{\label{fig:soft_refactorization_a}
\includegraphics[width=7.2cm]{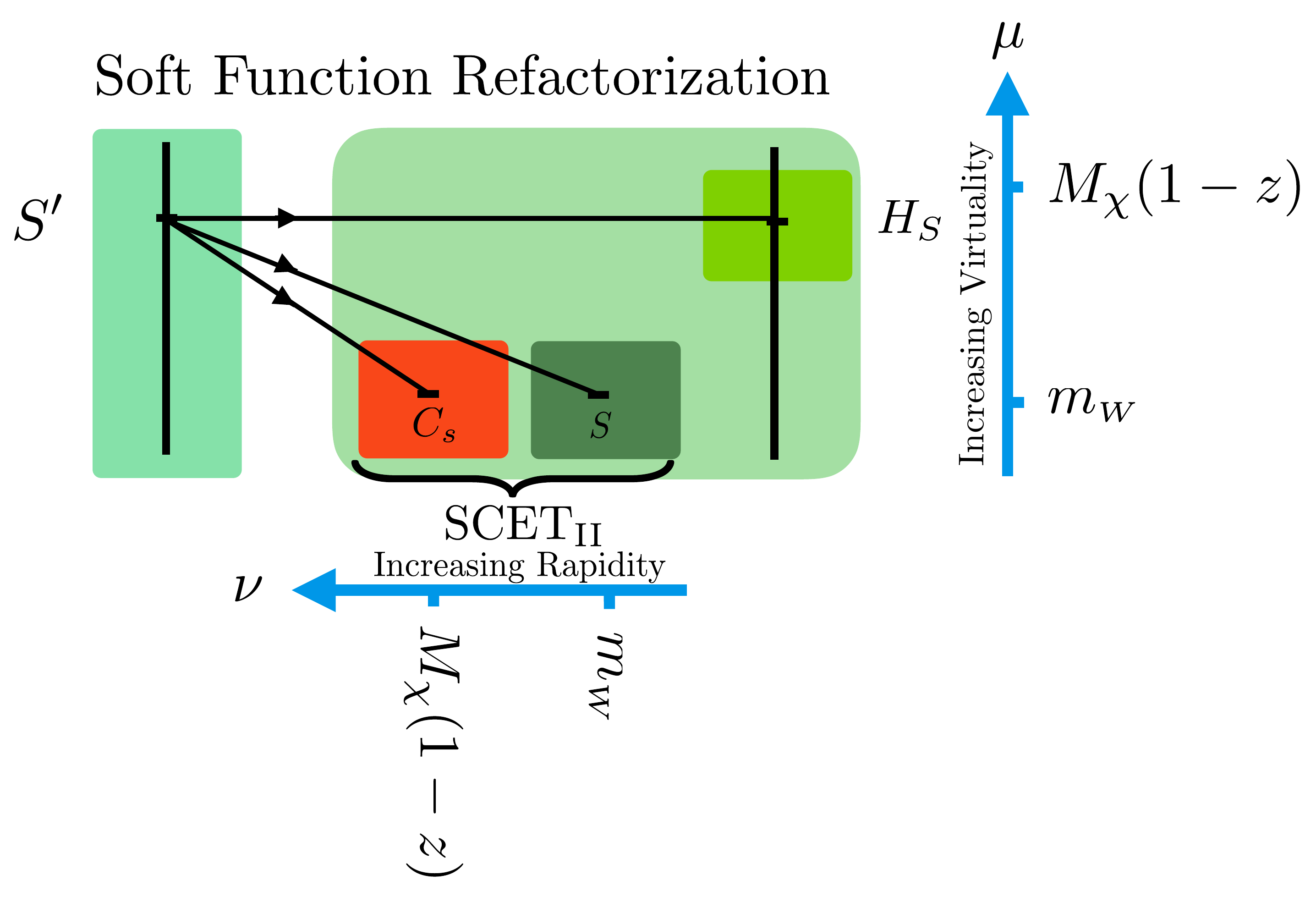}    
}\hspace{5pt}
\subfloat[]{\label{fig:soft_refactorization_b}
\includegraphics[width=7.0cm]{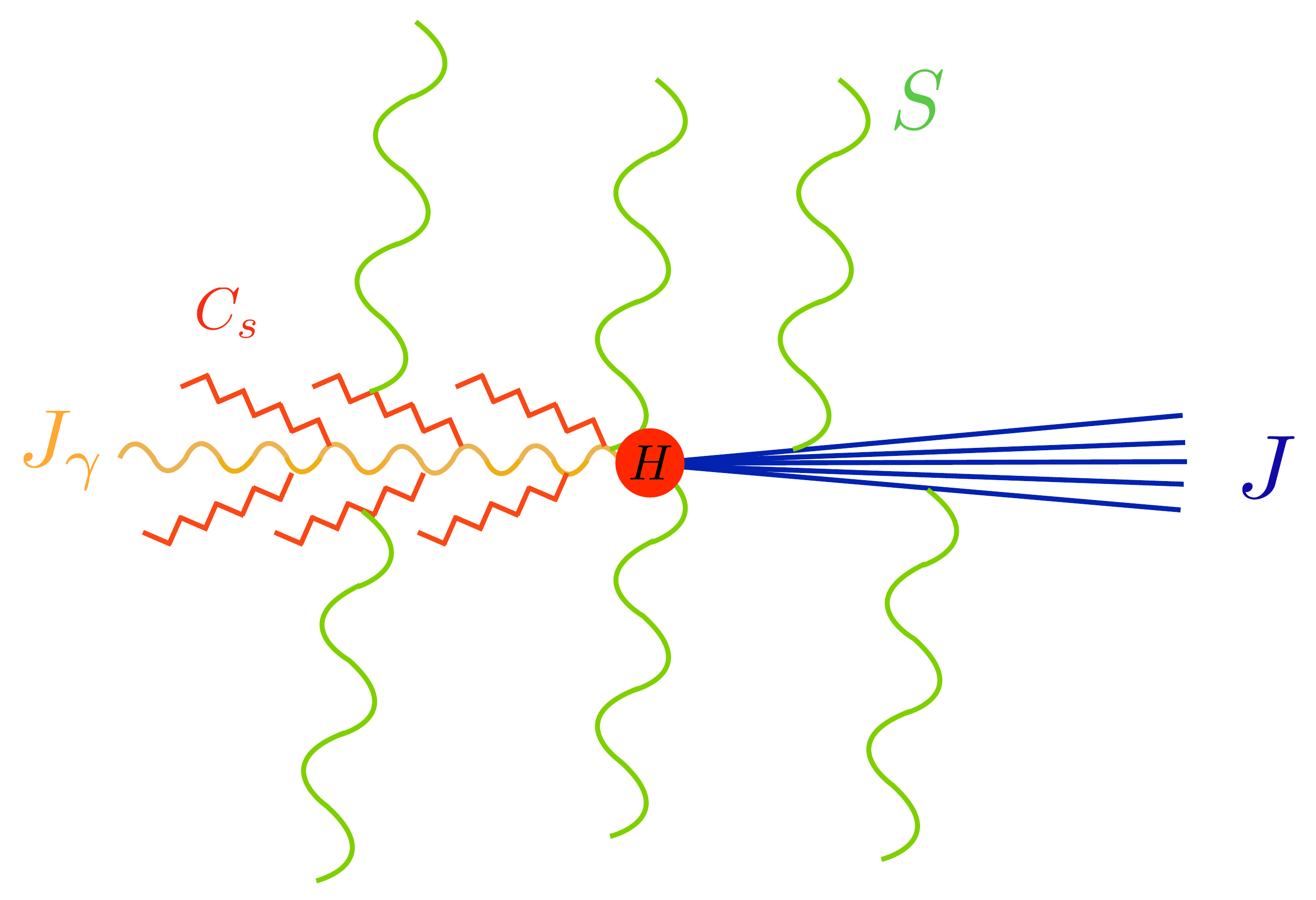}
}
\end{center}
\caption{(a) The refactorization of the soft function into collinear-soft and soft functions at different rapidity scales. (b) The relevant modes required for the refactorization of the soft function are collinear-soft modes, which are collimated along the direction of the photon, and wide angle soft modes, which are isotropic.
}
\label{fig:soft_refactorization}
\end{figure}

It is clear from the collapse of the Wilson lines that the modes $p_S$ are not sufficient to complete the picture. In particular, the divergences associated with $\mW$, for example in $S^\prime_{1}$, should be reproduced after factorization, but the function $S_{1}$ in \Eq{eq:S11collapse} does not have such a divergence. Interestingly, however, there is a second possibility for lowering the virtuality of the soft modes down to the scale $\mW$: keep their momentum component along the photon direction fixed, but decrease their angle (increase their collinearity) with respect to the photon. These modes are shown schematically in \Fig{fig:soft_refactorization}. Such modes then have the scaling 
\begin{align}
p_{c_S}\sim M_\chi (1-z)\big (1,\lambda^2,  \lambda\big)\,, \qquad \lambda = \frac{\mW}{ M_\chi (1-z)}\,.
\end{align}
These modes have a virtuality $\mu^2 \sim \mW^2$, but, like the original soft modes with momentum $p_S$, have a large momentum component $M_\chi (1-z)$.  This is an example of the collinear-soft modes discussed in \Sec{sec:scet}, which arise from the simultaneous presence of the two scales $M_\chi (1-z)$ and $\mW$.

These arguments imply that the Hilbert space of the soft sector factorizes into soft modes with uniform scaling and collinear-soft modes as 
\begin{align}
\big|X_{S^{\prime}} \big\rangle =  \big| X_S \big\rangle\, \big| X_{c_S} \big\rangle \,.
\end{align}
The soft modes do not contribute to the measurement, while the collinear-soft modes are sensitive to a measurement function
\begin{align}\label{eq:CSmeasurement}
\widehat{\cM}_{c_S}\, \big| X_{c_S} \big\rangle = \frac{1}{2\,M_\chi}  \sum\limits_{\,\,i\in X_{c_S}} \bn\cdot p_i \, \big| X_{c_S} \big\rangle \, .
\end{align}

The most interesting aspect of these collinear-soft modes is that they contribute to the measurement of the final state mass through their large component, which is independent of their virtuality.  To our knowledge, this type of collinear-soft mode has not previously appeared in the literature.  For example, in the case of thrust~\cite{Farhi:1977sg} or other $\SCETi$ event shapes, the definition of the measurement guarantees that it is always the small component of the momentum of a particle that is measured.

Using the measurement function in \Eq{eq:CSmeasurement}, the Wilson lines that make up the collinear-soft function do not collapse, but are instead expanded assuming the momentum scaling for the collinear-soft modes.  Since the collinear-soft modes are boosted along the photon's direction $n$, the $v$ and $\bn$ Wilson lines appear to collapse down to the $\bn$ direction. The collinear-soft function is therefore given as a product of Wilson lines
\begin{align}
C_S\big(M_\chi, 1-z_c, \mW,\mu,\nu\big)=\Big\langle0\Big| \big(X_n V_n\big)^\dagger \delta \Big((1-z_c) -\widehat M_{c_S}\Big) X_n V_n\Big|0 \Big\rangle\,,
\end{align}
where the $X$ and $V$ Wilson lines were defined in \Sec{sec:scet}, and implicitly include rapidity regulators. We have suppressed color indices for simplicity. Explicit expressions with color indices will be given below.  To regulate rapidity singularities for the collinear-soft Wilson lines, we do not expand the regulator,  using the full $|2\,k_z|^{-\eta}$ dependence.  Performing the naive power expansion of the regulator yields unregulated rapidity divergences in the collinear-soft sector.  This choice of regulator defines the zero-bin structure \cite{Manohar:2006nz} of the collinear-soft sector, and we find that non-trivial zero-bins are present, which must be correctly  incorporated to remove overlap. This is described in more detail in \App{sec:one_loop_app}. Strict power counting can be preserved by introducing a boost parameter $\beta$, and using the regulator $|2\,k_z|^{-\eta}\,\, \to\,\,|k_++\beta\, k_-|^{-\eta}$ \cite{Rothstein:2016bsq}. 

Having discussed the modes that are required to describe the physics at the scale $\mW$, we next explain how to refactorize the soft function into a matching coefficient that describes the dynamics at the scale $M_\chi (1-z)$, and a soft and jet function that describe the dynamics at the scale $\mW$. This is more complicated than for the jet function. The complication emerges due to the existence of a hierarchy in energy but not in angle for the homogeneous soft modes that live at the scales $\mW$ and $M_\chi (1-z)$. Hence, any emission at the scale $M_\chi (1-z)$, which can be at an arbitrary angle, eikonalizes from the perspective of the emissions at the scale $\mW$, and is described as a new Wilson line source.  In this way, an infinite number of operators is generated in the matching (although only a finite number appear at any order in $\alpha_{\scriptscriptstyle W}$). This situation is familiar from the case of NGLs~\cite{Dasgupta:2001sh}, where there exist multiple hierarchical soft scales. Due to the generation of these new sources, the resummation of NGLs is governed by the non-linear BMS equation \cite{Banfi:2002hw}. In the present case, however, the measurement function for the modes at the scale $\mW$ is expanded, and what is generated are Bloch-Nordsieck or KLN violating NGLs. We are not aware of these appearing previously in the literature. While it is possible that these take a simple form, or completely cancel, they first contribute at NLL order.  Here we restrict ourselves to LL accuracy, and so we will not discuss this higher order structure any further.   We leave the study of them using existing formulations of NGLs in factorization~\cite{Caron-Huot:2015bja,Larkoski:2015zka,Becher:2015hka,Becher:2016mmh,Larkoski:2016zzc} for future work.

At LL order, we do not need to consider the generation of additional Wilson lines in the matching. Nevertheless, the general structure of the refactorized function can become complicated since four Wilson lines appear in each of the soft and collinear-soft functions, and mixing between these color structures can be generated beyond tree-level.  In the most general case, the refactorization takes the form
\begin{align}\label{eq:genrefactorize}
S^{\prime \ aba'b'}_{i} \big(M_\chi, 1-z, \mW,\mu,\nu \big) = H_{S, ij}\big(M_\chi, 1-z, \mu\big) \,& \Big[ C_{S}\big(M_\chi, 1-z, \mW , \mu, \nu\big)  S\big( \mW, \mu\big) \Big]^{ab a'b'}_j \nn \\[5pt]
&\times \left[ 1+ \cO\left( \frac{\mW}{M_\chi (1-z)} \right) \right ]\,.
\end{align}
This refactorization, along with the scales of each of the functions, is shown in \Fig{fig:soft_refactorization}.
The functions $C_S$ and $S$ each carry eight color (triplet) indices. Two of these sixteen color indices are identified as carrying the quantum number of the photon, and the rest are contracted as to leave the overall indices $ab a'b'$, which are contracted with the initial state wavefunction factors. In \Eq{eq:genrefactorize}, we are using the notation introduced in \Eq{eq:1indexnotation}; the index $i$ enumerates the color structures in the soft function before refactorization, \emph{i.e.}, $i=1,2,3$. The index $j$ sums over the color structures in the soft function after refactorization. 

Instead of writing down a complete basis, we construct the color structures explicitly from the top down by explicitly refactorizing the soft function $S^\prime$.  This requires us to supplement the operators written in \Eq{eq:SprimeDefs} above with those that appear at one-loop, to ensure that the RG closes. Fortunately, only a limited basis of color structures is required at this order. The color structures are derived in \App{sec:app_refac_basis}. Here we simply state the results for the refactorization of the soft functions. We denote the combined collinear-soft and soft functions as 
\begin{align}
{\tilde S}_j^{\,aba'b'} = \big(C_S\, S\big)_j^{aba'b'}\,,
\label{eq:defSTilde}
\end{align}
and they are
\begin{align}\label{eq:cs_s_combined}
\!\!\!\!\tilde S^{\,aba'b'}_{1}  &=\Big \langle 0\Big| \left(X_n^{3f'}V_{n}^{dg'}\right)^{\dagger}\!\! \, \delta \Big((1-z_s)-\widehat M_{c_S}\Big)\left(X_n^{3f}V_{n}^{dg}\right)\!\, \Big|0 \Big \rangle\, \delta^{f'g'}\delta^{a'b'}\delta^{fg}\delta^{ab}\,,  \nn \\[8pt]
\!\!\!\!\tilde S_{2}^{\,aba'b'}&=  \Big \langle 0 \Big| \left( X_n^{ce} V_{n}^{Ae} \right)^\dagger \,\delta \Big( (1-z_s)-\widehat M_{c_S}\Big)X_n^{c'g'} V_{n}^{A'g'} \Big|0 \Big \rangle          \Big \langle 0 \Big|   \left[ S_n^{3c} S_n^{3c'}S_v^{a'A'}S_v^{aA}\right] \Big|0 \Big \rangle\,   \delta^{bb'}\,,  \nn \\[8pt]
\!\!\!\!\tilde S_{3}^{\,aba'b'} &=   \Big \langle 0 \Big|  \left( X_n^{ce} V_{n}^{B'e} \right)^\dagger \,\delta \Big((1-z_s)-\widehat M_{c_S}\Big)X_n^{c'g'} V_{n}^{A'g'}   \Big|0 \Big \rangle   \nn \\[5pt]
&\quad \times \left( \Big \langle 0 \Big|\left[ S_n^{3c} S_n^{3c'}S_v^{a'A'}S_v^{b'B'}\right] \Big|0 \Big \rangle  \,\delta^{ab} + \Big \langle 0 \Big|\left[ S_n^{3c} S_n^{3c'}S_v^{aA'}S_v^{bB'}\right] \Big|0 \Big \rangle  \,\delta^{a'b'} \right)  \, .
\end{align}
Here we have made the color structure explicit, but we have dropped the arguments and scale dependence of the functions for simplicity.
The collinear-soft function reproduces the $\mW$ dependent IR divergences of the soft function. Additionally, for the RG to close we will need the following operator 
\bea\label{eq:S3}
\tilde S^{\,aba'b'}_4 = \Big  \langle 0 \Big| \left(X_n^{3f'}V_{n}^{df'}\right)^{\dagger}\!\!\,\delta \Big((1-z_s)-\widehat M_{c_S}\Big)\left(X_n^{3f}V_{n}^{df}\right)\! \Big|0 \Big \rangle\, \delta^{a'a}\delta^{b'b}\,,
\eea
which has a vanishing tree-level matching coefficient, but will appear in the mixing that results as we RG evolve the functions. The refactorized functions ${\tilde S}^{\,aba'b'}_{1} $ and ${\tilde S}^{\,aba'b'}_{4} $ have a trivial soft sector, while the functions ${\tilde S}_{2}^{\,aba'b'} $ and ${\tilde S}_{3}^{\,aba'b'}$ have non-trivial collinear-soft and soft components. The final result is the factorization formula in \Eq{eq:genrefactorize} with index $j$ summed over $j=1,2,3,4$.  In Sec.~\ref{sec:LL_resum} the hard coefficients $H_S$ from tree-level matching will be given explicitly.  

\section{Leading Log Resummation for the Endpoint Region}\label{sec:LL_resum}

Having stated the factorization formula, and discussed the physical intuition that underlies it, this section tackles the resummation of large logarithms of $\mW/(M_\chi(1-z))$, $\mW/(M_\chi \sqrt{1-z})$, and $\mW/M_\chi$. In \Sec{sec:consis}, we present the one-loop anomalous dimensions obtained by computing the real and virtual corrections to the factorized functions presented in the previous section. We also check consistency conditions for these anomalous dimensions (namely that they sum to zero), thus verifying our factorization formula at the one-loop level.  In \Sec{sec:resum},  we describe a simplified resummation path sufficient for LL order and then solve the RGEs and collect all the resummation factors necessary for obtaining the final resummed cross section. The culmination of this work is \Eq{eq:resummed}. Explicit calculations are given in \App{sec:one_loop_app}. In \Sec{sec:limits}, we demonstrate that our result recovers both the exclusive and inclusive limits.

\subsection{One-Loop Anomalous Dimensions and Factorization Consistency}\label{sec:consis}

In the results for the anomalous dimensions presented below, we only keep the double log pieces that are required for resummation at LL accuracy.
The hard function $H_i(M_\chi, \mu)$ only has a $\mu$ anomalous dimension, 
\bea\label{eq:hardAD}
\gamma_{\mu, ij}^H = - 8\,C_A\, \ralpha \log \left(\frac{\mu^2}{(2\,M_\chi)^2\, z}\right) \delta_{ij} \,,
\eea
where $C_A$ is the SU$(2)_W$ quadratic Casimir invariant for the adjoint representation (explicitly $C_A=2$), $i,j = 1,2,3$, and the structure of the RGE is diagonal. Here, and throughout this section we will use $\ralpha=\aW/(4\pi)$ to simplify the results. Furthermore, we can set $z\to 1$ to leading power, so that the hard function is independent of the infrared measurement. The same anomalous dimension, but derived at the level of the amplitude, was obtained for exclusive heavy WIMP annihilation~\cite{Bauer:2014ula,Ovanesyan:2014fwa,Ovanesyan:2016vkk}.

The photon jet function $J_\gamma(\mW, \mu, \nu)$ consists of only virtual diagrams, and is computed in the broken theory. An example diagram is 
\begin{align}
\fd{4cm}{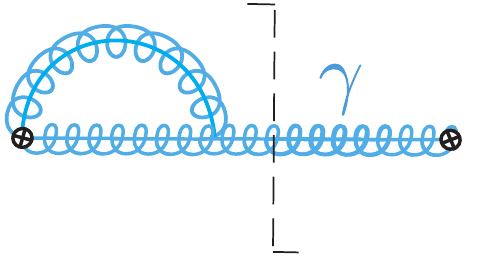} \ \ .
\end{align}
Here the dashed line indicates the final state cut, which puts the single identified photon on shell.
We find that the $\mu$ and $\nu$ anomalous dimensions are given by
\bea\label{eq:gamma_AD}
\gamma_{\mu}^{J_\gamma} =  8\,C_A\, \ralpha\log \left(\frac{\nu}{2\,M_\chi}\right)\,, \qquad \gamma_{\nu}^{J_\gamma} =   8\,C_A\, \ralpha\log\left(\frac{\mu}{\mW}\right)\,.
\eea

For the recoiling jet function $J_{\bn}(\mW, \mu, \nu)$, the low scale matrix element is fully inclusive.  Examples of real and virtual diagrams are
\begin{align}
\fd{4cm}{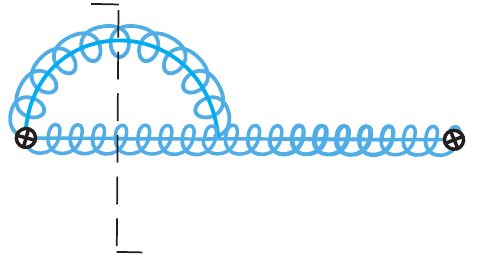} \ \ + \ \ \fd{4cm}{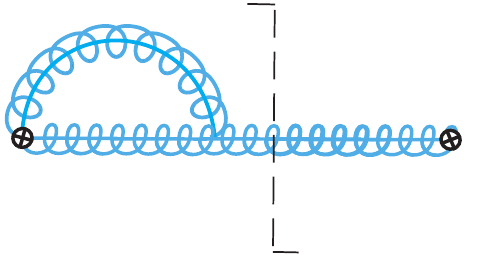} \ \ .
\end{align}
Due to its fully inclusive nature, we find that it has no anomalous dimension in $\mu$ or $\nu$. Instead, these dependences are entirely captured by the matching coefficient $H_{J_\bn}(M_\chi, 1-z, \mu)$, which is described by the same diagrams but at the high scale. The dashed line again represents the final state cut, which at NLO can contain one or two particles.
Since the one-loop correction to the jet function is a plus distribution, the RG evolution takes a simpler form in Laplace space. We will use $s$ to denote the Laplace variable conjugate to $M_\chi (1-z)$. We find its anomalous dimension to be
\bea\label{eq:hardjet_AD}
\gamma^{H_{J_\bn}}_{\mu}= 8\, C_A\,\ralpha \log \left(\frac{\mu^2\, s}{2\,M_{\chi}} \right) \,.
\eea

For the soft function, the relevant one-loop diagrams are represented by
\begin{align}
\fd{3.3cm}{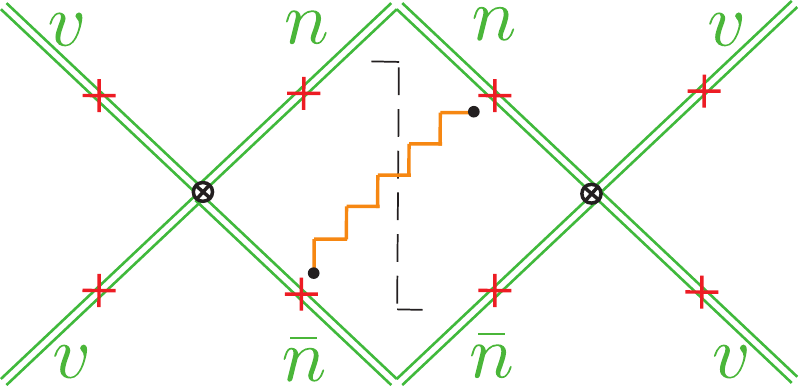}\,,
\end{align}
where the electroweak boson can attach to any of the crosses, and the double lines denote Wilson lines. We have drawn the two $v$ Wilson lines, which correspond to the annihilating heavy WIMPs, as distinct directions for visual clarity.  The collinear-soft function has a similar structure, except the incoming Wilson lines are contracted to lie in the same direction 
\begin{align}
\fd{4cm}{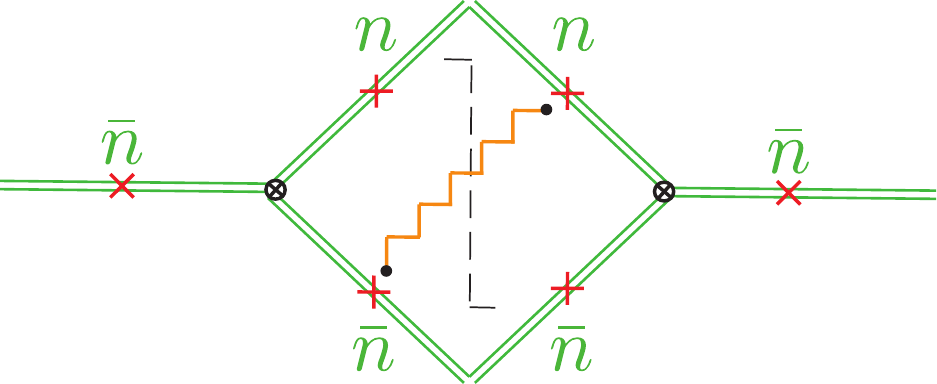}\,.
\end{align} 
As discussed in \Sec{sec:soft_refact}, the general case is complicated by a proliferation of color structures that mix beyond tree-level. For simplicity, we will consider, by top-down construction, only the functions that appear in our analysis at LL order. The $\mu$ RGE for the $\tilde S$ functions is a matrix equation
\begin{align}
\frac{\text{d}}{\text{d} \log\mu} \tilde S =\hat \gamma_\mu^{\tilde S} \,\tilde S\,,
\end{align}
where $\tilde S$ denotes the vector $\tilde S_i$. The explicit form of the anomalous dimension matrix at one-loop is given by
\begin{align}\label{eq:Stilde13RGE}
\hat \gamma_\mu^{\tilde S}=4\,C_A\,  \ralpha \left( \begin{array}{cccc}
-2  \log \nu\, s  \ & \  0 &\ 0\ & 0 \\[5pt]
0 \  \ & \ 3 \log \mu \,s-2\log \nu\, s  \ & 0\ &- \log \mu\, s \\[5pt]
- 2\log \mu \,s \  \ & \ 0& 3 \log \mu \,s-2\log \nu \,s &\ 0 \\[5pt]
0 \  \ & \ 0& \ 0&\ -2  \log \nu\, s 
\end{array} \right)\,,
\end{align}
which exhibits a non-trivial mixing structure. The $\nu$ RGE is given by
\begin{align}\label{eq:StildeNuRGE}
\frac{\text{d}}{\text{d} \log\nu} \tilde S =\hat \gamma_\nu^{\tilde S}\, \tilde S\,,
\end{align}
where the matrix is diagonal
\begin{align}\label{eq:StildeNuRGEresult}
\hat \gamma_\nu^{\tilde S}=  -8 \,C_A\, \ralpha \log \left(\frac{\mu}{\mW}\right) \id\,.
\end{align}

The interpretation of the scales appearing in the function $\tilde S= C_S S$ requires some care since this is a combined object. While both the $C_S$ and $S$ functions have a natural scale $\mu=\mW$ (see the $\nu$ anomalous dimension given in \Eq{eq:StildeNuRGEresult}), the scale $\mu=1/s$ appears in the logarithms of the $\mu$ anomalous dimension in \Eq{eq:Stilde13RGE}. This can be understood from the consistency of the RG, since the $\mu$ running of $C_S$ and $S$ must combine to yield the natural scale of $H_S$, namely $\mu = 1/s$. Despite its confusing appearance, this appearance of $1/s$ provides a non-trivial check on our refactorization.

One further important feature of the anomalous dimensions in \Eq{eq:StildeNuRGE} is that at LL order, all rapidity anomalous dimensions vanish for $\mu=\mW$. We will exploit this feature in \Sec{sec:resum} by choosing a resummation path where all rapidity evolution is done at the scale $\mu=\mW$, eliminating the need for a non-trivial rapidity evolution. 

For the matching coefficients $H_{S,ij}$ of the soft sector we have
\begin{align}
{\text{d} \over \text{d} \log\mu} {H_{S,ij}} = \gamma^{H_S}_{\mu, jk}\, {H_{S,ik}} \, ,
\end{align}
where the explicit results at one-loop order are
\begin{align}\label{eq:HS_AD}
{\text{d} \over \text{d} \log \mu} H_{S,11} &=0 \,,  \\[5pt]
{\text{d} \over \text{d} \log\mu} H_{S,22} &=  - 12\,C_A\, \ralpha \log (\mu\, s)\,  H_{S,22} \, , \qquad
{\text{d} \over \text{d} \log\mu} H_{S,24} =   4\,C_A\, \ralpha  \log (\mu\, s)\,  H_{S,22} \,,\nn\\[5pt]
{\text{d} \over \text{d} \log\mu} H_{S,31} &=   8\,C_A\, \ralpha  \log (\mu \,s) \, H_{S,33}  \, , \qquad
{\text{d} \over \text{d} \log\mu} H_{S,33} =  - 12\,C_A\, \ralpha \log (\mu\, s)\, H_{S,33} \, . \nn
\end{align}

Now we are in the position to verify our factorization formula by checking consistency relations among the anomalous dimensions. For the anomalous dimensions of the functions before the refactorization of the jet and soft functions, we have the relations
\begin{align}\label{eq:ADcheck1}
\gamma^{J_\gamma}_\nu + \frac13\, \gamma^{S^\prime}_{\nu, ii} = 0\,, \nn \\[5pt]
\frac13\, \gamma^H_{\mu, ii} + \gamma^{J_\gamma}_\mu + \gamma^{{J^\prime_\bn}}_\mu +\frac13\, \gamma_{\mu, ii}^{S^\prime} = 0 \, ,
\end{align}
which involves the anomalous dimensions for the soft and jet functions before refactorization, given by
\begin{align}\label{eq:stage1_AD}
\gamma^{S^\prime}_{\mu, ij} &= - 8\,C_A\, \ralpha \log(\nu\, s) 
\delta_{ij} \,,  \nn \\[5pt]
\gamma^{S^\prime}_{\nu, ij } &= -8\, C_A\, \ralpha \log \left( \mu \over \mW \right) \delta_{ij} \, , \nn \\[8pt]
 \gamma^{{J^\prime_\bn}}_\mu & = 8\, C_A\, \ralpha \log \left(\frac{\mu^2\, s}{2\,M_{\chi}} \right) \,.
\end{align}
As in the case of the hard function, the RG structure for the soft functions $S^\prime_i$ is diagonal. Using the anomalous dimensions in Eqs.~(\ref{eq:hardAD}), (\ref{eq:gamma_AD}), and (\ref{eq:stage1_AD}), one can check that the relations in \Eq{eq:ADcheck1} are indeed satisfied.

For the anomalous dimensions after refactorization, we have the consistency relations
\begin{align}
 \gamma^{{J^\prime_\bn}}_\mu &= \gamma^{H_{J_\bn}}_{\mu} \,, \nn \\[10pt]
\frac13\, \gamma^{S^\prime}_{\mu, ii}\, \delta_{kl}&=  \gamma^{{\tilde S}}_{\mu, kl} + \gamma^{H_S}_{\mu, lk} \,, \nn \\[10pt]
\frac13\,  \gamma^{S^\prime}_{\nu, ii}\, \delta_{kl} &=  \gamma^{{\tilde S}}_{\nu, kl} \,,\end{align}
where $k,l  = 1,2,3,4$.  One can check that these relations are satisfied using Eqs.~(\ref{eq:hardjet_AD}), (\ref{eq:Stilde13RGE}),  (\ref{eq:StildeNuRGE}), (\ref{eq:HS_AD}), and (\ref{eq:stage1_AD}).

\subsection{Analytic Resummation Formula}\label{sec:resum}

\begin{figure}
\begin{center}
\includegraphics[width=0.55\columnwidth]{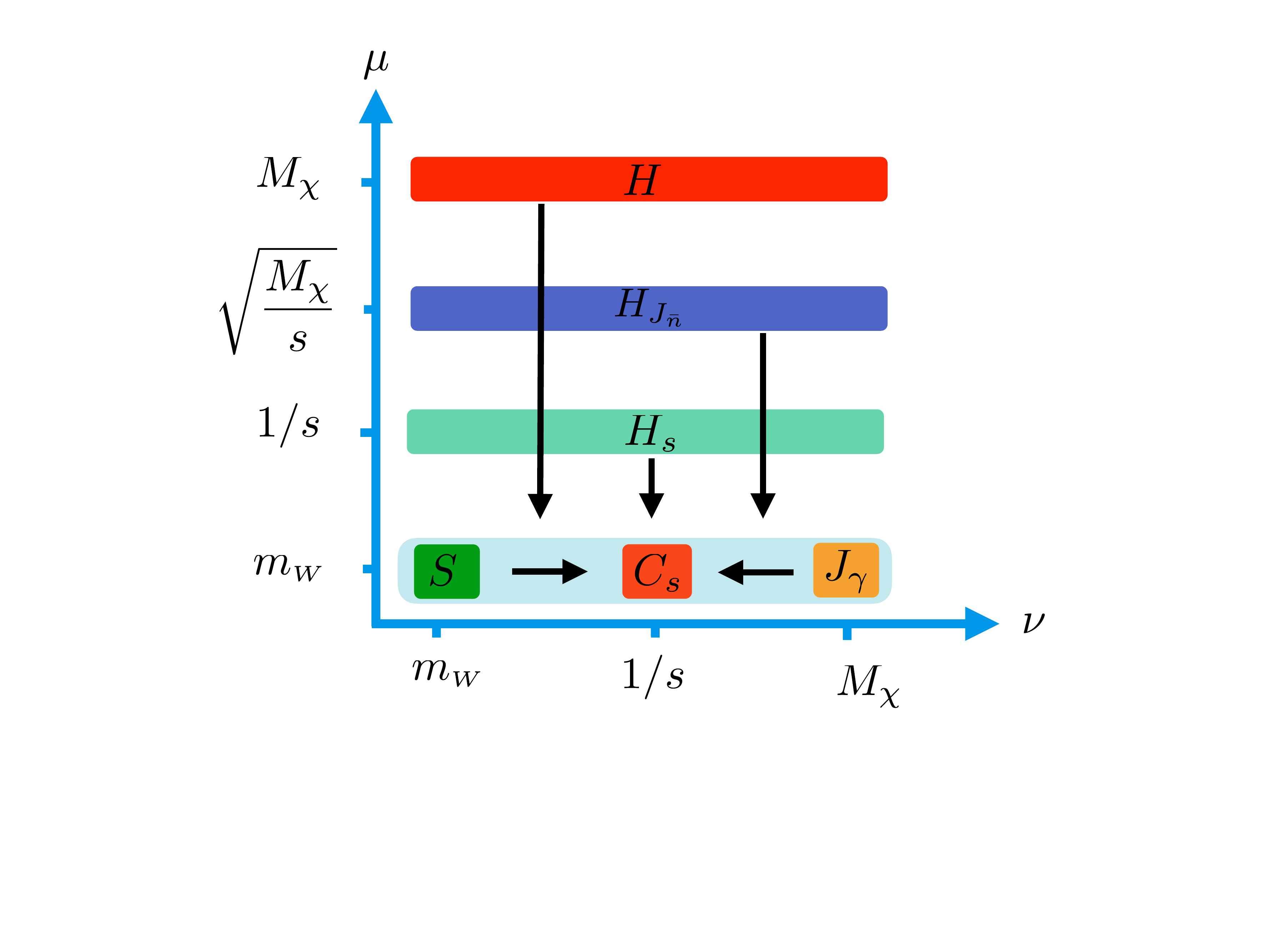}
\end{center}
\caption{A schematic of the resummation path in the $(\mu, \nu)$ plane used to perform the resummation. We choose to run all functions to $(\mu, \nu)=(\mW,1/s)$. This particular choice of path eliminates the need to separately run the soft and collinear-soft functions in rapidity at LL order. This independence in rapidity at the scale $\mW$ is depicted by the light blue box.} 
\label{fig:RG_path}
\end{figure}

We now have all the necessary ingredients to provide an analytic expression for the resummed spectrum at LL accuracy.   As discussed in \Sec{sec:scet}, the resummation can be simplified by making a judicious choice of path in the $(\mu, \nu)$ plane.  Our choice is illustrated in \Fig{fig:RG_path}. 

Due to the refactorization of the soft function $S^\prime$ into the soft and collinear-soft functions, each of which have a complicated color structure, and whose renormalization will involve color mixing, the renormalization group structure is quite complicated for a generic path. However, this can be avoided by noting that at $\mu=\mW$, the rapidity anomalous dimensions of the soft and collinear-soft functions given in \Eq{eq:StildeNuRGE} vanish at LL order.  Hence, we take the functions at their natural scale -- $H$ with $\mu=M_\chi$, $H_{J_\bn}$ with $\mu=\sqrt{2\,M_\chi/s}$, and $H_S$ with $\mu=1/s$ -- and run them all down to $\mu=\mW$.  Finally, at $\mu=\mW$, we can then trivially run the soft, collinear-soft, and jet functions to the same rapidity.  This choice of path provides a significant simplification since we can simply compute the $\mu$ anomalous dimensions for the functions $H$, $H_{J_\bn}$ and $H_S$.  Beyond LL accuracy, this is no longer possible, and the full factorization that we have developed in this paper must be utilized.

There is one additional subtlety regarding the evolution structure that has been glossed over in \Fig{fig:RG_path}, but that requires care to reproduce the correct behavior in the limit $z\to 1$. Recall that in deriving our factorization, which is summarized in \Fig{fig:matching}, we have assumed the hierarchy
\begin{align}\label{eq:scales_ref}
M_\chi (1-z) \gg M_\chi \sqrt{1-z} \gg \mW\,,
\end{align}
which allows us to factorize the dynamics at the scale $\mW$ from that at the scales $M_\chi \sqrt{1-z}$ and $M_\chi (1-z)$. However, at $z = 1-\mW/(2\,M_\chi)$ the soft scale hits the scale $\mW$ and at $z=1-\mW^2/(2\, M_\chi)^2$ the jet scale hits the scale $\mW$. In this small region near the endpoint, our EFT is technically speaking invalidated. Physically, the constraint on the final state becomes so restrictive that the jet is composed of a single boson.  Due to the intrinsic IR cutoff set by electroweak symmetry breaking, it is unphysical for these scales to go below the scale $\mW$. Instead, we must introduce a $\Theta$-function in the RG evolution, which ensures that the running only contributes in the region where the scales are above $\mW$. As we will see, with this modification, our EFT will correctly transition to the exclusive endpoint calculation. This choice of scales is implemented in $(1-z)$ space. Therefore, in Laplace space we take arbitrary scales $\mu_{H_{J_\bn}}$ and $\mu_{H_S}$ ($\mu_H$ can be set to its canonical value since it is $z$ independent) transform to cumulative space where we can implement our scale setting as a function of $(1-\zcut)$, and then differentiate to obtain the resummed spectrum. Note that in the following, we will always use $\zcut$ when discussing the cumulative space, as per the definition of \Eq{eq:cumulative_def_intro}.

The RG equations can now be solved in the usual manner. For the hard functions $H$ and $H_{J_\bn}$, we derive the evolution kernels
\begin{align}\label{eq:Uhard}
U_H \big( 2\,M_\chi  , \mW \big) &= \exp\left( - 8\,C_A\, \ralpha \log^2 \left( \mW \over  2\, M_\chi  \right) \right)  \,,   \\[10pt]
U_{H_{J_\bn}} \big( \mu_{H_{J_\bn}},  \mW \big) &=\exp\left( 8\,C_A\, \ralpha \left( \log^2 \left(  \mW \sqrt{s \over 2\, M_\chi} \right) - \log^2 \left(  \mu_{H_{J_\bn}} \sqrt{s \over 2\, M_\chi} \right) \right) \right) \,,\nn
\end{align}
where the first and second arguments of the kernels denote the scales we are running between, starting from the natural scale of the relevant function, and ending at $\mu \sim \mW$.
For the hard function $H_S$, we need to solve the system of RG equations in \Eq{eq:HS_AD} in order to run from $\mu = \mu_{H_S}$ down to $\mu = \mW$.  We find that
\begin{align}\label{eq:Usoftcollinear}
H_{S,11}(\mW)  &=  H_{S,11}(\mu_{H_S})  \,, \nn \\[10pt]
\left( \begin{array}{c} H_{S,33} (\mW) \\[5pt] H_{S,31} (\mW) \end{array} \right) &= \left( \begin{array}{cc}
U_{H_S}(\mu_{H_S},\mW) \  \ & \  0 \\[5pt]
2\,(1- U_{H_S}(\mu_{H_S},\mW))/3 \  \ & \ 1
\end{array} \right) \left( \begin{array}{c}  H_{S,33} (\mu_{H_S}) \\[5pt] H_{S,31} (\mu_{H_S}) \end{array} \right) \,, \nn \\[10pt]
\left( \begin{array}{c} H_{S,22} (\mW) \\[5pt] H_{S,24} (\mW) \end{array} \right) &= \left( \begin{array}{cc}
U_{H_S}(\mu_{H_S},\mW) \  \ & \  0 \\[3pt]
(1- U_{H_S}(\mu_{H_S},\mW))/3 \  \ & \ 1
\end{array} \right) \left( \begin{array}{c}  H_{S,22} (\mu_{H_S}) \\[5pt] H_{S,24} (\mu_{H_S}) \end{array} \right)  \,,
\end{align}
where
\begin{align}
 U_{H_S}\big(\mu_{H_S},\mW\big) = \exp \left( -6\, C_A \,\ralpha \left( \log^2 \left( \mW \,s \right) - \log^2 \left( \mu_{H_S} \,s \right)\right) \right)\,.
\end{align}
These kernels resum all leading double logarithms. 

To put together the resummed cross section, we need the tree-level values of the hard function $H$, see \Eq{eq:1indexnotation}, 
\begin{align}\label{eq:treevaluesH}
H_1^{\tree} = \frac{\pi^2\, \aW^2}{M_{\chi}^2} \,,\qquad\qquad  H_2^\tree = \frac{\pi^2\, \aW^2}{M_{\chi}^2} \,,\qquad\qquad  H_3^\tree =-\frac{\pi^2\, \aW^2}{M_{\chi}^2} \,,
\end{align}
the hard-soft functions $H_S$, see \Eq{eq:genrefactorize}, 
\begin{align}\label{eq:treevaluesHS}
H_{S,11}^\tree =1 \,, \qquad\qquad H_{S,22}^\tree = 2 \,, \qquad\qquad H_{S,33}^\tree =1 \,, 
\end{align}
and collinear-soft functions ${\tilde S}$, see \Eq{eq:defSTilde},
\begin{align}\label{eq:treevaluesStilde}
\begin{array}{ll}
\left( {\tilde S}_1^{aba'b'} \right)^\tree = \delta^{a'b'} \delta^{ab}\,, \qquad\qquad&  \left(  {\tilde S}_2^{aba'b'} \right)^\tree= \delta^{a3} \delta^{a' 3} \delta^{bb'} \,, \\[5pt]
\left( {\tilde S}_3^{aba'b'} \right)^\tree= \delta^{a3} \delta^{b3} \delta^{a'b'} + \delta^{a'3} \delta^{b'3} \delta^{ab} \,, \qquad\qquad& \left( {\tilde S}_4^{aba'b'} \right)^\tree = \delta^{a'a} \delta^{bb'}\, .
\end{array}
\end{align}

In order to express the final result, we need to include one final piece, the Sommerfeld enhancement which is encoded in the wavefunction factor $F^{a'b'ab}$ introduced in \Eq{eq:Lfunction}. The required contractions are
\begin{align}
\left( {\tilde S}_1^{aba'b'} \right)^\tree F^{a'b'ab} &= 16 \,M_{\chi}^2\,\big|\sqrt{2}\, s_{00}+2\, s_{0 \pm}\big|^2 \,, \nn \\[5pt]
\left({\tilde S}_2^{aba'b'} \right)^\tree F^{a'b'ab} &= 32\,M_{\chi}^2\,\big | s_{00}\big|^2  \,, \nn \\[5pt]
\left({\tilde S}_3^{aba'b'} \right)^\tree F^{a'b'ab} &= 16 \,M_{\chi}^2\, \big(\sqrt{2}\,s_{00}+2 \, s_{0 \pm}\big)^*  \times \sqrt{2} \,s_{00}+ \text{c.c.} \,, \nn \\[5pt]
\left({\tilde S}_4^{aba'b'} \right)^\tree F^{a'b'ab} &= 32\,M_{\chi}^2\,\big| s_{00}|^2 + 32\,M_{\chi}^2\,\big| s_{0 \pm}|^2 \,, 
\label{eq:LSommerfeldFactors}
\end{align}
where we have used the tree-level values of the functions ${\tilde S}_i$ and the expressions for the wavefunction factor $F^{a'b'ab}$ in terms of the Sommerfeld factors $s_{00}$ and $s_{0\pm}$ (see \Eq{eq:sdefn} in \Sec{sec:sommerfeld}). Upon expanding the product $H_i (\mW)\, H_{S,ij} (\mW)\, {\tilde S}_j (\mW)$ in terms of the evolution kernels in \Eq{eq:Uhard} and \Eq{eq:Usoftcollinear} and using the tree-level results in Eqs.~(\ref{eq:treevaluesH}), (\ref{eq:treevaluesHS}), (\ref{eq:treevaluesStilde}), we find
\begin{align}
{1 \over z} {\text{d}\sigma^{\text{LL}} \over \text{d} z} &= {\pi\, \aW^2 \sin^2 \thetaW \over M_\chi\, v}  \, \text{LP}^{-1} \Bigg \{  U_H(2M_\chi,\mW) \,U_{H_{J_\bn}}(\mu_{H_{J_\bn}},\mW)  \nn \\ 
&\bigg( \frac{4}{3}\, |s_{00}|^2  \big(1-U_{H_S}(\mu_{H_S},\mW) \big) +2\, |s_{0 \pm} |^2\big (1+ U_{H_S}(\mu_{H_S}, \mW)\big ) \nn \\
&\hspace{3cm}+ {2\,\sqrt{2} \over 3} (s_{00}\, s^*_{0\pm} + s^*_{00}\, s_{0\pm}  ) \big(1-U_{H_S}(\mu_{H_S},\mW)\big) \bigg) \Bigg \}\,.
\end{align}
Here $ \text{LP}^{-1}$ denotes the inverse Laplace transform. The prefactors are determined by tree-level matching to full theory, and we have suppressed the arguments of the evolution kernels.

At LL accuracy, the cumulative distribution,
\begin{align}
\sigma^{\text{LL}}(\zcut) = \int\limits_{\zcut}^{1}\text{d}z \frac{\text{d} \sigma^{\text{LL}}}{\text{d}z}\,,
\end{align}
can be obtained setting $s=1/(2M_\chi(1-\zcut))$ in the Laplace space expression for the cross section, and inserting a $1/(2M_\chi)$ for the measure. At the level of the cumulative, we can now explicitly set our canonical scales as
\begin{align}
\mu_{H_{J_\bn}}&=2\,M_\chi \sqrt{1-\zcut} ~\Theta\Big(2\,M_\chi \sqrt{1-\zcut}-\mW   \Big)+\mW\, \Theta\Big(\mW -2\,M_\chi \sqrt{1-\zcut} \, \Big) \,, \nn \\[5pt]
\mu_{H_S}&= 2\,M_\chi (1-\zcut) ~\Theta\Big(2\,M_\chi (1-\zcut)-\mW   \Big)+\mW\, \Theta\Big(\mW -2\,M_\chi (1-\zcut)  \Big)  \,.
\end{align}
This implements the physical constraint that the jet and soft scales never go below the scale $\mW$. For a more sophisticated analysis, smooth transition functions could be used instead of $\Theta$-functions. This is often done to transition from resummation to fixed order, where the smooth transition functions are referred to as profiles \cite{Abbate:2010xh}. Here we content ourselves with this simple choice of scales. This simple choice of profiles also allows us to give a closed form analytic result for the differential spectrum involving the $\Theta$-functions. With this choice of scale, the evolution kernels appearing in the cross section, now also explicitly involve the $\Theta$-functions that cut off their evolution as appropriate. For example, for the jet function evolution kernel, we have
\begin{align}
U_{H_{J_\bn}} \left( \mu_{H_{J_\bn}},  \mW \right)\,\, =\,\, &\exp\left( 8\,C_A\, \ralpha \log^2 \left( \mW \over 2 \,M_\chi \sqrt{1-\zcut} \right) \right) ~\Theta\!\left(2\,M_\chi \sqrt{1-\zcut}-\mW   \right)  \nn \\[5pt]
&+ \Theta\!\left(\mW -2\,M_\chi \sqrt{1-\zcut} \, \right) \,,
\end{align}
which becomes unity for $\mW \geq 2\,M_\chi \sqrt{1-\zcut}$. The soft function evolution kernel is completely analogous.

Combining all the ingredients, we arrive at the final expression for the cumulative cross section at LL accuracy
\begin{align}\label{eq:cumulant_expression}
\sigma^{\text{LL}}(\zcut) &= 4\, |s_{0 \pm}|^2 \sigma^{\tree} e^{-2\, \Gamma_0\, \TaW\, L^2_{\chi}} \,\Theta(1-\zcut)\nn\\[3pt]
&+  \sigma^{\tree} e^{-2\, \Gamma_0 \, \TaW \, L^2_{\chi}} \Bigg\{ \left(-F_0+F_0\,e^{2\, \Gamma_0 \, \TaW \, L_J^2(\zcut)} \right)\Theta\! \left(1-\frac{\mW^2}{4\,M^2_{\chi}}-\zcut\right) \nn\\[3pt]
&+ \left(-F_1+F_1\, e^{2\, \Gamma_0 \, \TaW\, L_J^2(\zcut)} \right)\Theta\! \left(\zcut-1+\frac{\mW}{2\,M_{\chi}}\right)\Theta\! \left(1-\frac{\mW^2}{4\,M_{\chi}^2}-\zcut\right)\nn\\[3pt]
&+ \left(-F_1+F_1\, e^{2\, \Gamma_0  \,\TaW \left(L_J^2(\zcut)-\frac{3}{4}L_S^2(\zcut)\right)} \right)\Theta\! \left(1-\frac{\mW}{2\,M_{\chi}}-\zcut\right)\Bigg\}\,.
\end{align}
Here the $\Theta$-functions explicitly enforce that none of the functions are RG evolved below the scale $\mW$, as emphasized above, and are a crucial part of the final result. Each of the functions appearing in this expression, as well as their physical significance will be defined shortly.

We can now obtain the differential spectrum by taking the derivative of \Eq{eq:cumulant_expression} with respect to $(1-\zcut)$. The differentiation of the cumulative result must be performed carefully due to the presence of the $\Theta$-functions, which when differentiated give rise to $\delta$-functions. However, all the $\delta$-functions explicitly cancel, except for the $\delta$-function for the fully exclusive contribution. Carefully performing the differentiation, we obtain the final result for the differential spectrum:
\begin{mdframed}[linewidth=1.5pt, roundcorner=10pt]
\vspace{-5pt}
\begin{align}\label{eq:resummed}
{\text{d} \sigma^{\text{LL}} \over \text{d}z} &=
  4\, |s_{0\pm}|^2\, \sigma^{\text{tree}} ~e^{\,-2\,\Gamma_0\, \TaW\, L_\chi^2 }     \,\,\delta(1-z)  \nn   \\[3pt]
&+\, 4\,\sigma^{\text{tree}} ~e^{\,-2\,\Gamma_0\,\TaW\, L_\chi^2 } \,\bigg\{ C_A\, \TaW\, F_1  \Big( 3\,\cL_1^S(z) - 2\,\cL_1^J(z) \Big)  \, e^{\, 2\,\Gamma_0\,\TaW \, \big(\Theta_J L_J^2(z)-\frac{3}{4}\Theta_S L^2_S(z)  \big) } \nn \\[3pt]
&  \hspace{120pt}   - 2\, C_A \,\TaW \, F_0 \,\cL_1^J(z) \,e^{\,2\,\Gamma_0\,\TaW \, L_J^2(z)} \bigg\} .
\end{align}
\end{mdframed}\vspace{0.3cm}
This simple formula provides the resummation of all logarithmically enhanced terms to the spectrum at LL accuracy.

As before, to simplify the notation, we have written this expression with $\TaW =\aW/(4\,\pi)$.  This result is composed of several pieces with clear physical significance, each of which we now explain. The tree-level cross section  
\begin{align}
\sigma^{\text{tree}}= {\pi \,\aW^2\, \sin^2 \thetaW \over 2\,M_\chi^2\, v}\,,
\end{align}
appears as an overall multiplicative factor, as does the standard massive Sudakov form factor with logarithm
\begin{align}
L_\chi=\log \left( \frac{\mW}{2\,M_\chi}  \right)\,.
\label{eq:Lchi}
\end{align}
The double logarithmic asymptotics is governed by the cusp anomalous dimension \cite{Korchemsky:1987wg}, in this case at one-loop, 
\begin{align}\label{eq:cusp}
\Gamma_0=4\,C_A\,,
\end{align}
where we recall that $C_A$ is the Casimir of the adjoint representation of SU$(2)$. Explicitly, in our normalization, $C_A=2$. In \Eq{eq:resummed} we have written $\Gamma_0$ in the exponent to emphasize that it is the cusp that controls the anomalous dimensions, but used the explicit form of \Eq{eq:cusp} in the prefactors.

The first term in the \Eq{eq:resummed} is localized at $z=1$. Only the Sommerfeld factor $|s_{0\pm}|^2$ appears since the tree-level process is the annihilation of the charged states $\chi^\pm$. The second term describes the non-trivial $z$ dependence. Here the combination of Sommerfeld factors
\begin{align}
F_0 &= \frac43 \,\big|s_{00}\big|^2 + 2 \,\big|s_{0\pm}\big|^2 + {2\, \sqrt{2} \over 3}\big (s_{00} \,s^*_{0\pm} + s^*_{00} \,s_{0\pm} \big) \,, \nn \\ 
F_1 &= - \frac43 \,\big|s_{00}\big|^2 + 2\,\big |s_{0\pm}\big|^2 - {2 \sqrt{2} \over 3} \big(s_{00}\, s^*_{0\pm} + s^*_{00}\, s_{0\pm}\big ) \,,
\end{align}
appear. The perturbative dynamics are controlled by the two logarithms
\begin{align}
L_J(z)= \log \left({\mW \over 2 \,M_\chi \sqrt{1-z} }\right) \,, \qquad L_S(z)= \log \left({\mW \over 2 \,M_\chi (1-z) }\right) \,,
\label{eq:LJLS}
\end{align}
associated with the jet and soft scales, respectively. For convenience, we have also defined $\Theta$ functions associated with the range of the soft and collinear scales
\begin{align}
\Theta_J=\Theta\!\left(1-\frac{\mW^2}{4\, M_\chi^2}-z \right)\,, \qquad  \Theta_S=  \Theta\!\left( 1-\frac{\mW}{2\,M_\chi}-z \right)\,.
\end{align}
In addition to the Sudakov logarithms, the $z$ dependence is controlled by the functions  
\begin{align}
\cL_1^J(z) =  \frac{L_J}{1-z} \, \Theta_J\,, \qquad \cL_1^S(z)=\frac{L_S}{1-z} \, \Theta_S\,,
\label{eq:LogPlusFn}
\end{align}
which capture the power divergence in $1-z$, and the subscript is standard notation denoting that these contain a single power of the logarithm.  The presence of the $1/(1-z)$ factor gives the expected leading power scaling for the cross section. The power divergence for the soft logarithm is cutoff at $z = 1-\mW/(2\,M_\chi)$ and for the jet logarithm at $z=1-\mW^2/(2\, M_\chi)^2$. These physical cutoffs arise from the value of $z$ at which the soft and jet scales hit the scale $\mW$, where the running must be turned off, as has been discussed above. We note that in the massless theory, the power law divergences would be regulated as plus distributions. Instead, here $\mW$ explicitly cuts off the divergence at a finite distance from the endpoint. 

There is a physical interpretation for each of the different terms in \Eq{eq:resummed}. The first term, which is localized at the endpoint, corresponds to the fully exclusive cross section, while the other terms describe deviations from the endpoint associated with either soft or collinear radiation. With this understanding of the correct treatment of the scales as we transition to the fully exclusive endpoint, and how they are implemented in our final factorization formula, in the next section we show that our LL expression in the endpoint region correctly reproduces the LL in both the exclusive and OPE regions. Firstly, however, note that expanding \Eq{eq:resummed} to fixed order,  setting the Sommerfeld factor to its tree-level result $|s_{00}|^2 = 1$, and dropping $\Theta$-functions, we find
\begin{align}
\frac{\text{d} \sigma}{\text{d}z}  = \frac{4\,\aW^3\,\sin^2\thetaW}{M_\chi^2\,v} \frac{\log\left(\frac{2\,M_\chi\,(1-z)}{\mW}\right)}{1-z} + \cO(\aW^4) \,.
\end{align}
This result agrees with the $\cO(\aW^3)$ logarithm derived in the fixed order calculation of~\cite{Bergstrom:2005ss}.

\subsection{Reproducing the Exclusive and Inclusive Cross Sections}\label{sec:limits}
In this section, we demonstrate that our EFT acts as a mother theory which includes both the exclusive ($\zcut \to 1$) and inclusive ($\zcut \to 0$) results as limiting cases of our resummed expression~\Eq{eq:resummed}.  It is important to note that the expansions performed here differ from previous calculations such that power corrections are not expected to be identical.  However, this complication is avoided here due to the simple structures that are present at LL order.  The focus of this section will be showing how to take these two limits analytically. \Sec{sec:num_transition} will provide a numerical study of the theoretical error that results from scale variation. 

\subsubsection{Inclusive Limit}\label{sec:inc_limit}
To obtain the inclusive limit of the total cross section, we simply integrate the differential cross section given in \Eq{eq:resummed} from $z=0$ to the endpoint $z =1$. Explicitly,
\begin{align}
\sigma^{\text{incl}}=&\int\limits_{0}^1 \text{d}z {\text{d} \sigma \over \text{d}z} =  \int\limits_{0}^1 \text{d}z\, 4\, |s_{0\pm}|^2\, \sigma^{\text{tree}} ~e^{\,-2\,\Gamma_0\, \TaW\, L_\chi^2 }     \,\,\delta(1-z)  \nn   \\[3pt]
&\hspace{-0.5cm}+ \int\limits_{0}^1 \text{d}z\, 4\,\sigma^{\text{tree}} ~e^{\,-2\,\Gamma_0\,\TaW\, L_\chi^2 } \,\bigg\{ C_A\, \TaW\, F_1  \Big( 3\,\cL_1^S(z) - 2\,\cL_1^J(z) \Big)  \, e^{\, 2\,\Gamma_0\,\TaW \, \big( \Theta_J L_J^2(z)-\frac{3}{4}\Theta_S L^2_S(z)  \big) } \nn \\[3pt]
&  \hspace{140pt}   - 2\, C_A \,\TaW \, F_0 \,\cL_1^J(z) \,e^{\,2\,\Gamma_0\,\TaW \, L_J^2(z)} \bigg\} \,.
\end{align}
Performing the integral, we have
\begin{align}
\sigma^{\text{incl}}&=\sigma^{\text{tree}} \left(  F_0+ F_1 e^{-\frac{3}{2}\, \Gamma_0\, \TaW\, L_\chi^2} \right) \nn \\[5pt]
&= \sigma^{\text{tree}} \left(    \frac{4}{3} |s_{00}|^2f_{-}+ 2|s_{0\pm}|^2f_++\frac{2 \sqrt{2}}{3}\left(s_{00}\,s_{0\pm}^*+c.c \right) f_- \right)\,,
\end{align}
where in the last line we have introduced the notation of~\cite{Baumgart:2014vma}
\bea
f_{\pm} =  1 \pm e^{-\frac{3}{2}\, \Gamma_0\, \TaW\, L_\chi^2}\,.
\eea
This is precisely the result obtained in~\cite{Baumgart:2014vma,Baumgart:2014saa,Baumgart:2015bpa}, demonstrating that we reproduce the inclusive limit to LL order.

\subsubsection{Exclusive Limit}\label{sec:exc_limit}

Note that the signature of interest for experiments like HESS, where the experimental resolution has a width $\sigma \gg \mW^2/(4\,M_\chi^2)$, includes a contribution from the exclusive line and the endpoint spectrum.  It is therefore important that we are also able to reproduce the resummed fully exclusive cross section from our factorization. This can be accomplished by integrating \Eq{eq:resummed} from $z=1-\mW^2/(4\,M_\chi^2)$ to $z=1$, which corresponds to a kinematic requirement such that only the exclusive final state is possible since both the jet and soft scales are set by the electroweak boson mass.\footnote{Note that for $z>1-\mW^2/(4\,M_\chi^2)$, \Eq{eq:resummed} is proportional to a delta function for exclusive production, namely $\delta(1-z)$.  It is important to note that we have power expanded away any mass dependence that would lead to kinematic differences between the $\gamma\,\gamma$ and $\gamma\, Z$ final states. We therefore are implicitly assuming that the finite resolution function sufficiently smears these differences such that they are not experimentally relevant.}  This demonstrates that for the case where the experimental resolution has a width $\delta \gg \mW^2/(4\,M_\chi^2)$, we have provided the complete description as relevant experimentally (with the additional caveats discussed in Appendix~\ref{sec:continuum}). 

When integrating from $z =1- \mW^2/(4\,M^2_{\chi})$ to the endpoint both $\cL_1^J$ and $\cL_1^S$ are zero.  Therefore, we can we trivially integrate the $\delta(1-z)$ dependent term to find
\begin{align}
\sigma^{\text{excl}}=\int\limits_{1-\frac{\mW^2}{4\,M^2_{\chi}}}^1 \text{d}z \,\frac{\text{d}\sigma}{\text{d}z} =  4\,|s_{0\pm}|^2\, \sigma^{\text{tree}} \,e^{ \,-2\,\Gamma_0 \,\TaW \,L_\chi^2 }\,.
\label{eq:LLexclusive}
\end{align}
This agrees with the exclusive calculation at leading log accuracy performed in \cite{Ovanesyan:2014fwa,Bauer:2014ula}. The fact that we reproduce this result makes it straightforward to convolve the resummed photon spectrum with the experimental resolution -- no merging between different results is required. In this sense, our EFT acts as a mother theory that completely describes the photon spectrum for heavy WIMP annihilation at LL order.

\section{Numerical Results and Scale Variation}\label{sec:num_transition}

In this section, we provide a numerical study of our final prediction for the spectrum by evaluating \Eq{eq:resummed} for wino DM.  This allows us to explore the relative contributions from the line annihilation and the endpoint spectrum for different choices of the DM mass.  We will also show the cumulative spectra, as given to LL accuracy in \Eq{eq:cumulant_expression}, which provides intuition for the finite bin effects that are relevant to realistic experiments.  Then in Sec.~\ref{sec:results} we will provide a mock reanalysis of the HESS line search, and will convolve our predicted spectrum with the Gaussian line shape assumed by HESS.  

As shown explicitly in \Sec{sec:limits}, our resummed spectrum analytically reproduces the fully exclusive and the fully inclusive limits, so that we can additionally study the transition between these approximations.  This clarifies the disparate conclusions that have been drawn using these different approaches. In particular, the exclusive calculations of~\cite{Bauer:2014ula, Ovanesyan:2014fwa, Ovanesyan:2016vkk} claimed a reduction factor of $\sim2.2$ when compared with the tree-level cross section for a $3$ TeV wino.  For contrast, the inclusive calculation of~\cite{Baumgart:2014vma,Baumgart:2014saa,Baumgart:2015bpa} found a reduction of only a few percent.  Physically, this results from the fact that an increasingly exclusive constraint on the final state implies there will be less cancellation between the virtual and real corrections (for discussions in the context of electroweak logarithms, see \emph{e.g.}~\cite{Bell:2010gi,Manohar:2014vxa}).  The proper interpretation of the experimental limits depends on how rapidly the transition between the exclusive and inclusive cross sections occurs.  Our EFT analysis provides a complete and decisive resolution of this issue. Interestingly, we find that the experimental values of current interest to the HESS line search, $\zcut \sim 0.8$-$0.9$, lies right in a transition region between the two limiting cases.  This emphasizes the need to properly treat the impact of finite resolution, as we will do in the next section.  However, before moving to our mock reanalysis of the HESS search, we will provide some numerical results along with an estimate of the impact of scale uncertainty.

\begin{figure}[t!]
\begin{center}
\includegraphics[width=0.65\columnwidth]{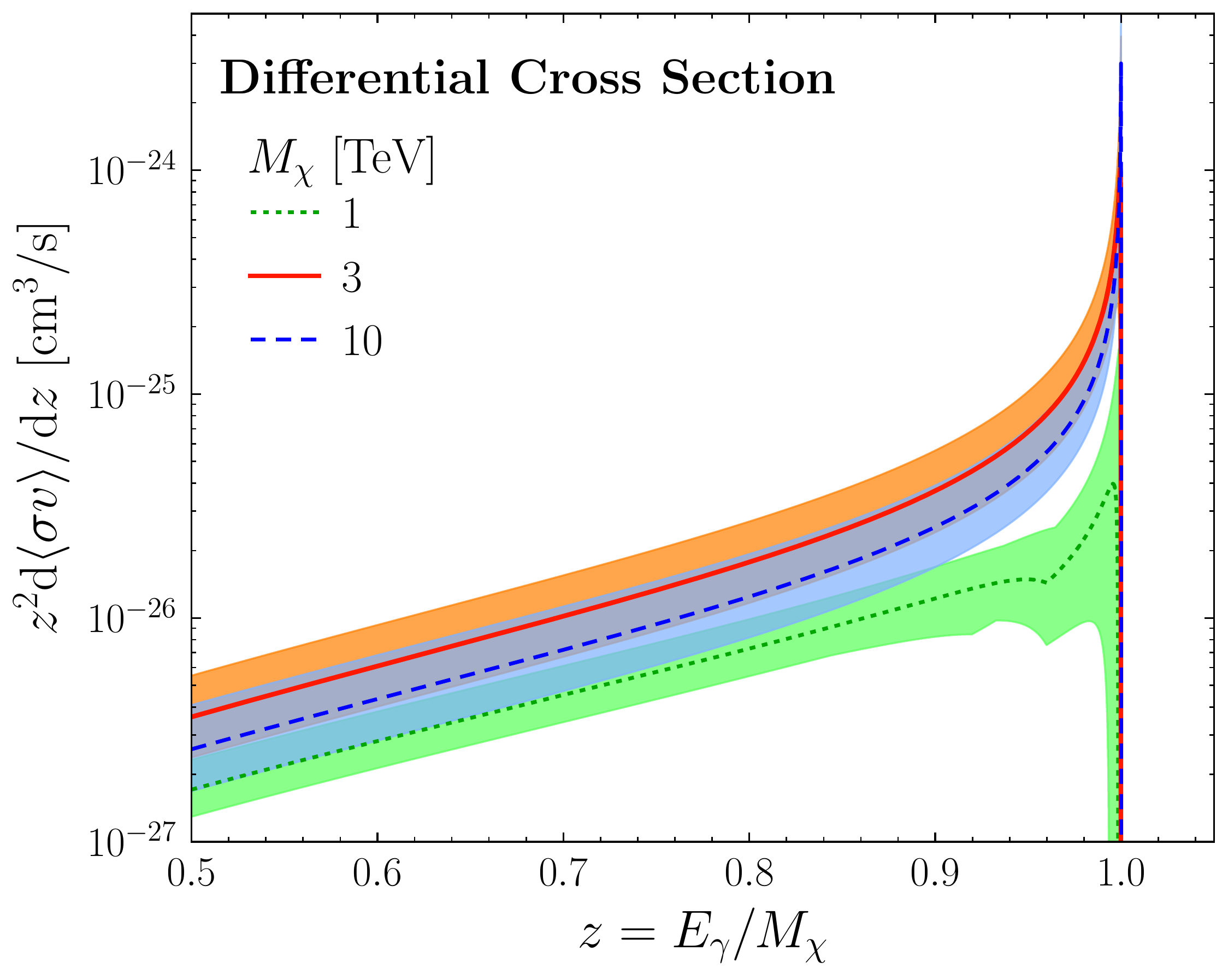}    
\caption{The $z^2$ weighted differential endpoint cross section as a function of $z$ for three choices of the wino mass.  Note that the delta function contribution due to the exclusive annihilation process is not included for clarity of presentation.  The error bands are due to scale variation as discussed in the text.}
\label{fig:differential_zcut}
\end{center}
\end{figure}

\begin{figure}[t!]
\begin{center}
\includegraphics[width=0.65\columnwidth]{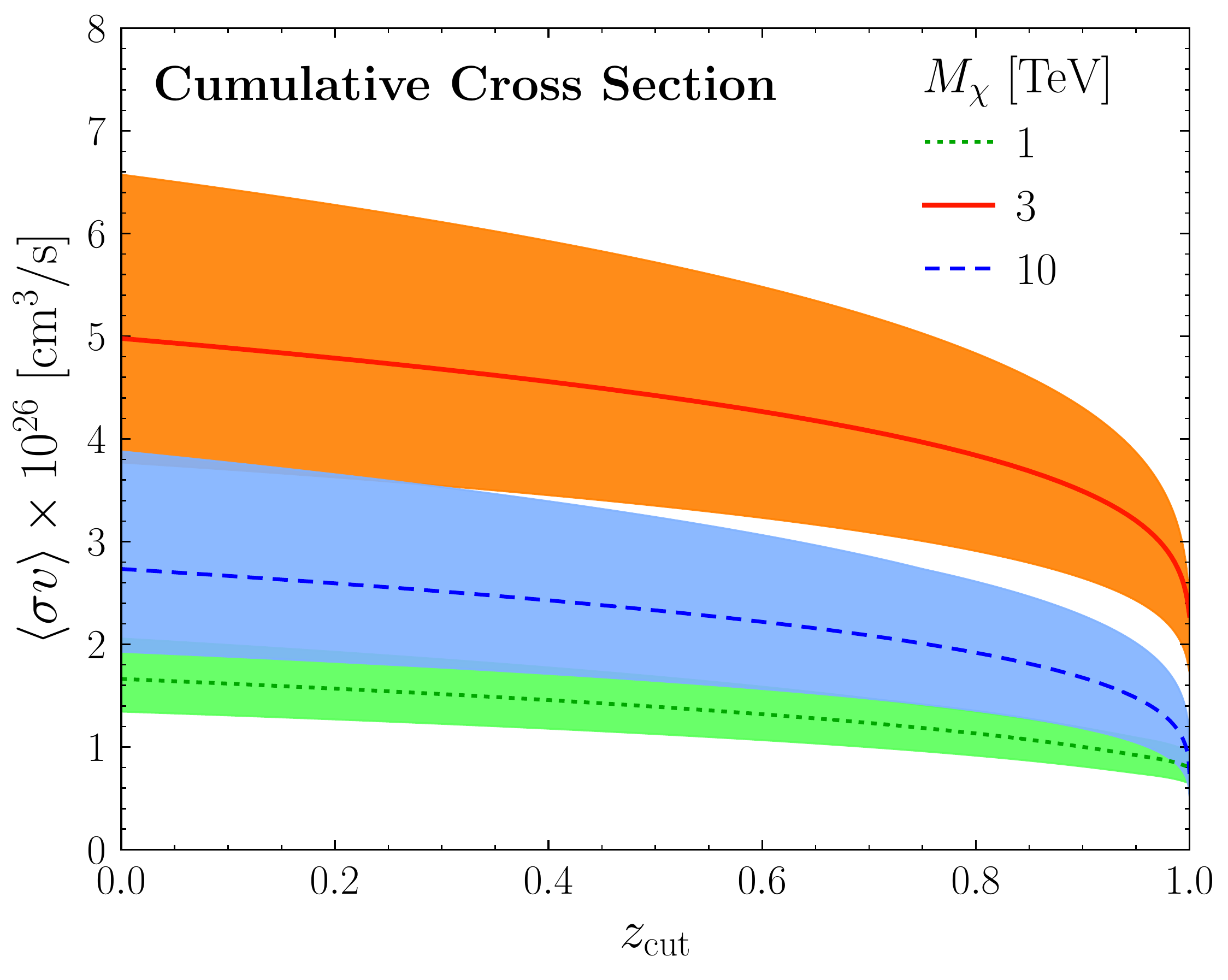}
\caption{The cumulative cross section as a function of $\zcut$ for three choices of the wino mass.  The exclusive contribution is included here.  The error bands are due to scale variation as discussed in the text.}
\label{fig:cumulative_zcut}
\end{center}
\end{figure}

In \Fig{fig:differential_zcut} we show the differential spectrum $z^2\, \text{d}\langle \sigma v\rangle/\text{d}z$ for several values of the DM mass.  The delta function contribution from the exclusive process is not included.  We see that the endpoint tracks the mass of the DM as expected.  Furthermore, the contribution from the resummed continuum grows as the DM mass is increased.  However, this effect is mitigated by the strong mass dependence of the overall cross section, both due to Sommerfeld enhancement and the overall $1/M_\chi^2$ scaling, which explains why the 3~TeV result lies above both the 1~TeV and 10~TeV results. The kink in the $1$ TeV distribution is a result of the $\Theta$-functions appearing in the choice of scales, as discussed in \Sec{sec:resum} (in reality, there are kinks in all the distributions, but they are only visible by eye for the $1$ TeV distribution). This kink is ultimately unphysical and could be removed by a smooth choice of scales, but is well within our uncertainty bands.

The uncertainty bands in \Fig{fig:differential_zcut} are the result of varying the renormalization scales corresponding to the natural scales of the functions appearing in our factorization.  Due to our choice of renormalization path, we simply vary the $\mu$ scale of the different functions by a factor of two about their natural scales.

An alternative numerical representation of our results is provided in~\Fig{fig:cumulative_zcut}, where we plot the cumulative cross section as a function of the $\zcut$, for several values of the DM mass.  Here we do include the delta function contribution that yields the exclusive annihilation process, which accounts for the finite value when $\zcut = 1$.  The uncertainty bands are computed using the same prescription for the scale variation performed for Fig.~\ref{fig:differential_zcut}.  

The two endpoints, namely $\zcut=1$ and $\zcut=0$, correspond to the fully exclusive and fully inclusive limits, respectively.  Interestingly, for the experimentally relevant range $\zcut \sim 0.8$-$0.9$, the cumulative cross section takes an intermediate value approximately midway between the two extremes. This implies that for these values of $\zcut$, logarithms of the resolution are playing an important role, in keeping with the conclusions of the fixed order calculation in the inclusive limit~\cite{Baumgart:2015bpa}.  Theoretically robust results require the all-orders resummation of logarithms from finite bin effects, as has been done here for the first time.

\section{Impact on Indirect Detection Constraints}\label{sec:results}

The resummed photon spectra derived above have clear implications for heavy DM line searches. In particular, thermal wino annihilations would produce TeV scale photons.\footnote{Another case where a careful treatment of endpoint contributions will be relevant is Higgsino DM, as demonstrated in~\cite{Baumgart:2015bpa}. We leave this study to future work.} When these photons strike the Earth's atmosphere, they initiate a detectable shower of particles that persists to the surface. Exactly reconstructing the energy of the incident photon from the resultant shower is impossible, and as such any real instrument will need to account for finite energy resolution effects associated with the spread of possible reconstructed energies given a single true energy.

As discussed at the outset, the strongest constraints on the wino are due to HESS observations of the Galactic Center~\cite{Abramowski:2011hc,Abramowski:2013ax}; updated limits are expected shortly involving the full HESS~I dataset~\cite{Rinchiuso:2017kfn,Rinchiuso:2017pcx}.  Line searches are typically designed to be model-independent, and thus assume that only the line emission is relevant (although some specific non-line hard spectra have also been tested \cite{Abramowski:2013ax,Lefranc:2016fgn}). As we demonstrated in Fig.~\ref{fig:cumulative_zcut} above, photons away from the endpoint contribute to a finite bin at a non-trivial level. This is especially true for HESS, where the effective $\zcut \sim 0.8-0.9$ depending on the incident energy. Furthermore, the line analysis of HESS is not a bin-based counting experiment but requires subtraction of an unknown background, which is modeled by a smooth function. The presence of signal photons at even lower energies may bias the data-driven background model if this signal spectrum is not correctly modeled, further modifying the limit.

The goal of this section is to estimate how much including the correct shape and normalization of the resummed spectrum would be expected to change the HESS~limit, relative to the case of a pure line.

It is important to emphasize that the results presented in this section are approximate, and should not be taken as updated limits on the wino.  At issue is that the full dataset HESS~used to construct their limits in Ref.~\cite{Abramowski:2013ax} is not public.  What we will show are results from a simplified mock version of that analysis, using a Gaussian likelihood rather than the full likelihood, which has been validated to yield comparable limits when assuming exclusive line emission.  We can then explore how the various conclusions are modified when we include the endpoint emission spectrum.  The conclusion is that the additional emission should strengthen limits on the wino by a $\mathcal{O}(1)$ factor.  This provides motivation for future experimental analyses to include these contributions when determining limits.

This section contains three parts. First, we review how to map from DM model parameters, including the relevant astrophysical inputs, to a prediction for the number of photons that HESS~would observe.  Then we apply this formalism to demonstrate the range of parameters that HESS~can constrain.  Finally, we outline our mock analysis procedure and present approximate results showing the impact of our resummed spectra on current constraints.

\subsection{Predicting the Indirect Detection Flux}\label{sec:IDreview}

In order to determine the sensitivity to wino DM, we need a prediction for the number of photons that should arrive at an experiment as a function of the DM parameters. This can be derived using the canonical indirect detection formula, which specifies the differential energy flux arriving at the detector,
\begin{equation}
\frac{1}{\Omega_{\rm ROI}} \frac{\text{d}\Phi_{\gamma}}{\text{d}E} = J\, \frac{\langle \sigma v \rangle}{8\, \pi\, M_{\chi}^2} \,\frac{\text{d}N_{\gamma}}{\text{d}E}\,,
\label{eq:IDflux}
\end{equation}
where $\Omega_{\rm ROI} \equiv \int_{\rm ROI} \text{d} \Omega$.

The particle physics contribution $\langle \sigma v \rangle / (8\, \pi\, M_{\chi}^2) \, \text{d}N_{\gamma}/\text{d}E$ depends on the velocity averaged total annihilation cross section $\langle \sigma v \rangle$, which is summed over all final states involving a photon, and the average photon spectrum per annihilation,  $\text{d}N_{\gamma}/\text{d}E$, which can be written as\footnote{This is sometimes defined as the spectrum per DM particle, which differs by a factor of 2.}
\begin{equation}
\frac{\text{d}N_{\gamma}}{\text{d}E} = \sum_f {\rm Br}_f \,\frac{\text{d}N^f_{\gamma}}{\text{d}E} \,,
\end{equation}
where the $f$ index refers to the different final states with associated branching fractions ${\rm Br}_f$ and photon spectra $\text{d}N^f_{\gamma}/\text{d}E$. Since the spectrum here is the result of resumming multiple electroweak final states (not including the photons that result from decay of unstable $W^\pm$ and $Z$ bosons, see Appendix~\ref{sec:continuum} for a discussion), we will only refer to the total averaged quantity $\text{d}N_{\gamma}/\text{d}E$ for the remainder of this section.

The remaining ingredient is the so-called $J$-factor, which is an astrophysical input.  It is determined by the distribution of the DM along the line of sight in the region of interest (ROI) under consideration.  It additionally accounts for the fact that two particles must find each other for for annihilation to occur; the $J$-factor depends on the number density squared as
\begin{equation}
J = \frac{\int_{\rm ROI} \text{d}s\,\text{d}\Omega\, \rho^2_{\rm DM}(s, \Omega)}{\Omega_{\rm ROI}}\,,
\label{eq:Jfactor}
\end{equation}
where $\rho_{\rm DM}$ is the Milky Way DM mass distribution, $s$ is the distance from Earth along the line of sight, and $\Omega$ gives the coordinates on the sky within the ROI. Note that as written, the $J$-factor has units of ${\rm TeV}^2\,\cdot\,{\rm cm}^{-5}$, and in particular there are no units of ${\rm sr}$ due to the denominator in Eq.~(\ref{eq:Jfactor}).  We caution, however, that a number of other conventions are in use.\footnote{For a recent review of the conventions used for indirect detection, see Appendix A of~\cite{Lisanti:2017qoz}.}

For a fixed ROI, $J$ is then in principle determined by the Milky Way DM profile. Unfortunately, the shape of $\rho_{\rm DM}$ is very uncertain near the Galactic Center, see \emph{e.g.}~\cite{Pato:2015tja}, and in particular within the ROI of the HESS search of Ref.~\cite{Abramowski:2013ax}. For the case of the wino, once the mass is fixed the cross section is fully specified.  Therefore, one can translate limits on wino annihilations into a constraint on $J$, as done in Fig.~\ref{fig:DM_limits_a} below.  

It is also of interest to fix a prototypical value for $J$ and then set a limit on the annihilation cross section, since this is how these constraints are typically presented.  For this purpose we adopt the Einasto profile, the default profile assumed in the HESS~analyses, which is given by
\begin{equation}
\rho_\text{\tiny Einasto}(r) \propto \exp \left[ - \frac{2}{\alpha} \left( \left( \frac{r}{r_s} \right)^{\alpha} - 1 \right) \right]\,,
\label{eq:Einasto}
\end{equation}
where $r$ is the distance from the center of the halo, and following~\cite{Pieri:2009je}, by default we take $\alpha=0.17$, $r_s = 20$ kpc, and then normalise the profile so that we reproduce the local DM density of $0.39~{\rm GeV}~{\rm cm}^{-3}$ at our location which is 8.5 kpc from the Galactic Center. Another frequently invoked DM distribution is the Navarro-Frenk-White (NFW) profile~\cite{Navarro:1996gj}, which takes the form
\begin{equation}
\rho_\text{\tiny NFW}(r) \propto \frac{1}{(r/r_s)(1+r/r_s)^2}\,,
\label{eq:NFW}
\end{equation}
where again we take $r_s = 20$ kpc. We will also make use of the NFW profile (including the possibility of a non-trivial core) when interpreting our results below.

Finally, putting this all together results in the differential energy flux arriving at the detector, $\frac{1}{\Omega_{\rm ROI}} \frac{\text{d}\Phi_{\gamma}}{\text{d}E}$, which has units of ${\rm photons}\,\cdot\,{\rm cm}^{-2}\,\cdot\,{\rm s}^{-1}\,\cdot\,{\rm TeV}^{-1}\,\cdot\,{\rm sr}^{-1}$. This  quantity can be converted into a predicted number of photons (per unit area per unit time) arriving at the experiment from DM annihilation by first multiplying by the solid angle of the considered ROI, $\Omega_{\rm ROI}$, and then integrating over the energy range determined by the experimental search.  This photon flux $\Phi_{\gamma}$ has units of ${\rm photons}\,\cdot\,{\rm cm}^{-2}\,\cdot\,{\rm s}^{-1}$. Converting this to the actual number of photons depends on the experimental effective area and time over which the ROI is observed; a larger detector and longer observations will result in more observed photons. For HESS, the effective area is $\sim 10^9$~cm$^2$ at 1 TeV and current searches make use of 112 hours of observations of the Galactic Center, yielding sensitivity to fluxes $\sim 10^{-14}~{\rm cm}^{-2}~{\rm s}^{-1}$. We can then constrain the DM model using this prediction for the number of photons as an input to a likelihood analysis.

\subsection{From Predictions to Constraints}\label{sec:IDconstraint}

Before we give the details of and results from our mock analysis, it is useful to discuss how we are mapping from the theory prediction to the experimental constraints.  The subtlety arises because the original search was performed under the assumption that the annihilation signature is a line; in this case, by definition all photons have the same energy.  The spectrum of a typical WIMP can be decomposed into two contributions
\begin{equation}
\frac{\text{d}N_{\gamma}}{\text{d}E} \sim~{\rm line}~+~{\rm continuum}\,.
\end{equation}
The line is due to exclusive annihilations to $\gamma\,\gamma$ and $\gamma\,Z$.  Since our interest here is in heavy WIMPs, we will neglect the fact that the finite $Z$ mass causes $E_\gamma=M_\chi -\mW^2/(4M_\chi) < M_\chi$ for the photons that result from the $\gamma\,Z$ process, and will combine these line contributions using
\begin{equation}
\langle \sigma v \rangle_{\rm line} \equiv \langle \sigma v \rangle_{\gamma \gamma} + \frac{1}{2} \langle \sigma v \rangle_{\gamma Z} \,,
\label{eq:BRGamma}
\end{equation}
with $E_\gamma = M_\chi$ for all line photons.

The continuum receives many contributions.  In the DM literature, this is usually separated into photons from ``internal bremsstrahlung''~\cite{Beacom:2004pe, Birkedal:2005ep, Bergstrom:2004cy, Bergstrom:2005ss, Bringmann:2007nk}, as well as final and initial state radiation, on one hand, and those photons that result from the cascade decay chain of unstable particles on the other hand.  The decay processes can yield many final state photons with a broad energy spectrum.  Our endpoint calculation for winos resums the non-decay perturbative processes, and as such it does not include the additional continuum photons that result from the decay of the $W^\pm$ and $Z$.  However, this contribution is demonstrated to have little impact on the limits for heavy winos in Appendix~\ref{sec:continuum}.  This conclusion is intuitive since the photons from the $W^\pm/Z$ cascade decays are much lower energy than the exclusive and endpoint contributions.  Therefore, we model the continuum as only being due to the endpoint contributions, which we denote with $\mathcal{E}(E)$, and using \Eq{eq:resummed} the LL result is given explicitly by
\begin{align}
\mathcal{E}^\text{LL}(E) = &\frac{1}{\langle \sigma v \rangle_{\rm line}} \frac{\text{d}\langle\sigma v\rangle}{\text{d}E} - 2\, \delta\big(E-M_\chi\big) \label{eq:epspec}\\[10pt]
= &\frac{2}{|s_{0\pm}|^2\, M_{\chi}}\bigg\{ C_A\, \TaW\, F_1  \Big( 3\,\cL_1^S(z) - 2\,\cL_1^J(z) \Big)  \, e^{\, 2\,\Gamma_0\,\TaW \, \big( L_J^2(z)-\frac{3}{4}(\cL_1^S(z))^2 (1-z)^2 \big) }  \nn \\
&\hspace{7cm}- 2\, C_A \,\TaW \, F_0 \,\cL_1^J(z) \,e^{\,2\,\Gamma_0\,\TaW \, L_J^2(z)} \bigg\}\,, \nonumber
\end{align}
where as usual, $z=E/M_\chi$.
The resulting spectrum per annihilation is
\begin{equation}
\frac{\text{d}N_{\gamma}}{\text{d}E} = \frac{\langle \sigma v \rangle_{\rm line}}{\langle \sigma v \rangle} \Big( 2\,\delta(E-M_{\chi}) +\mathcal{E}(E) \Big)\,,
\end{equation}
such that $\langle \sigma v \rangle_{\rm line} / \langle \sigma v \rangle$ is the branching fraction to line photons. Note that our calculation predicts not only the shape of the endpoint contribution, but also the relative normalization of this with respect to the line spectrum. Putting these details together, we arrive at the theory prediction
\begin{equation}
\left(\frac{\text{d}\Phi_{\gamma}}{\text{d}E}\right)_{\!\text{ideal}} = \frac{J\,\Omega_{\rm ROI}\,\langle \sigma v \rangle_{\rm line}}{8\, \pi\, M_{\chi}^2} \Big[ 2\,\delta(E-M_{\chi}) + \mathcal{E}(E)\Big]\,,
\end{equation}
which is idealized in the sense that it neglects experimental effects.

As such we are still missing one ingredient, which is the fact that we need to convolve this with the experimental energy resolution. We can describe the energy resolution via a convolution function $\Sigma(E-E')$, where $E'$ is the true photon energy and $E$ is the reconstructed value, and the spectrum an experiment would measure is
\begin{equation}
\left(\frac{\text{d}\Phi_{\gamma}}{\text{d}E}\right)_{\!\text{smeared}} = \frac{J\,\Omega_{\rm ROI}\,\langle \sigma v \rangle_{\rm line}}{8\, \pi\, M_{\chi}^2} \int_0^{M_{\chi}} \text{d}E'\,\Sigma\big(E'-E\big)\Big[ 2 \,\delta\big(E'-M_{\chi}\big) + \mathcal{E}\big(E'\big)\Big]\,.
\label{eq:FluxTheory}
\end{equation}
The HESS collaboration has published a model for $\Sigma(E-E')$ which we use here, a Gaussian that is peaked near the true energy with a width that varies from 17\% at 0.5 TeV and 11\% at 10 TeV. We interpolate in between these values using the log of the energy, and find a width $\sim$$15$\% at 3 TeV.

HESS can constrain the overall normalization of \Eq{eq:FluxTheory}; in terms of the theory prediction, this can be interpreted as a constraint on the quantity
\begin{equation}
C_{\rm HESS} = \frac{J\,\Omega_{\rm ROI}\,\langle \sigma v \rangle_{\rm line}}{8\, \pi\, M_{\chi}^2}\,.
\label{eq:CHESS}
\end{equation}
However, it is critical to specify the assumed energy spectrum $\mathcal{E}(E)$ (in addition to a line contribution) when deriving a HESS constraint on the cross section.  For the following comparisons, we will use the LL endpoint spectrum computed in this work, so that
\begin{equation}
\left(\frac{\text{d}\Phi_{\gamma}}{\text{d}E}\right)_{\!\text{HESS}} = C_\text{HESS} \int_0^{M_{\chi}} \text{d}E'\,\Sigma\big(E'-E\big)\Big[ 2 \,\delta\big(E'-M_{\chi}\big) + \mathcal{E}^\text{LL}\big(E'\big)\Big]\,,
\label{eq:FluxHESS}
\end{equation}
where $C_\text{HESS}$ is the coefficient that is constrained using the HESS data, we take $\mathcal{E}^\text{LL}\big(E'\big)$ from \Eq{eq:epspec}, and $\Sigma(E'-E)$ is as discussed above. In the next section, we will interpret the HESS data as a constraint on $C_\text{HESS}$ using a mock analysis, and will then convert this into an approximate constraint on winos using \Eq{eq:CHESS}.  We will either use \Eq{eq:LLexclusive} to predict $\langle \sigma v \rangle_{\rm line}$ for a given mass in order to set a constraint on $J$, or we will assume the Einasto profile which gives us $J$ and then constrain the cross section $\langle \sigma v \rangle_{\rm line}$.   We will also provide a constraint on the core size, using the NFW profile modified to include a core.

Note that we can test the effects of ignoring the non-line endpoint contributions by simply setting $\mathcal{E}(E)=0$; up to the approximations in our analysis required by not having the full likelihood available, this should reproduce the limits stated in Ref.~\cite{Abramowski:2013ax}. This allows us to directly compare constraints on the line only and the line plus endpoint spectrum, thereby highlighting the impact of our main result \Eq{eq:resummed}.  The next section outlines the details of our mock analysis and provides approximate constraints on either the cross section or the $J$-factor. 

\subsection{Approximate Constraints}\label{sec:mockHESS}

Using the procedure described in the previous section, one can in principle interpret the HESS data as a constraint on wino DM annihilations.  As the data collected by the instrument is not public, we are not able to provide a full and precise update of the constraints on winos.  Instead, we will perform a simplified mock version of the HESS analysis in order to estimate the impact of the corrections calculated here on the resulting limits.  Our mock analysis can roughly recover the published line limits in the case where we take $\mathcal{E}(E) = 0$ above.  We will then extend the analysis to include the endpoint contributions, demonstrating that they strengthen the limits by an $\mathcal{O}(1)$ factor.

Our mock analysis is based on a simplified version of the analysis performed in Ref.~\cite{Abramowski:2013ax}. Figure~1 of that work provides the measured flux and the associated uncertainty as a function of energy in their ROI near the Galactic Center. We digitized this dataset and used it as the input to a Gaussian likelihood analysis. We note that since HESS is an instrument that counts the number of incident photons, the Poisson likelihood should in principle be used. However, the number of counts cannot be exactly reconstructed from the publicly released flux data. The non-Gaussian nature of this dataset is made manifest by the asymmetric error bars that are particularly clear at higher energies. We approximately included the asymmetry in the likelihood by using the upper error bars to determine the likelihood contribution from bins where our model prediction exceeded the data, and the lower error bars for bins where our model prediction fell below the data. We found that this approach gave better agreement with the published HESS results than symmetrizing the error bars. 

The dataset $d_i$ is defined using energy bins with associated index $i$, where the digitized HESS flux gives a central value $\mu_i$ and error $\sigma_i$, chosen (between the upper and lower error bars) in the manner described above. The prediction $m_i(\boldsymbol{\theta})$ is a function of the model parameters $\boldsymbol{\theta}$. The DM-signal contribution to the model is computed using Eq.~\eqref{eq:FluxTheory}. We will treat this theory flux as being a function of the DM mass, $M_{\chi}$, the line photon cross section, $\langle \sigma v \rangle_{\rm line}$, and the $J$-factor.  As emphasized above, given $M_{\chi}$ we can either calculate $\langle \sigma v \rangle_{\rm line}$ and then constrain $J$, or assume a value of $J$ and turn this into a constraint on $\langle \sigma v \rangle_{\rm line}$.

Even in the most optimistic DM scenario, the events collected by HESS will not be solely due to DM annihilation. Firstly, there is a substantial flux of cosmic rays colliding with the atmosphere, which can mimic gamma-ray signals. Secondly, there will be genuine gamma-rays due to high-energy astrophysical processes, such as protons in the inner galaxy colliding with gas and producing energetic neutral pions which decay to gamma-rays. The expected flux from cosmic-rays and astrophysical sources of gamma-rays is not well understood in the HESS energy range, and as such Ref.~\cite{Abramowski:2013ax} parametrized the background contribution using the following seven parameter model:
\begin{equation}\begin{aligned}
\left(\frac{\text{d}\Phi_{\gamma}}{\text{d}E} \right)_{\rm bkg} &= a_0 \left( \frac{E}{1~{\rm TeV}} \right)^{-2.7} \left[ P \left( \log_{10} \left[ \frac{E}{1~{\rm TeV}} \right] \right) + \beta\, G \left( \log_{10} \left[ \frac{E}{1~{\rm TeV}} \right] \right) \right]\,, \\[10pt]
P(x) &\equiv \exp(a_1 \,x + a_2\, x^2 + a_3\, x^3)\,, \\[10pt]
G(x) &\equiv \frac{1}{\sqrt{2\,\pi\, \sigma_x^2}} \exp \left[ - \frac{(x-\mu_x)^2}{2\,\sigma_x^2} \right]\,.
\end{aligned}\end{equation}
The background is then specified by the seven parameters $\boldsymbol{\theta}_{\rm bkg} = \{ a_0, a_1, a_2, a_3, \beta, \mu_x, \sigma_x \}$.

Combining the signal and background, we arrive at our full model prediction of
\begin{equation}
m_i(\boldsymbol{\theta}) =\left.\left[ \left(\frac{\text{d}\Phi_{\gamma}}{\text{d}E}\big( M_{\chi}, \langle \sigma v \rangle_{\rm line}, J \big) \right)_{\rm Smeared}  + \left(\frac{\text{d}\Phi_{\gamma}}{\text{d}E}\left( \boldsymbol{\theta}_\text{bkg} \right) \right)_{\rm bkg} \right]\right|_{E=E_i}\,,
\end{equation}
so that the model is specified by three signal and seven background parameters. From here, given the HESS dataset described above, $d=\{d_i\}=\{\mu_i,\sigma_i\}$, we can write down our assumed Gaussian likelihood function as
\begin{equation}
\mathcal{L}\big(d \vert \boldsymbol{\theta}\big) = \prod_i \frac{1}{\sqrt{2\,\pi \,\sigma_i^2}} \exp \left[ - \frac{(m_i(\boldsymbol{\theta})-\mu_i)^2}{2\,\sigma_i^2} \right]\,.
\end{equation}
In order to restrict our likelihood to be a function of only the signal parameters, we eliminate the nuisance parameters using the profile likelihood method,
\begin{equation}
\mathcal{L}\big(d \vert \boldsymbol{\theta}_{\rm sig}\big) = \mathcal{L}\big(d \vert \boldsymbol{\theta}_{\rm sig}, \hat{\boldsymbol{\theta}}_{\rm bkg}\big)\,,
\end{equation}
where the hat indicates evaluating the function at the values of $\boldsymbol{\theta}_{\rm bkg}$ that maximize the likelihood (see~\cite{Rolke:2004mj} for a review).

\begin{figure}[t!]
\begin{center}
\includegraphics[width=0.65\columnwidth]{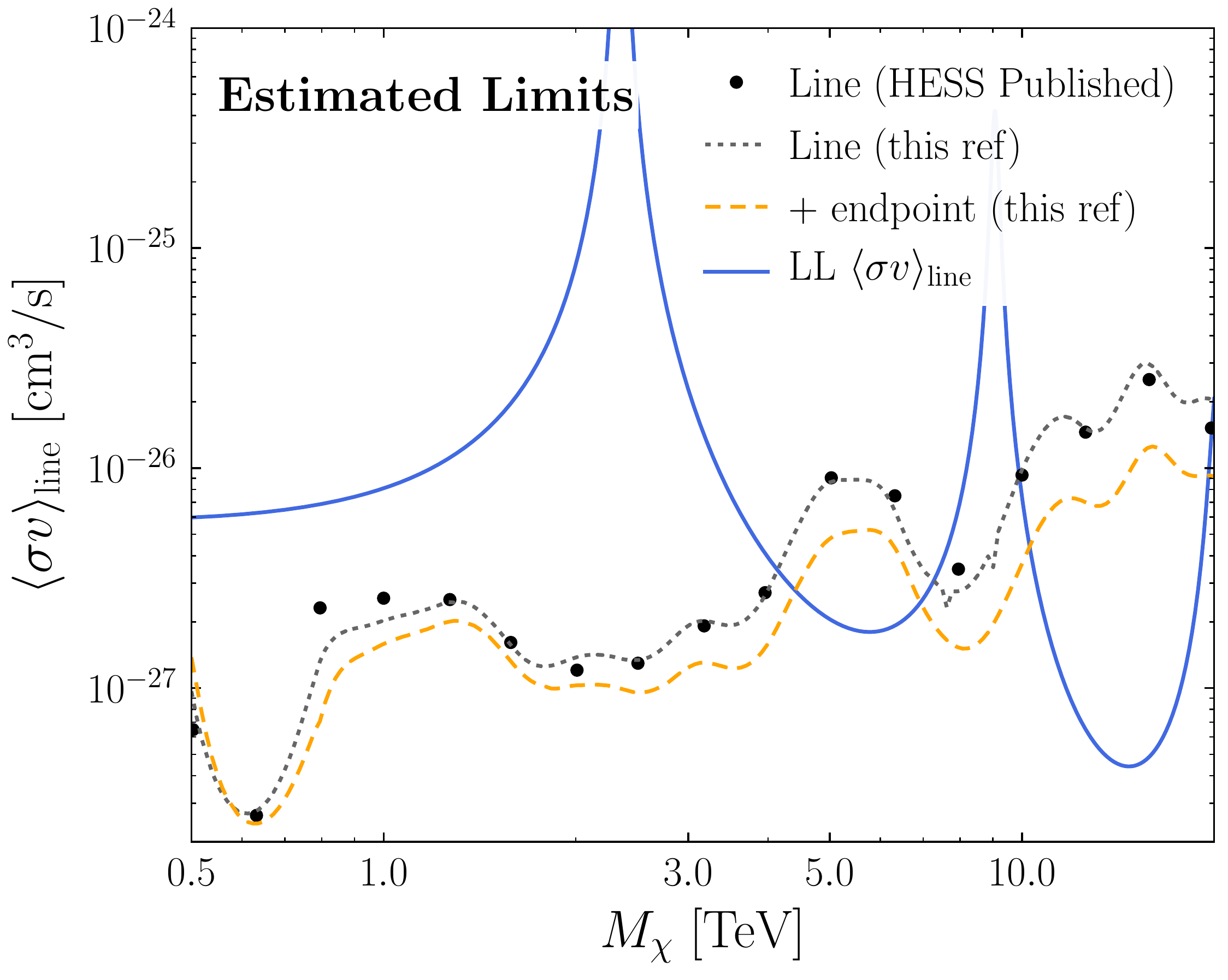}
\end{center}
\caption{
The approximate constraints on the line annihilation cross section as a function of the DM mass  for the Einasto profile using our mock reanalysis of the HESS line search.  The dotted line assumes the line-only spectrum and the dashed line assumes the full endpoint + line spectrum.  We additionally provide the LL resummed prediction (including the Sommerfeld enhancement) for the line annihilation.  Under these assumptions, the wino would be excluded when the LL prediction is above the HESS full constraint.  We also show the published HESS line limit in dots to demonstrate the extent to which our line-only analysis reproduces their result.
}
\label{fig:DM_limits_b}
\end{figure}

\begin{figure}[h!]
\begin{center}
\includegraphics[width=0.65\columnwidth]{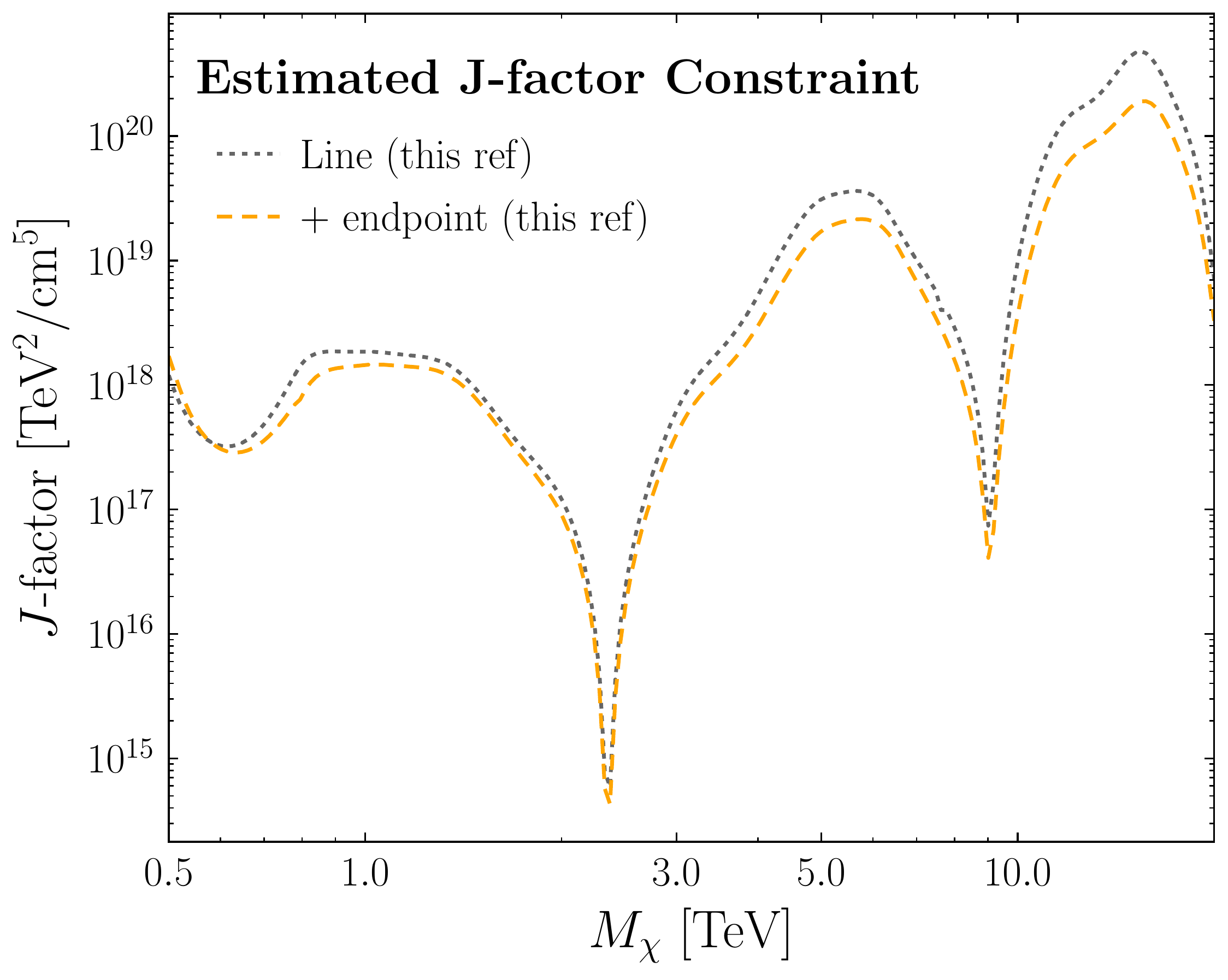}    
\end{center}
\caption{The approximate constraints on the $J$-factor as a function of the DM mass, assuming the line only spectrum and the full endpoint + line spectrum, as derived from our mock reanalysis of the HESS line search.}
\label{fig:DM_limits_a}
\end{figure}

With this reduced likelihood, we can then define a test statistic for upper limits as a function of $M_{\chi}$ on either $\langle \sigma v \rangle_{\rm line}$ or $J$. To begin with, we can fix $J$ and set a limit on $\langle \sigma v \rangle_{\rm line}$. To determine the fixed value of $J$, we use Eq.~\eqref{eq:Jfactor} assuming an Einasto profile as given in Eq.~\eqref{eq:Einasto}. The ROI for this dataset was a 1$^{\circ}$ circle around the Galactic Center, with the Galactic plane masked for latitudes less than 0.3$^{\circ}$, which yields
\begin{equation}
J \simeq 7.39 \times 10^{18}~{\rm TeV}^2~{\rm cm}^{-5}\,.
\end{equation}
Fixing this value, we define the test statistic as
\begin{equation}
\hspace{-7pt}q_{\langle \sigma v \rangle_{\rm line}} \left( M_{\chi} \right) \equiv \left\{
\begin{array}{ll} 2 \left[ \log \mathcal{L}(d \vert M_{\chi}, \langle \sigma v \rangle_{\rm line}) - \log \mathcal{L}(d \vert M_{\chi}, \widehat{\langle \sigma v \rangle}_{\rm line}) \right] & \langle \sigma v \rangle_{\rm line} \geq \widehat{\langle \sigma v \rangle}_{\rm line} \\[10pt]
0 & \langle \sigma v \rangle_{\rm line} < \widehat{\langle \sigma v \rangle}_{\rm line}
\end{array}
\right.\!,
\end{equation}
where again a hat denotes the value that maximizes the likelihood. Using this test statistic, the 95\% limit on $\langle \sigma v \rangle_{\rm line}$ is then determined by solving for $q_{\langle \sigma v \rangle_{\rm line}} \left( M_{\chi} \right) = -2.71$, and is shown in Fig.~\ref{fig:DM_limits_b}. In this figure we have also shown the prediction for the wino cross section -- \emph{if} these were exact limits and \emph{if} the DM distribution followed an Einasto profile in the inner galaxy, then the wino would be excluded when this prediction is above the mock limit curve.
 
This figure also contains the published HESS limits, taken from Fig.~4 of~\cite{Abramowski:2013ax}. The extent to which our line-only limits disagree with the published values highlights that our mock analysis is not exact and thus should not be taken as the true limit on wino DM. Nevertheless the figure does make it clear that the addition of the endpoint contributions can lead to a non-trivial enhancement on the sensitivity. For this reason, the effects calculated in this work represent an important contribution that should be included in future searches for heavy DM annihilation.

Alternatively, for limits on $J$, we fix $\langle \sigma v \rangle_{\rm line}$ to the exclusive wino prediction appropriate for that mass using \Eq{eq:resummed}, and in a similar notation to~\cite{Cowan:2010js}, define our test statistic as
\begin{equation}
q_J \left( M_{\chi} \right) \equiv \left\{
\begin{array}{ll} 2 \left[ \log \mathcal{L}\big(d \vert M_{\chi}, J\big) - \log \mathcal{L}\big(d \vert M_{\chi}, \hat{J}\big) \right] & J \geq \hat{J} \\[10pt]
0 & J < \hat{J}
\end{array}
\right.\,.
\end{equation}
As for the cross section, this test statistic allows us to establish the 95\% limit at a given mass through the relation $q_J \left( M_{\chi} \right) = -2.71$, and the result is shown in Fig.~\ref{fig:DM_limits_a}. In this case we have repeated the analysis with and without the endpoint contributions calculated in this work, with the impact of our calculation being as much as a factor of 3 improvement in the limit, and a factor of $\sim$1.5 at the thermal mass.

\begin{figure}[t!]
\begin{center}{
\includegraphics[width=0.65\columnwidth]{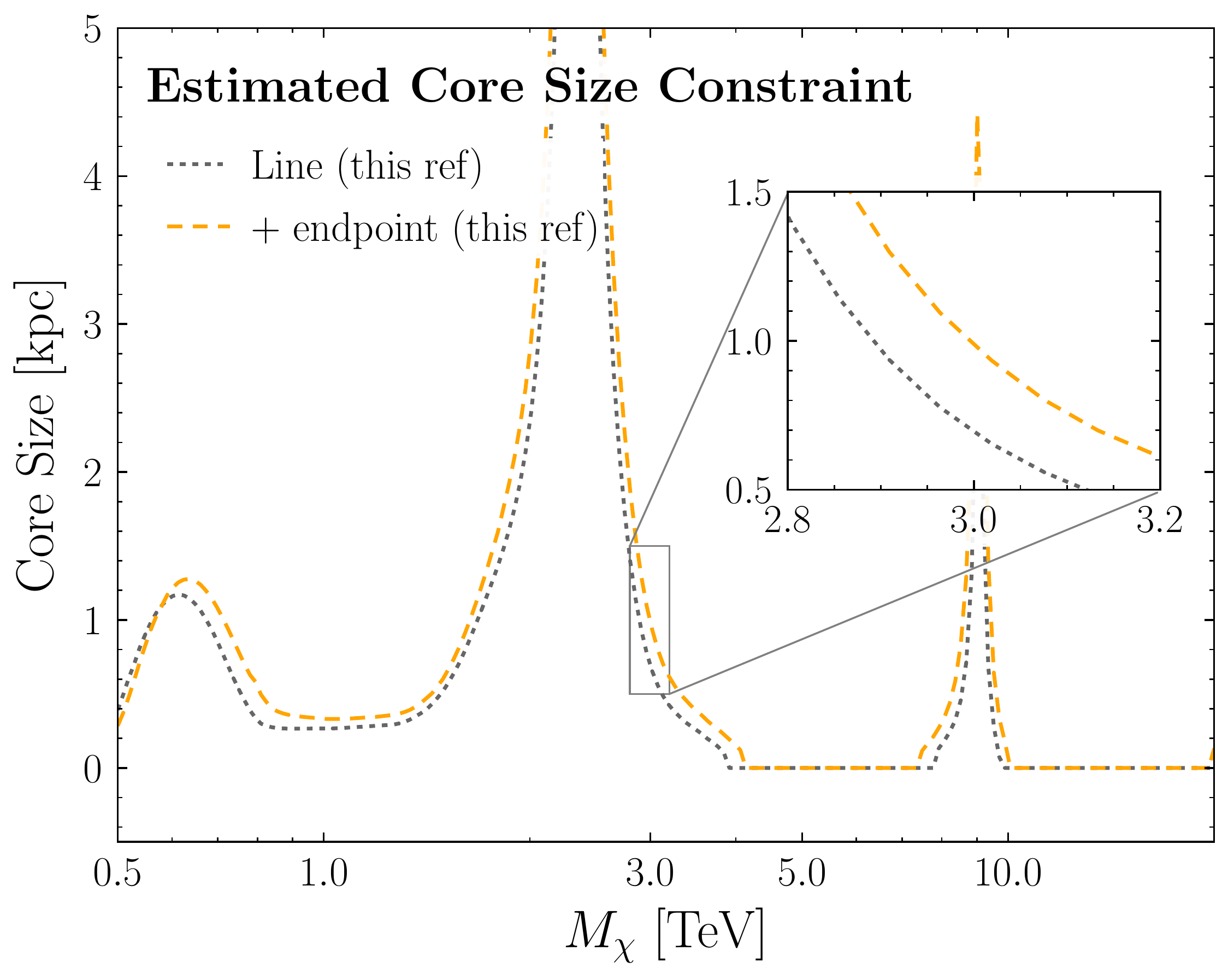}    
}
\end{center}
\caption{The NFW core size required to save the wino as derived using our mock analysis. This figure follows from the $J$-factor constraints given in Fig.~\ref{fig:DM_limits_a}.  At a given mass, the constraint on $J$ can be converted into a core size limit by calculating the corresponding cored NFW $J$-factor in the HESS analysis ROI.  For a thermal wino at 3 TeV the estimated constraints improve from 0.70 to 0.99 kpc when including the endpoint contributions.  This core size is beginning to be probed in both numerical and astrophysical settings.  We again emphasize that these constraints should be only taken as an estimate.
}
\label{fig:DM_core_limits}
\end{figure}

The results above demonstrate that updating the wino limits to include the endpoint contribution can easily lead to $\mathcal{O}(1)$ improvements in the limits on $\langle \sigma v \rangle_{\rm line}$ or the Galactic Center $J$-factor. Finally, we emphasize that the search for the wino is reaching a level of sensitivity such that $\mathcal{O}(1)$ factors are important.  One way to see this, is by converting the limits into a statement on how large a core in the Milky Way DM density profile is required to save the wino from the HESS constraints.

For concreteness, we use a cored version of the NFW profile, following~\cite{Fan:2013faa}.  For $r > r_{\rm core}$, we take the NFW profile as defined in \Eq{eq:NFW}; when $r \leq r_{\rm core}$, we set the profile to a constant value $\rho_\text{\tiny NFW}(r_{\rm core})$, such that the density profile is flat within the core radius. Restricting ourselves to cores smaller than 8.5 kpc, the presence of a core reduces the associated $J$-factor of the halo. In this way we can directly convert $J$-factor limits into a corresponding constraint on $r_{\rm core}$, which we show in Fig.~\ref{fig:DM_core_limits}. From this, we can see that for a thermal wino at exactly 3 TeV, the estimated core constraint increases from 0.70 kpc to 0.99 kpc when including the additional photons from the endpoint spectrum.

The values constrained in Fig.~\ref{fig:DM_core_limits} turn out to be at the edge of the core sizes that are beginning to be probed by a combination of numerical simulations and data. On the numerical side, it was shown that recent simulations of Milky Way-like halos in simulations including the effects of baryons, can potentially contain cores up to $\mathcal{O}(1)$ kpc~\cite{Chan:2015tna}. The total DM mass in the Galactic Bulge region can be estimated from observations of stars in the Bulge \cite{2015MNRAS.448..713P}, and disfavors a canonical NFW profile with a core size larger than $\sim$2 kpc~\cite{Hooper:2016ggc}. The core sizes needed for the thermal relic wino to survive indirect detection bounds are thus beginning to be constrained by stellar observations; accounting for the detailed endpoint spectrum is an important component when drawing this conclusion.

\section{Conclusions}\label{sec:conc}

In this paper we have developed a comprehensive effective field theory framework to compute the photon spectrum for annihilating (or decaying) DM.  We provided a new factorization formula, which allows for a resummation of all large logarithmic contributions, properly treating the effects due to electroweak symmetry breaking, the experimental resolution on the $\gamma+X$ final state, and the Sommerfeld enhancement.  We have computed the relevant one-loop anomalous dimensions, showing the consistency of the factorization formula at this order.  We have shown that the contribution from the spectrum has a numerically important effect for experimental searches of interest, \emph{e.g.} gamma-ray line searches from the HESS telescope.   Our final result is a compact analytic expression for the differential spectrum at LL accuracy, which can easily be convolved with experimental resolution functions to provide realistic predictions.

Our EFT can be interpreted as a mother theory that includes as particular limits the fully exclusive and fully inclusive cases.  The framework presented here correctly describes the transition between these two limits, allowing us to understand how Sudakov double logarithms impact the spectrum as a function of the experimental resolution.  It also allows us to assess the extent to which a fully exclusive or fully inclusive approximation, as had been previously considered in the literature,  is appropriate. Interestingly, we find that for the range of resolution parameters applicable for current and near future experiments, the result is intermediate between the fully exclusive and fully inclusive predictions.  This resolves the differing conclusions obtained in the literature, and provides a unifying picture of the importance of Sudakov resummation for indirect detection searches.  We have estimated the impact on the interpretation of current searches by providing a mock reanalysis of the HESS data, and we find that we are probing core sizes in a region where precise calculations of the particle physics components are relevant. 

Now that this paper has established an EFT framework for describing the photon spectrum resulting from DM annihilation, one can extend this work in a  number of future directions.  It would be of formal interest to understand the structure of the factorization and resummation at higher logarithmic order. Although the electroweak couplings are small, significantly improved uncertainties have been observed at NLL~\cite{Bauer:2014ula,Ovanesyan:2014fwa,Ovanesyan:2016vkk}, implying that NLL is likely the highest order that is relevant.  Additionally, the explicit NLO calculations provided in~\cite{Bauer:2014ula,Ovanesyan:2016vkk} demonstrate that higher order terms that are not logarithmically enhanced are numerically unimportant, justifying our choice to neglect them.  

There are also additional phenomenological applications.  One could extend these results to other heavy DM models, \emph{e.g.} a thermal Higgsino, a mixed wino-higgsino, or minimal DM candidates.  In many of these cases, the constraints can be different~\cite{Fan:2013faa,Beneke:2016jpw,Krall:2017xij}, implying that a dedicated analysis is warranted.  From the point of view of extending the work presented here, the underlying EFT is unchanged, but one must modify the Sommerfeld calculation and the explicit values for the hard matching coefficients and anomalous dimensions.

A simple heavy DM candidate provides a viable and phenomenologically relevant explanation for the observed relic abundance that could show up in current or future indirect detection searches.  This work casts the prediction for the photon spectrum that can result from this class of models in a theoretically robust setting, where perturbation theory can be maintained by performing resummation of all large double logarithms.  If a signal of heavy DM annihilation appears, this work will be critical to interpreting it.

\begin{acknowledgments}

We thank Martin Bauer, Marco Cirelli, Richard Hill, Emmanuel Moulin, Duff Neill, and Lucia Rinchiuso for useful discussions. 
MB is supported by the U.S. Department of Energy, under grant number DE-SC0003883.
TC is supported by the U.S. Department of Energy, under grant number DE-SC0018191. 
IM is supported by the U.S. Department of Energy, under grant number DE-AC02-05CH11231 and the LBNL LDRD program.
NLR and TRS are supported by the U.S. Department of Energy, under grant numbers DE-SC00012567 and DE-SC0013999.
MPS is supported by the U.S. Department of Energy, under grant number DE-SC0011632. 
IWS is supported by the Office of Nuclear Physics of the U.S. Department of Energy under the Grant No. DE-SCD011090 and by the Simons Foundation through the Investigator grant 327942.
VV was supported by the Office of Nuclear Physics of the U.S. Department of Energy under the Grant No.  Contract DE-AC52-06NA25396 and through the LANL LDRD Program.

\end{acknowledgments}

\appendix
\addcontentsline{toc}{section}{\protect\numberline{}Appendices}%
\addtocontents{toc}{\protect\setcounter{tocdepth}{1}}
\section*{Appendices}
\section{One-Loop Calculations}
\label{sec:one_loop_app}
In this Appendix we provide details of the calculation of the one-loop anomalous dimensions for the different functions appearing in the factorization formula, or provide references where they can be obtained from known results. Details of the refactorization are provided, and relevant integrals used in the calculation are also collected.

\subsection{One-Loop Calculation and Anomalous Dimensions: Intermediate EFT}\label{sec:app_int_EFT}

We begin by giving details related to the calculation of the anomalous dimensions for the intermediate EFT, before refactorization. This will help to make clear how these anomalous dimensions,  and the associated divergences, are split in the refactorized description.

\vspace{0.5cm}
\noindent{\bf{Hard Function}}
\vspace{0.3cm}

The hard function is independent of the infrared measurement made on the final state. It can therefore be extracted directly from the literature. Although we will only consider the LL resummation in this paper, we give the NLL anomalous dimension for completeness. The anomalous dimension matrix for $(C_{1}\ C_2)^T$ can be written in terms of a diagonal and a non-diagonal component as 
\begin{eqnarray}
&& \hat\gamma= 2\, \gamma_{W_T} \,\id + \hat\gamma_S \,.
\end{eqnarray}
Explicit results for $\gamma_{W_T}$ and $\hat\gamma_S$ were given in \cite{Ovanesyan:2014fwa}, namely
\begin{align} \label{eq:anomdimWT}
\gamma_{W_T}^{\rm NLL}=\frac{\aW}{4\,\pi} \Gamma_0 \log\left(\frac{2\,M_\chi}{\mu}\right) -\frac{\aW}{4\,\pi}\, b_0 + \Big(\frac{\aW}{4\,\pi}\Big)^2 \Gamma_1 \log\left(\frac{2\,M_\chi}{\mu}\right),
\end{align}
and
\begin{eqnarray}
\hat \gamma_{S}^{\rm NLL}=\frac{\aW}{\pi}(1-i\pi)
\bigg( \begin{array}{ccc}
2 & \,\,\,\,\,\,\,\,1 \\
0 & \,\,\,\,-1  \end{array} \bigg)
 - \frac{2\,\aW}{\pi} 
\bigg( \begin{array}{ccc}
1 & \,\,\,\,\,\,0 \\
0 & \,\,\,\,\,\,1  \end{array} \bigg)
 \,.
\end{eqnarray}
The constants appearing in these expressions are the  SU$(2)$ Casimir $C_A=2$, the one-loop $\beta$-function $b_0=19/6$, and the relevant cusp anomalous dimensions are $\Gamma_0=4 \,C_A$ and $\Gamma_1 = 8\left(\frac{70}{9}-\frac{2}{3}\,\pi^2\right)$.

\vspace{0.5cm}
\noindent{\bf{Photon Jet Function}}
\vspace{0.3cm}

The photon jet function, which has a single photon as its final state, is defined in \Eq{eq:JDefs} as 
\begin{align}
J_\gamma =  \Big\langle 0\Big| B_{n \perp}^{c}(0)\Big| \gamma \Big\rangle \Big\langle \gamma \Big| B_{n \perp}^c(0) \Big|0 \Big\rangle\,.
\end{align}
Evaluating this function at one-loop yields
\begin{align}
J_\gamma =&-2 -2\,C_A\, \frac{\aW}{\pi} \left(\frac{\mu}{\mW}\right)^{2\, \epsilon} \left(\frac{\nu}{2\, M_\chi}\right)^{\eta}\frac{\Gamma(\epsilon)}{\eta}\nn\\[5pt]
 &+ \frac{\aW}{2\, \pi} \Gamma(\epsilon) \left( \frac{\mu}{\mW}\right)^{2\, \epsilon} \left[ \frac{11}{3} \,C_A -\frac{4}{3} \,n_f\, C(R)\right] + \mathcal{O}\big(\aW^2\big)\,,
\end{align}
where $\mu$ and $\nu$ are the virtuality and rapidity renormalization scales respectively. Here $n_f$ denotes the number of fermion flavors. We take $n_f=5$ in our numerical results.  The $\mu$ and $\nu$ anomalous dimensions can immediately be extracted from this result, and we find, 
\begin{align}
\gamma_{\mu}^{n} &=  2\,C_A \frac{\aW}{\pi}\log \left(\frac{\nu}{2\,M_\chi}\right)\,, \\[10pt]
\gamma_{\nu}^{n} &=   2\,C_A \frac{\aW}{\pi}\log\left(\frac{\mu}{\mW}\right)\,.
\end{align}

\vspace{0.5cm}
\noindent{\bf{Recoiling Jet Function}}
\vspace{0.3cm}

When computing the recoiling jet function, all IR divergences are explicitly regulated by the measurement of the final state mass. This is unlike the photon jet function, where the scale $\mW$ acts as a regulator. To compute the anomalous dimensions, it is therefore sufficient to expand away the scale $\mW$ from the beginning, simplifying the calculation. The inclusive recoiling jet function is then defined as 
\begin{align}
  J'_{\bar n}(k^+) &= \sum_{X_{C}} \Big\langle 0\Big| B_{\bar n\perp}^{d}\!(0)\,\delta\big(k^+ - \mathcal{P}^+\big)\,\delta\big(M_{\chi} - \mathcal{P}^-/2\big)\,\delta^2\big(\vec{\mathcal{P}}_{\perp}\big)\Big |X_C \Big\rangle\Big \langle X_C\Big |B_{\bar n\perp}^d\!(0)\Big|0 \Big\rangle\,.
\end{align}
We can rewrite this jet function as 
\bea
J'_{\bar n}(p) = \sum_{X_{C}}\int \frac{\text{d}^4x}{(2\,\pi)^4}\, e^{i\, p \cdot x}\,\Big\langle 0\Big| B_{\bar n\perp}^{d}\!(0)\Big |X_C\Big\rangle \Big\langle X_C\Big |B_{\bar n\perp}^d\!(x)\Big|0 \Big\rangle\,.
\eea
with $p = (2\,M_{\chi} ,k^+,0)$ in order to enforce the $\delta$-function measurement constraints.  Written in this form, the function is completely inclusive.  Therefore, we can use the optical theorem to write this as the imaginary part of the forward scattering amplitude 
\bea
J'_{\bar n}(p)  = \frac{1}{2} \,\text{Im}\int\! \frac{\text{d}^4x}{(2\,\pi)^4} \,e^{i\, p \cdot x}\,\Big\langle 0\Big|\, T\,{B_{\bar n\perp}^{d}\!(0) B_{\bar n\perp}^d\!(x)}\Big|0 \Big\rangle\,.
\eea
This jet function has been evaluated in the literature~\cite{Bauer:2003pi,Bosch:2004th,Becher:2006qw}; the one-loop result is  
\bea
 J'_{\bar n}(p^2)= \delta(p^2) + \frac{\aW}{4\,\pi} \left( \frac{4\, C_A \log (p^2/\mu^2) - b_0}{p^2}\right)_+ + \mathcal{O}\big(\aW^2\big)\,,
\eea
where the subscript plus denotes a plus distribution, see \emph{e.g.}~\cite{Ligeti:2008ac} for an extensive review of its properties.  The kinematics for heavy DM annihilation imply that $p^2= 2\,M_{\chi}\,k^+$, so that 
\bea
  J'_{\bar n}(k^+)= \delta\big(2\,M_{\chi}\,k^+\big) + \frac{\aW}{4\,\pi} \frac{1}{\mu^2} \left( \frac{4 \,C_A \log \big(2\,M_{\chi}\,k^+/\mu^2\big) - b_0}{2\,M_{\chi}\,k^+/\mu^2}\right)_+ + \mathcal{O}\big(\aW^2\big)\,.
\eea
To expose the simple renormalization group structure, we transform to Laplace space, where the Laplace conjugate variable of $k^+$ is taken to be $s$.  Keeping only the leading log term, we find
\bea
J'_{\bar n}(s) = \frac{1}{2\,M_{\chi}} +  2 \,C_A\frac{\aW}{4\,\pi}\frac{1}{2\, M_{\chi}}\log^2\left(\frac{\mu^2 \,s \,e^{\gamma_E}}{2\,M_{\chi}}\right)+ \mathcal{O}\big(\aW^2\big)\,,
\eea
where $\gamma_E$ is the Euler-Mascheroni constant. Finally, we extract the $\mu$ anomalous dimension 
\bea\label{eq:mudim_jet}
\gamma_{\mu}^{\bar n} = 2\, C_A\frac{\aW}{\pi} \log \left(\frac{\mu^2\, s\, e^{\gamma_E}}{2\,M_{\chi}} \right) \,.
\eea
This function has no rapidity anomalous dimension as it is a $\SCETi$ type function.

\vspace{0.5cm}
\noindent{\bf{Ultrasoft Function}}
\vspace{0.3cm}

There are four operators that contribute to the ultrasoft function in the EFT: $S'_{12}$, $S'_{21}$, $S'_{11}$, $S'_{22}$, see \Eq{eq:SprimeDefs} above.  Using the expressions below, we can then extract the LL $\mu$ and $\nu$ anomalous dimensions.  We will find that each operator yields the same result, 
\begin{align}
\gamma_{\mu}^{S'} &= - 2 \,C_A\,\frac{\aW}{\pi}  \log \big(\nu\, s\big)\,,\nn\\[10pt]
\gamma_{\nu}^{S'} &= - 2 \,C_A\,\frac{\aW}{\pi} \log \left(\frac{\mu}{\mW} \right)\,.
\label{eq:AppendixSoftAnnDim}
\end{align}
This calculation will also expose additional IR divergent contributions, which is the sign that refactorization is necessary.  

The one-loop results will be expressed in terms of several integrals, denoted in bold and labeled with $\bf{V}$ and $\bf{R}$ for virtual and real respectively, which are defined and evaluated below. These integrals are evaluated using dimensional regularization as an IR regulator, and with the rapidity regulator as defined in \Sec{sec:scet}.  The integrals that we will require in our calculation are defined as follows.  The ${\bf n\bar n}$ integrals are 
\label{sec:Int}
\bea
\label{IRnnbar}
{\bf I^{R}_{n \bar n }} &=& -\gW^2 \int \frac{\text{d}^d\ell}{(2\,\pi)^{d-1}} \frac{2\, \delta^+\big(\ell^2-\mW^2\big) \big|\ell^+-\ell^-\big|^{-\eta/2}\delta(q^+-\ell^+)}{\ell^+\, \ell^-} \nn\\[5pt]
&=& -\frac{\aW}{2\,\pi} \left(\frac{\mu}{\mW}\right)^{2\,\epsilon}\left(\frac{\nu\, q^+}{\mW^2}\right)^{\eta/2} \frac{\Gamma[\epsilon+\eta/2]}{q^+} \,,
\eea
\bea
\label{Ivnnbar}
{\bf I^V_{n \bar n}}&=&   \delta\big(q^+\big)\,\gW^2\, \mu^{2\, \epsilon}\, \nu^{\eta/2} \int \frac{\text{d}^d \ell}{(2\,\pi)^d} \frac{-2\,i}{\left(\ell^2-\mW^2+i0\right)} \frac{\big|\ell^+-\ell^-\big|^{-\eta/2} }{\left(\ell^+ +i0 \right)(\ell^--i0)}\nn\\[5pt]
&=&- \delta\big(q^+\big)\frac{2\, \aW}{\pi} \left(\frac{\mu}{\mW}\right)^{2\,\epsilon}\left(\frac{\nu}{\mW}\right)^{\eta/2}\frac{2^{-\eta/2}}{\eta}\frac{\Gamma[\epsilon+\eta/4]\Gamma[1/2-\eta/4]}{\Gamma[1/2]} \nn\\[5pt]
&&+i \pi\,\delta\big(q^+\big)\frac{\aW}{2\,\pi} \left(\frac{\mu}{\mW}\right)^{2\,\epsilon} \Gamma[\epsilon]\,.
\eea
The $i\pi$ term in this expression arises from a Glauber contribution, and as it does not contribute at LL we will not consider it further, although we include it for completeness as it is relevant at NLL~\cite{Baumgart:2018yed}. Note the expression here agrees with~\cite{Beneke:2019vhz}. Continuing, the ${\bf nv}$ integrals are 
\bea
 {\bf I^V_{n v}} &=& \gW^2\, \mu^{2\, \epsilon}\,\nu^{\eta/2} \int \frac{\text{d}^d \ell}{(2\,\pi)^d} \frac{-2\,i}{\left(\ell^2-\mW^2+i0\right)} \frac{\big|\ell^+\big|^{-\eta/2} }{\left(\ell^++\ell^- -i0 \right)(\ell^--i0)} \nn\\[5pt]
 &=&-\frac{\aW}{\pi} \left(\frac{\mu}{\mW}\right)^{2\,\epsilon}\left(\frac{\nu }{\mW}\right)^{\eta/2} \frac{\Gamma[\epsilon+\eta/4]\Gamma[1-\eta/4]}{\eta}\,,
\eea
\bea
{\bf I^R_{n v}} &=& -\gW^2 \int \frac{\text{d}^d\ell}{(2\,\pi)^{d-1}} \frac{ 2 \delta^+\big(\ell^2-\mW^2\big) \delta\big(q^+-\ell^+\big)}{\left(\ell^+ + \ell^-\right) \ell^-}\nn\\[5pt]
&=& -\frac{\aW}{\pi} \frac{1}{q^+} \log \left(\frac{\sqrt{(q^+)^2+\mW^2}}{\mW}\right) \,,
\eea
and the ${\bf \bar{n} v}$ integrals are 
\bea
\label{nbarv}
{\bf I^V_{\bar n v}} &=& \delta\big(q^+\big)\, \gW^2\, \mu^{2\, \epsilon}\nu^{\eta/2} \int \frac{\text{d}^d \ell}{(2\,\pi)^d} \frac{-2\,i}{\left(\ell^2-\mW^2+i0\right)} \frac{\big|\ell^+-\ell^-\big|^{-\eta/2}}{\left(\ell^+ -i0 \right)(\ell^++\ell^--i0)} \nn\\[5pt]
&=& -\delta\big(q^+\big)\frac{\aW}{\pi} \left(\frac{\mu}{\mW}\right)^{2\, \epsilon} \left(\frac{\nu}{\mW}\right)^{ \eta/2}\frac{\Gamma[\epsilon+\eta/4] \Gamma[1-\eta/4]}{\eta} \,,
\eea
\bea
{\bf I^R_{\bar n v}} &=& -\gW^2 \int \frac{\text{d}^d\ell}{(2\,\pi)^{d-1}} \frac{ 2\, \delta^+\big(\ell^2-\mW^2\big) \delta\big(q^+-\ell^+\big)\big|\ell^+-\ell^-\big|^{-\eta/2}}{\left(\ell^+ + \ell^-\right) \left(\ell^+\right)} \nn\\[5pt]
&=&  {\bf I^R_{n \bar n}}- {\bf I^R_{nv}}\,.
\eea
Next, we consider each of the four ultrasoft functions in turn.  First we provide the operator definition, followed by the tree-level and one-loop evaluation in order to compute the anomalous dimensions for the different color structures of the ultrasoft function. Since we are doing this in the EFT before refactorization, we will refer to these as ultrasoft functions and the corresponding states as $|X_{US}\rangle$. These ultrasoft functions will ultimately be refactorized. 

\begin{itemize}[leftmargin=*]
\item $S'_{11}$ is defined as  
\begin{align}
S^{\prime\ aba'b'}_{11} =  \sum_{X_{US}} &\Big\langle 0\Big| \left(Y_n^{3f'}\,Y_{\bar n}^{dg'}\right)^{\dagger}\!\!(0)\Big|X_{US}\Big\rangle\nn\\[5pt] &\times \Big\langle X_{US}\Big| \delta(q^+ - \mathcal{P}^+)\,\left(Y_n^{3f}\,Y_{\bar n}^{dg}\right)\!(0) \Big|0 \Big\rangle \,\delta^{f'g'}\delta^{a'b'}\delta^{fg}\delta^{ab}\,.
\end{align}
Evaluating at tree-level in Laplace space yields 
\begin{align}
\left(S^{\prime\ aba'b'}_{11}\right)^\text{tree} &= \delta^{ab}\delta^{a'b'}\, ,
\end{align}
and at one-loop yields
\begin{align}
\left(S^{\prime \ aba'b'}_{11}\right)^\text{1-loop}=&-\delta^{ab}\delta^{a'b'}\,2\,C_A \,\Big({\bf I^{R}_{n \bar n}}-{\bf I^{V}_{n \bar n}}\Big)\nn\\[10pt]
\xrightarrow[\text{Laplace}]{}\hspace{5pt}&-\delta^{ab}\delta^{a'b'}\,2 \,C_A\,\frac{\aW}{2\,\pi} \left(\frac{\mu}{\mW}\right)^{2\, \epsilon} \Gamma[\epsilon]\left(\frac{2}{\eta} + \log \big(\nu \,s\big)\right)\,,
\end{align}
where the second line is expressed in Laplace space.  Extracting the LL anomalous dimensions from these results yields \Eq{eq:AppendixSoftAnnDim}.
\item $S'_{12}$ and $S'_{21}$ are defined as  
\begin{align}
S^{\prime\ aba'b'}_{12} &= \sum_{X_{US}} \Big\langle 0\Big| \left(Y_n^{3f'}Y_{\bar n}^{dg'}\right)^{\dagger}\!\!(0) \,\delta\big(q^+ - \mathcal{P}^+\big)\Big|X_{US}\Big\rangle\nn\\[5pt]
&\hspace{30pt}\times \Big\langle X_{US} \Big|\left(Y_n^{3g}Y_{\bar n}^{df}Y_v^{ag}Y_v^{bf}\right)\!(0) \Big|0 \Big\rangle\, \delta^{f'g'}\delta^{a'b'}\,,\nn\\[10pt]
S^{\prime\ aba'b'}_{21} &= S'^{\,a'b'ab}_{12}\,.
\end{align}
Evaluating at tree-level in Laplace space yields
\bea
\left(S^{\prime\ aba'b'}_{12}\right)^\text{tree} = \delta^{b3} \delta^{a3} \delta^{a'b'}\,,
\eea
and at one-loop yields
\begin{align}
\left(S^{\prime\ aba'b'}_{12}\right)^\text{1-loop} =& -\delta^{a'b'}\Big[ \big(-\delta^{a3} \delta^{b3}-\delta^{ab}\big)\Big({\bf I^V_{n \bar n}}-{\bf I^R_{n \bar n}}\Big)\nn\\[5pt]
&\hspace{35pt}+ \big(\delta^{ab}-3\delta^{a3}\delta^{b3}\big)\Big({\bf I^V_{nv}}+{\bf I^R_{nv}} +{\bf I^V_{\bar n v}}-{\bf I^R_{\bar n v}}\Big)\Big] \nn \displaybreak[4]\\
\xrightarrow[\text{Laplace}]{}\hspace{5pt}& -\delta^{a'b'}\delta^{a3}\delta^{b3}\,2\, C_A\, \frac{\aW}{2\,\pi}\, \mu^{2\, \epsilon}\, \Gamma[\epsilon]\left( \frac{2}{\eta}+ \log \big(\nu\, s\big)\right) \nn\\[5pt]
&+ \delta^{a'b'}\big(\delta^{ab}-3\,\delta^{a3}\delta^{b3}\big)\frac{\aW}{\pi} \log^2\big(\mW\,s\big)\,,
\end{align}
where the second line is expressed in Laplace space.  Extracting the LL anomalous dimensions from these results yields \Eq{eq:AppendixSoftAnnDim}.

\hspace{10pt} This result manifests the same UV virtuality and rapidity divergences as in the case of the $S'_{11}$ operator which is why it yields the same anomalous dimension as $S'_{11}$.  However, we see an additional IR divergence appears in the form of $\log^2\big(\mW\,s\big)$.  This results from the non-singlet nature of this operator.  In order to factorize this new double log, we need to match this ultrasoft operator onto an EFT below the scale $s$.  This allows us to separate the scales $s$ and $\mW$, yielding our final fully factorized result.

\item $S'_{22}$ is defined as
\begin{align}
S^{\prime\ aba'b'}_{22}=\sum_{X_{US}} &\Big\langle 0\Big| \left(Y_n^{3f'}Y_{\bar n}^{dg'}Y_v^{a'f'}Y_v^{b'g'}\right)^{\dagger}\!\!(0)\,\delta\big(q^+ - \mathcal{P}^+\big)\Big|X_{US}\Big\rangle\nn\\[5pt] 
&\times \Big\langle X_{US} \Big|\left(Y_n^{3f}Y_{\bar n}^{dg}Y_v^{af}Y_v^{bg}\right)\!(0) \Big|0 \Big\rangle\,.
\end{align}
Evaluating at tree-level in Laplace space yields 
\bea
\left(S^{\prime\ aba'b'}_{22}\right)^\text{tree}= \delta^{a3}\delta^{a'3}\delta^{bb'} \,,
\eea
and at one-loop yields
\begin{align}
\left(S^{\prime\ aba'b'}_{22}\right)^\text{1-loop} =&  \left(-\delta^{a3} \delta^{b3} \delta^{a'b'} +\delta^{a3}\delta^{b'3}\delta^{a'b}- \delta^{a'3}\delta^{b'3}\delta^{ab}+\delta^{b3}\delta^{a'3}\delta^{ab'}\right)\Big({\bf I^V_{n \bar n}}-{\bf I^R_{n \bar n}}\Big)\nn\\[5pt]
&+ \Big(\delta^{a3}\Big\{ -2\, \delta^{a'3}\delta^{bb'}-\delta^{b'3}\delta^{a'b}+\delta^{b3}\delta^{a'b'} \Big\}\nn\\[5pt]
&\hspace{20pt}+ \delta^{a'3} \Big\{-2 \,\delta^{a3}\delta^{bb'}-\delta^{b3}\delta^{ab'}+\delta^{b'3}\delta^{ab}\Big\} \Big)\Big({\bf I^V_{\bar n v}}-{\bf I^R_{\bar n v}}+ {\bf I^V_{n v}}\Big)\nn\\[10pt]
\xrightarrow[\text{Laplace}]{}\hspace{5pt}&\, 2\,C_A\, \delta^{a3}\delta^{a'3}\delta^{bb'} \,\frac{\aW}{2\,\pi}\, \mu^{2\, \epsilon}\, \Gamma[\epsilon]\left( \frac{2}{\eta} + \log \big(\nu\, s\,\big)\right)\nn\\[5pt]
&+ \delta^{bb'}\big(\delta^{aa'}-3\,\delta^{a3}\delta^{a'3}\big)\frac{\aW}{\pi} \log^2\big(\mW\,s\big)\,.
\end{align}
where the second line is expressed in Laplace space.  Extracting the LL anomalous dimensions from these results yields \Eq{eq:AppendixSoftAnnDim}. Note that although the result appears not to be symmetric in the color structure, the wavefunction $F^{a'b'ab}$ defined in \Eq{eq:LSommerfeldFactors} is symmetric under the interchange $ a, a' \leftrightarrow b,b'$.

\end{itemize}

\subsection{Calculations in the Refactorized Theory}\label{sec:app_refac_basis}

Having presented the calculations for the anomalous dimensions in the intermediate EFT, in this section we discuss some details related to the refactorization that were skipped in the text, and present the anomalous dimensions in the refactorized theory.

\vspace{0.5cm}
\noindent{\bf{Photon Jet Function}}
\vspace{0.3cm}

The photon jet function $J_\gamma$ is only sensitive to a single scale $\mW$, and therefore is unmodified under the refactorization procedure.

\vspace{0.5cm}
\noindent{\bf{Recoiling Jet Function}}
\vspace{0.3cm}

As discussed in \Sec{sec:app_int_EFT}, although in the intermediate theory the recoiling jet function is sensitive both to the scale $M_\chi \sqrt{1-z}$ set by the final state measurement, as well as to the scale $\mW$, the final state measurement regulates all singularities, and could therefore be expanded to begin with. Combining this result with the structure of the factorization
\begin{align}
J'_\bn\big(M_\chi, \sqrt{1-z}, \mW, \mu\big) = H_{J_\bn}\big(M_\chi, \sqrt{1-z}, \mu\big)\, J_\bn\big(\mW, \mu, \nu\big)+\cO\left( \frac{\mW}{M_\chi \sqrt{1-z}} \right)\,,
\end{align}
we find that the one-loop result for the matching coefficient in Laplace space ($M_\chi \sqrt{1-z} \to s$) is given by
\bea
H_{J_\bn}(M_\chi,s,\mu) = \frac{1}{2\,M_{\chi}} +  2 \,C_A\frac{\aW}{4\,\pi}\frac{1}{2\, M_{\chi}}\log^2\left(\frac{\mu^2 \,s \,e^{\gamma_E}}{2\,M_{\chi}}\right)+ \mathcal{O}\big(\aW^2\big)\,.
\eea
This then implies that
\begin{align}
\frac{\text{d}}{\text{d}\log\mu} J'_\bn\big(M_\chi, \sqrt{1-z}, \mW, \mu\big)= \frac{\text{d}}{\text{d}\log\mu}   H_{J_\bn}\big(M_\chi, \sqrt{1-z}, \mu\big)\,,
\end{align}
which is given in \Eq{eq:mudim_jet},
and
\begin{align}
\frac{\text{d}}{\text{d}\log\mu} J_\bn\big(\mW, \mu, \nu\big)= 0\,.
\end{align}
To the order that we work, we need just the tree level value for $J_\bn$, which is 
\begin{align}
J^{\text{tree}}_\bn\big(\mW,\mu, \nu)=1\,.
\end{align}

\newpage
\noindent{\bf{Anomalous Dimensions for the Refactorized Ultrasoft Function}}
\vspace{0.3cm}

Unlike for the jet functions, the refactorization of the ultrasoft function is significantly more involved. As given in the text, the general form of the refactorization is
\begin{align}
S^{\prime \, aba'b'}_{i} \big(M_\chi, 1-z, \mW,\mu, \nu \big) = H_{S, ij}\big(M_\chi, 1-z, \mu\big) \,& \Big[ C_{S}\big(M_\chi, 1-z, \mW , \mu, \nu\big)  S\big( \mW, \mu\big) \Big]^{ab a'b'}_j \nn \\[5pt]
&\times \left[ 1+ \cO\left( \frac{\mW}{M_\chi (1-z)} \right) \right ]\,.
\end{align}
The goal of this section will be to describe this refactorization in more details, and derive the required anomalous dimensions.

Before considering the structure of the anomalous dimensions, we must first derive the color structures of the collinear-soft and soft functions, which were only stated without derivation in the main body of the text. The structure of the Wilson lines in the soft and collinear-soft functions can be derived  by performing the BPS field redefinition iteratively. We therefore return to the two  
amplitude level operators (see \Eq{eq:hardscattering} above) 
\bea
\cO_1 = \Big(\chi_v^{aT}\, i\sigma_2\, \chi_v^b\Big)\,\cB^c_{n \perp}\cB^d_{\bar n \perp}\, \delta^{ab} \delta^{cd}\,,\nn\\[10pt]
\cO_2 = \Big(\chi_v^{aT}\, i\sigma_2\, \chi_v^b\Big)\,\cB_{n \perp }^c \cB^d_{\bar n \perp}\, \delta^{ac}\delta^{bd}\,.
\eea
Next we iteratively perform the BPS field redefinition for both the collinear-soft modes and refactorized soft modes,
\bea
\cO_1 &=& \Big[\delta^{AB}\,V_{n}^{Dc}X_{n}^{Cc}\Big] \left[S_v^{\bar A A}S_{v}^{\bar B B}S_{\bar n}^{\bar D D} S_n^{\bar C C}\right] \Big(\chi_v^{\bar A T}\, i\sigma_2\, \chi_v^ {\bar B}\Big)\cB^{\bar C}_{n \perp}\cB^{\bar D}_{\bar n \perp}\,,\nn\\[10pt]
\cO_2   &=& \Big[\delta^{BD}\,V_{n}^{Ac}X_{n}^{Cc}\Big] \left[S_v^{\bar A A}S_{v}^{\bar B B}S_{\bar n}^{\bar D D} S_n^{\bar C C}\right]\Big(\chi_v^{\bar A T}\, i\sigma_2 \,\chi_v^{\bar B}\Big)\cB_{n \perp }^{\bar C} \cB^{\bar D}_{\bar n \perp}\,. 
\eea
We can now derive the soft and collinear-soft functions in the standard way, by squaring the amplitude level operators and setting $ \bar D = \bar D', \bar C = \bar C'=3$.
We find
\begin{align}
 \tilde S_{11} &= \Big\langle 0\Big| \left[V_{n}^{dc}X_{n}^{Cc}\right]\left[V_{n}^{dc'}X_{n}^{C'c'}\right]\times  \left[S_n^{3 C'}S_n^{3 C}\right]\delta^{\bar A \bar B}\delta^{\bar A' \bar B'}\Big|0\Big\rangle\,, \nn\\[10pt]
 \tilde S_{12} +\tilde S_{21}&=\Big\langle 0\Big|   \left[V_{n}^{B'c}X_{n}^{Cc}\right] \left[V_{n}^{A'c'}X_{n}^{C'c'}\right] \times \left[ S_n^{3 C}S_v^{\bar A' A'}S_{v}^{\bar B' B'} S_n^{3 C'}\right]\delta^{\bar A \bar B}+ \big\{ \bar A, \bar B  \leftrightarrow \bar A', \bar B' \big\}\Big|0\Big\rangle\,,\nn\\[10pt]
\tilde S_{22} &=\Big\langle 0\Big|  \left[V_{n}^{B'c}X_{n}^{Cc}\right] \left[V_{n}^{A'c'}X_{n}^{C'c'}\right] \times \left[ S_n^{3 C}S_v^{\bar A' A'}S_{v}^{\bar A B'} S_n^{3 C'}\right]\delta^{\bar B \bar B'} \Big|0\Big\rangle\,.
\end{align}
To simplify the notation and focus on the color structures, we have suppressed the measurement function.

One additional complication that arises in the refactorization of the ultrasoft function, is that there are non-trivial zero-bins \cite{Manohar:2006nz} that must be correctly incorporated. We therefore briefly discuss the structure of these zero-bins, and their dependence on our choice of regulator, showing through two examples how the factorization correctly reproduces the structure of integrands once the zero bin is included. We consider one example of a virtual integral and one example of a real integral, arising from the $S'_{11}$ integrand.
\begin{itemize}[leftmargin=*]
\item Consider the \emph{virtual} integral  (see \Eq{Ivnnbar} for the evaluation of the unexpanded integral)
 \bea
\label{Virtual}
{\bf I^{V}_{n \bar n}}&=&   \delta\big(q^+\big)\,\gW^2\, \mu^{2\, \epsilon}\, \nu^{\eta/2} \int \frac{\text{d}^d \ell}{(2\,\pi)^d} \frac{-2\,i}{\left(\ell^2-\mW^2+i0\right)} \frac{\big|\ell^+-\ell^-|^{-\eta/2} }{\left(\ell^+ +i0 \right)(\ell^--i0)}\,.
\eea
Let us now consider the collinear-soft limit ($\ell^+ \gg \ell^-$) of this integral. It would appear that according to the power counting the only effect is to drop $\ell^-$ from the rapidity regulator term $| \ell^+-\ell^-|^{\eta/2}$. Since the rest of the integrand is unchanged, this would lead to an unregulated divergence as $\ell^- \rightarrow \infty$. We would then be forced to introduce a new regulator to counter this divergence.  While there are several ways to do this (a $\Delta$-regulator \cite{Chiu:2009yx}, for instance), the simplest way is just to keep the original form of the rapidity regulator. The choice of the regulator we use will affect the zero-bin subtraction that will be needed.  \\ \\
If we do not expand out the regulator, then the collinear-soft and soft limits of \Eq{Virtual} are identical to the full US integral. 
The soft-bin subtraction is implemented in the collinear-soft (CS) sector by subtracting out the soft limit of the CS integral. With this subtraction 
\begin{align}
{\bf I_{n \bar n}^{V,CS}}  &= 0\,, \nn\\[10pt]
 {\bf I_{n \bar n}^{V,S}}  &= {\bf I_{n \bar n}^{V}} \,,
\end{align}
so that we recover the full US virtual contribution. 

\item Now, consider the \emph{real} emission integral (see \Eq{IRnnbar} for the evaluation of the unexpanded integral)
\bea
 {\bf I^{R}_{n \bar n }} &=& -\gW^2 \int \frac{\text{d}^d\ell}{(2\,\pi)^{d-1}} \frac{2\, \delta^+\big(\ell^2-\mW^2\big) \big|\ell^+-\ell^-\big|^{-\eta/2}\delta\big(q^+-\ell^+\big)}{\ell^+\, \ell^-}\,.
\eea
The soft limit is 
\bea
 {\bf I^{R,S}_{n \bar n }} &=& - \delta\big(q^+\big)\, \gW^2 \int \frac{\text{d}^d\ell}{(2\,\pi)^{d-1}} \frac{2\, \delta^+\big(\ell^2-\mW^2\big) \big|\ell^+-\ell^-\big|^{-\eta/2}}{\ell^+\, \ell^-}\,. 
\eea
The CS limit is identical to the full integral. Applying the zero-bin subtraction to this (which turns out to be the same as the soft integral),  we are left with 
\bea
 {\bf I^{R,CS}_{n \bar n }} &=& {\bf I^{R}_{n \bar n }}-{\bf I^{R,S}_{n \bar n }}\,.
\eea
 Thus, once again we recover the full US real contribution adding together the CS and soft limits. 
The zero-bin subtraction scheme then is to simply subtract the soft limit of the CS integrals from the CS sector. 
\end{itemize}
The analysis of these integrals provides a non-trivial check that our factorization is indeed correct, and that the infrared is completely reproduced by our factorized description.

Having understood the operator basis and the structure of the zero-bin subtractions, we can now compute the anomalous dimensions of the functions arising after the factorization of the ultrasoft function. Here we can considerably simplify the calculation by using the choice of resummation path described in \Sec{sec:resum} and shown in \Fig{fig:RG_path}. In particular, for this path it is not necessary to separately run the collinear-soft and soft functions. We can therefore simplify our refactorization to
\bea
 S_{ij}^{\,aba'b'}= H_{S, ijkl}\Big(C_{S,k}^{{A}}\,S_l^{{B}}\Big)^{aba'b'}=H_{S, ijkl}\left(\tilde S_{kl}\right)^{aba'b'}\,,
\eea
and only compute the anomalous dimensions for the functions $H_{S, ijkl}$ and $\big(\tilde S_{kl}\big)^{aba'b'}$. This drastically simplifies the calculation, since the structure of the color mixing for the collinear-soft and soft operators is quite involved. In the remainder of this appendix we give the explicit results for the anomalous dimensions for  $H_{S, ijkl}$ and $\big(\tilde S_{kl}\big)^{aba'b'}$ for all relevant color channels appearing in our factorization.

For ease of notation, as in the body of the text, we will define our ultrasoft operators as, see \Eq{eq:1indexnotation}, 
\bea
 S'_1  \equiv S'_{11} \ \ \ \ \  S'_2 \equiv S'_{22}, \ \ \ \ \ S'_3 \equiv S'_{12}+S'_{21}\,.
\eea 
In this notation, the refactorization of the ultrasoft function is given by, see \Eq{eq:genrefactorize},
\bea
 S_{i}^{\prime\,aba'b'}= H_{S, ikl}\left(C_{S,k}^{{A}}S_l^{{B}}\right)^{aba'b'}=H_{S, ij}\left(\tilde S_{j}\right)^{aba'b'}\,.
\eea
The tree-level, and one-loop results, along with the $\mu$ and $\nu$ anomalous dimensions for the $H$ and $\tilde S$ functions appearing in the factorization are as follows:

\begin{itemize}[leftmargin=*]
\item $\tilde S_{1}$ is defined as 
\begin{align}
\tilde S^{\,aba'b'}_{1}  =& \sum_{X_{c_S}} \Big\langle 0\Big| \left(X_n^{3f'}V_{n}^{dg'}\right)^{\dagger}\!\!(0)\Big|X_{c_S}\Big\rangle \nn\\[5pt]
&\hspace{25pt}\times \Big\langle X_{c_S}\Big|\delta\big(q^+ - \mathcal{P}^+\big)\left(X_n^{3f}V_{n}^{dg}\right)\!(0) \Big|0\Big \rangle \,\delta^{f'g'}\delta^{a'b'}\delta^{fg}\delta^{ab}\,,
\end{align}
where the soft sector Wilson lines have contracted to the identity.  By inspection, the anomalous dimension for this operator is identical to $S_{1}^{aba'b'}$, implying that $H_{S,11} =1 $ is the only non-zero matching coefficient.

\item $\tilde S_{3}$ is defined as
\begin{align}
\tilde S_{3}^{\,aba'b'} &=  \sum_{X_{c_S}} \Big\langle 0\Big| \left[ X_n^{ce} V_{n}^{B'e} \delta( q^+ - \mathcal{P}^+)\Big|X_{c_S}\Big\rangle \Big\langle X_{c_S}\Big |X_n^{c'g'} V_{n}^{A'g'}\right]\left[ S_n^{3c} S_n^{3c'}S_v^{a'A'}S_v^{b'B'}\right] \delta^{ab}   \Big|  0  \Big\rangle \,\nn\\[5pt]
& \hspace{40pt}+ \big( a,b \leftrightarrow a',b'\big)\,.
\end{align}
At tree-level in Laplace space we have 
\bea
 \left(\tilde S_{3}^{\,aba'b'}\right)^{\text{tree}} = \delta^{a'3}\delta^{b'3}\delta^{ab}+ \big( a,b \leftrightarrow a',b'\big)\,,
\eea
and at one-loop in Laplace space, we have
\begin{align}
\Big( \tilde S_{3}^{\,aba'b'}\Big)^\text{1-loop} = & - 4\,\delta^{a'3} \delta^{b'3} \delta^{ab}\,\frac{\aW}{2\,\pi}\left(\frac{\mu}{\mW}\right)^{2\, \epsilon} \frac{1}{\epsilon} \,\log\big(\nu\, s\big) \nn\\[5pt]
	&+ \delta^{ab}\big( \delta^{a'b'} -3\, \delta^{a'3} \delta^{b'3} \big)\,\frac{\aW}{\pi}\left(-2\,\log\left(\frac{\mu}{\mW}\right)\log\big(\mu \, s\big) +\log^2\left(\frac{\mu}{\mW}\right)\right)\nn\\[5pt]
	&+ \big ( a,b \leftrightarrow a',b'\big)\,.
\end{align}
The second line here is essentially the IR piece of the term $\log^2\big(\mW\,s\big)$. Extracting the LL anomalous dimensions yields
\begin{align}
\frac{\text{d}}{\text{d}\log\mu} \tilde S_{3}^{\,aba'b'} =&  \left(- 2\,C_A\, \frac{\aW}{\pi} \log\big(\nu\, s\big) + 3\, C_A\, \frac{\aW}{\pi} \log\big(\mu\, s\big) \right) \tilde S_{3}^{\,aba'b'} \nn\\
 &\hspace{5pt}- 2C_A\, \frac{\aW}{\pi} \log\big(\mu\, s\big) \, \tilde S_{1}^{\,aba'b'}\,,
\end{align}
which shows a mixing between $\tilde S_{3} $ and $\tilde S_{1}$, along with
\bea
\frac{\text{d}}{\text{d}\log\nu} \tilde S_{3}^{\,aba'b'} &=& -2 \,C_A \frac{\aW}{\pi} \log \left(\frac{\mu}{\mW}\right)\tilde S_{3}^{\,aba'b'} \,.
\eea

We can now read off the matching coefficients  
\begin{align}
H_{S, 33}   &= 1 - 3\frac{\aW}{\pi}\log^2\left(\mu\, s\right) \,, \nn\\
H_{S, 31} &=  2\frac{\aW}{\pi}\log^2\left(\mu \,s\right)\,,
\end{align}
which immediately tells us that 
\begin{align}
\frac{\text{d}}{\text{d} \log\mu}H_{S,33} &= -3\, C_A\frac{\aW}{\pi}\log\big(\mu \,s\big)H_{S,33} \,,\nn\\
\frac{\text{d}}{\text{d} \log\mu}H_{S,31} &= 2C_A\frac{\aW}{\pi}\log\big(\mu \,s\big)H_{S,33}\,.
\end{align}

\item $\tilde S_{2}$ is defined as 
\begin{align}
\tilde S_{2}^{\,aba'b'}=  \sum_{X_{c_S}} \Big\langle 0\Big| \left[ X_n^{ce} V_{n}^{Ae} \delta( q^+ - \mathcal{P}^+)\Big|X_{c_S}\Big\rangle \Big\langle X_{c_S} \Big|X_n^{c'g'} V_{n}^{A'g'}\right]\left[ S_n^{3c} S_n^{3c'}S_v^{a'A'}S_v^{aA}\right] \delta^{bb'}  \Big|  0  \Big\rangle\,.
\end{align}
At tree-level in Laplace space we have
\begin{align}
 \left(\tilde S_{2}^{\,aba'b'}\right)^{\text{tree}} = \delta^{bb'}\delta^{a'3} \delta^{a3}\,,
\end{align}
and at one-loop in Laplace space, we have
\begin{align}
 \Big(\tilde{S}_{2}^{\,aba'b'}\Big)^\text{1-loop} = & - 4\,\delta^{bb'}\delta^{a'3} \delta^{a3}\,\frac{\aW}{2\,\pi}\left(\frac{\mu}{\mW}\right)^{2 \epsilon} \frac{1}{\epsilon} \log\big(\nu\, s\big)\\[5pt]
	&+ \delta^{bb'}\big( \delta^{aa'} -3\, \delta^{a'3} \delta^{a3}\big )\frac{\aW}{\pi}\left(-2\,\log\left(\frac{\mu}{\mW}\right)\log\big(\mu\, s\big) +\log^2\left(\frac{\mu}{\mW}\right)\right)\,.\nn
\end{align}
From the color structure of this result, it is clear that another operator has been induced at loop level, namely 
\bea
\hspace{-10pt} \tilde S^{\,aba'b'}_4 =  \sum_{X_{c_S}} \Big\langle 0\Big| \left(X_n^{3f'}V_{n}^{df'}\right)^{\dagger}\!\!(0)\Big|X_{c_S}\Big\rangle\Big \langle X_{c_S}\Big| \delta\big(q^+ - \mathcal{P}^+\big) \left(X_n^{3f}V_{n}^{df}\right)\!(0) \Big|0 \Big\rangle\, \delta^{aa'}\delta^{bb'}\,,\hspace{15pt}
\eea
which is similar to $\tilde S_{1}^{\,a'b'ab}$ but with a different color structure.  Evaluating this operator at tree-level in Laplace space yields 
\begin{align}
 \Big(\tilde S^{\,aba'b'}_4\Big)^\text{tree} = \delta^{aa'}\delta^{bb'}\,,
 \end{align}
 and at one-loop in Laplace space yields
 \begin{align}
\Big(\tilde S^{\,a'b'ab}_4\Big)^\text{1-loop} = -\delta^{aa'}\delta^{bb'}\,2\,C_A\frac{\aW}{2\,\pi}\left(\frac{\mu}{\mW}\right)^{2\, \epsilon} \frac{1}{\epsilon}\left(\frac{2}{\eta}+\log\big(\nu\, s\big) \right)\,.
\end{align}
Recall that the matching coefficient for this operator is $0$ at tree-level, since it did not appear in our original basis. Extracting the LL anomalous dimensions for this operator yields
\begin{align}
\frac{\text{d}}{\text{d}\log \mu} \tilde S_{4}^{\,aba'b'} &= - 2\, C_A\, \frac{\aW}{\pi} \log \big(\nu\, s\big)\, \tilde S_{4}^{aba'b'}\,, \nn\\
\frac{\text{d}}{\text{d}\log \nu} \tilde S_{4}^{\,aba'b'} &= -2 \,C_A\, \frac{\aW}{\pi} \log \left(\frac{\mu}{\mW}\right)\, \tilde S_{4}^{aba'b'} \,.
\end{align}
We can use these results to extract the anomalous dimension for $\tilde S_{2}$,
\begin{align}
\frac{\text{d}}{\text{d}\log\mu} \tilde S_{2}^{\,aba'b'} =&  \left(- 2\,C_A\, \frac{\aW}{\pi} \log\big(\nu\, s\big) + 3\, C_A\, \frac{\aW}{\pi} \log\big(\mu\, s\big) \right) \tilde S_{2}^{\,aba'b'} \nn\\[5pt]
 &\hspace{10pt}- C_A \frac{\aW}{\pi} \log(\mu\, s)  \tilde S_{4}^{\,aba'b'}\,,\nn\\[10pt]
\frac{\text{d}}{\text{d}\log\nu} \tilde S_{2}^{\,aba'b'} =& -2\, C_A\, \frac{\aW}{\pi} \log \left(\frac{\mu}{\mW}\right)\tilde S_{2}^{\,aba'b'}\,.
\end{align}

We can then extract the Wilson coefficients,
\begin{align}
H_{S, 22}  &= 1 - 3\,\frac{\aW}{\pi}\log^2\big(\mu \,s\big)\,,\nn\\[5pt]
H_{S, 24} &=  \frac{\aW}{\pi}\log^2\big(\mu\, s\big)\,,
\end{align}
and their anomalous dimensions
\begin{align}
\frac{\text{d}}{\text{d} \log \mu} H_{S,22} = -6\,\frac{\aW}{\pi}\log\big(\mu\, s\big)H_{S,22}\,, \nn\\[5pt]
\frac{\text{d}}{\text{d} \log \mu} H_{S,24} = 2\frac{\aW}{\pi}\log\big(\mu \, s\big)H_{S,22}\,.
\end{align}

\end{itemize}

This provides the complete set of ingredients required for the LL resummation in the endpoint region. 

\section{Impact of Continuum Photons from Cascade Decays}\label{sec:continuum}

\begin{figure}
\begin{center}
\subfloat[]{\label{fig:Smoothed_Spectra_3TeV}
\includegraphics[width=0.65\columnwidth]{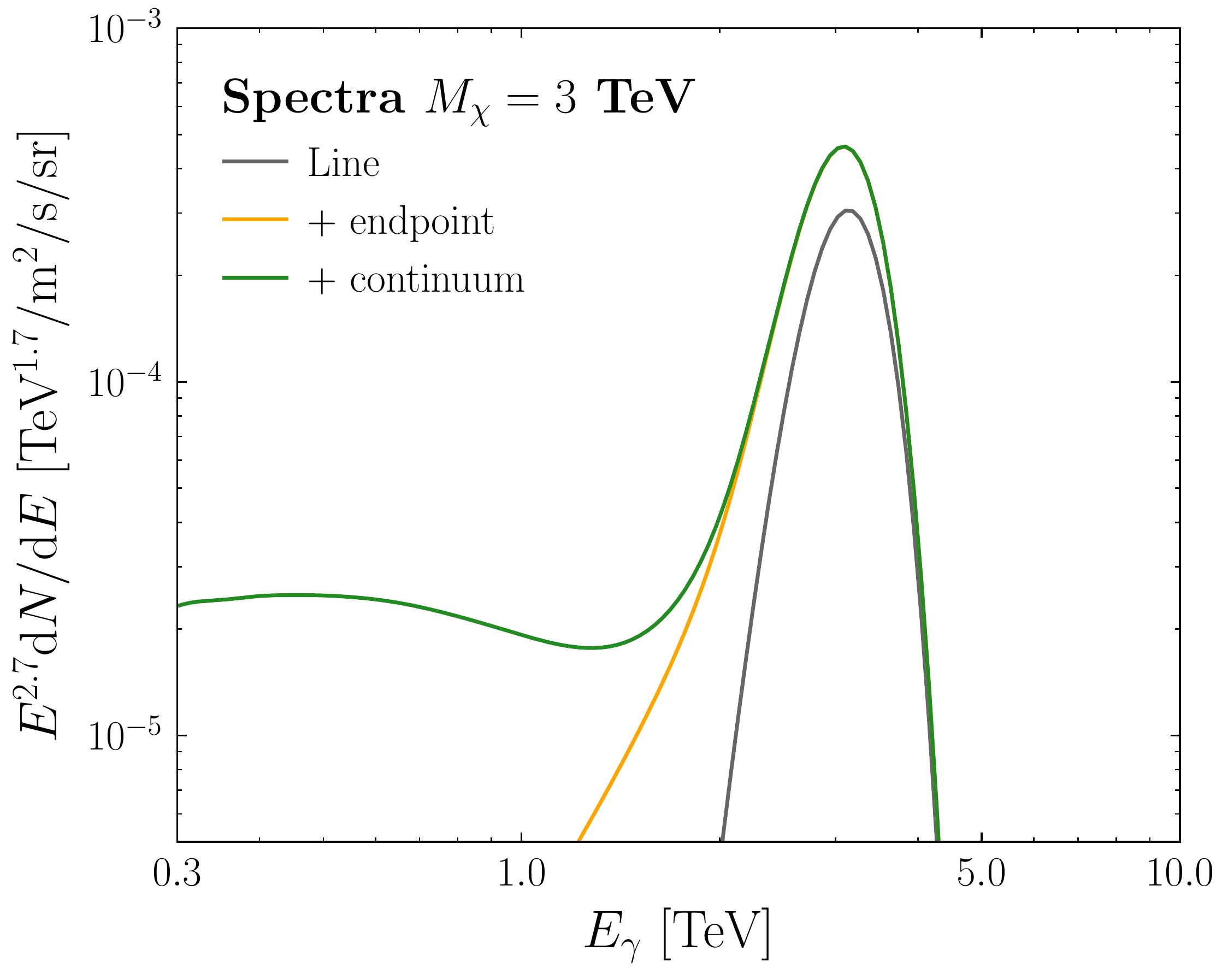}    
}
\vspace{10pt}
\subfloat[]{\label{fig:Smoothed_Spectra_10TeV}
\includegraphics[width=0.65\columnwidth]{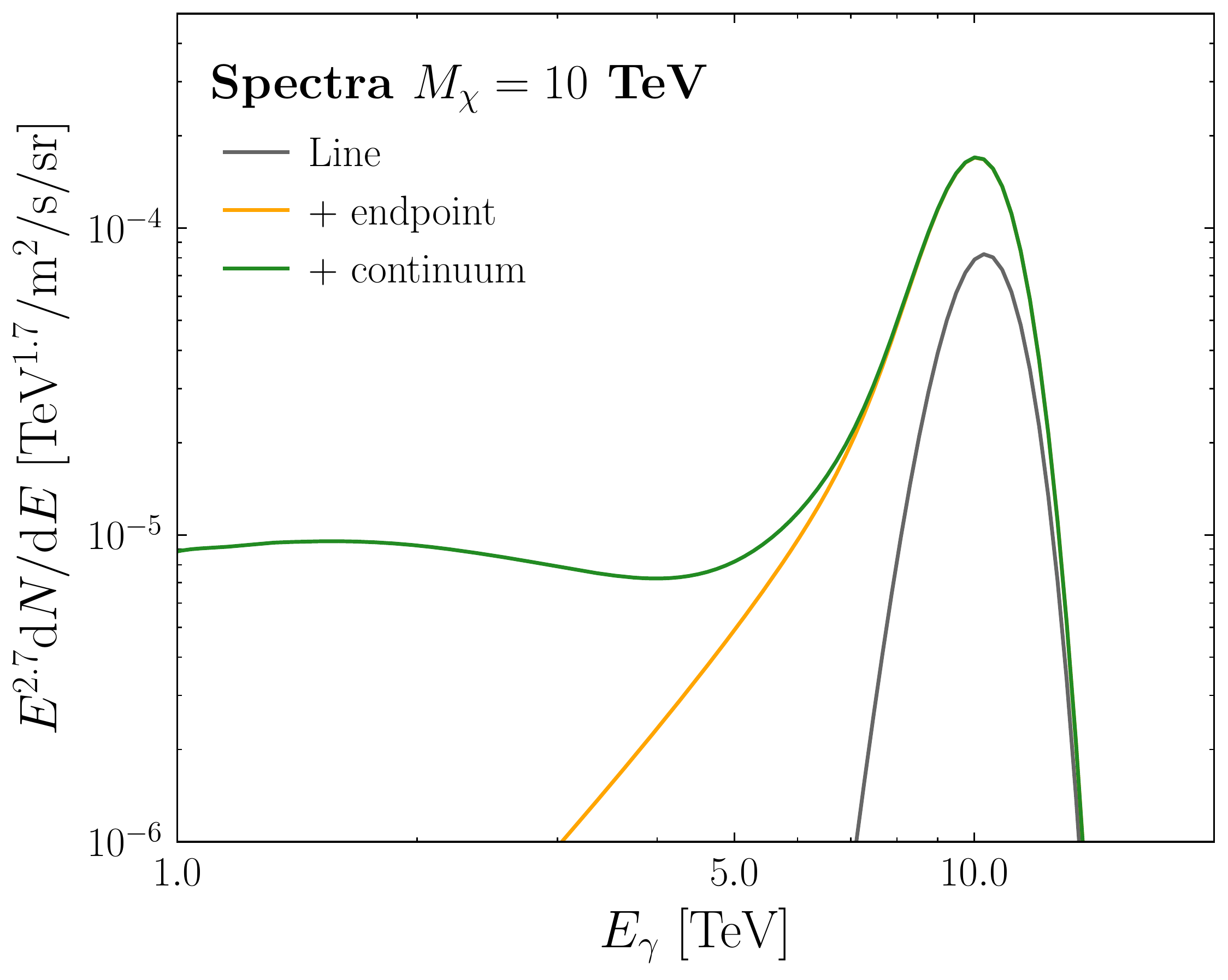}
}
\end{center}
\caption{The differential photon flux observed at HESS for the wino at (a) 3 TeV; and (b) 10 TeV. In each case we show, progressively, the contribution from the line only case, the endpoint contribution calculated in the main body, and finally the continuum arising from the decay of the produced $W$ and $Z$ bosons. In all cases the contributions have been smeared by the HESS energy resolution.
}
\label{fig:Smoothed_Spectra}
\end{figure}

\begin{figure}
\begin{center}
\subfloat[]{\label{fig:DM_limits_cont_b}
\includegraphics[width=0.65\columnwidth]{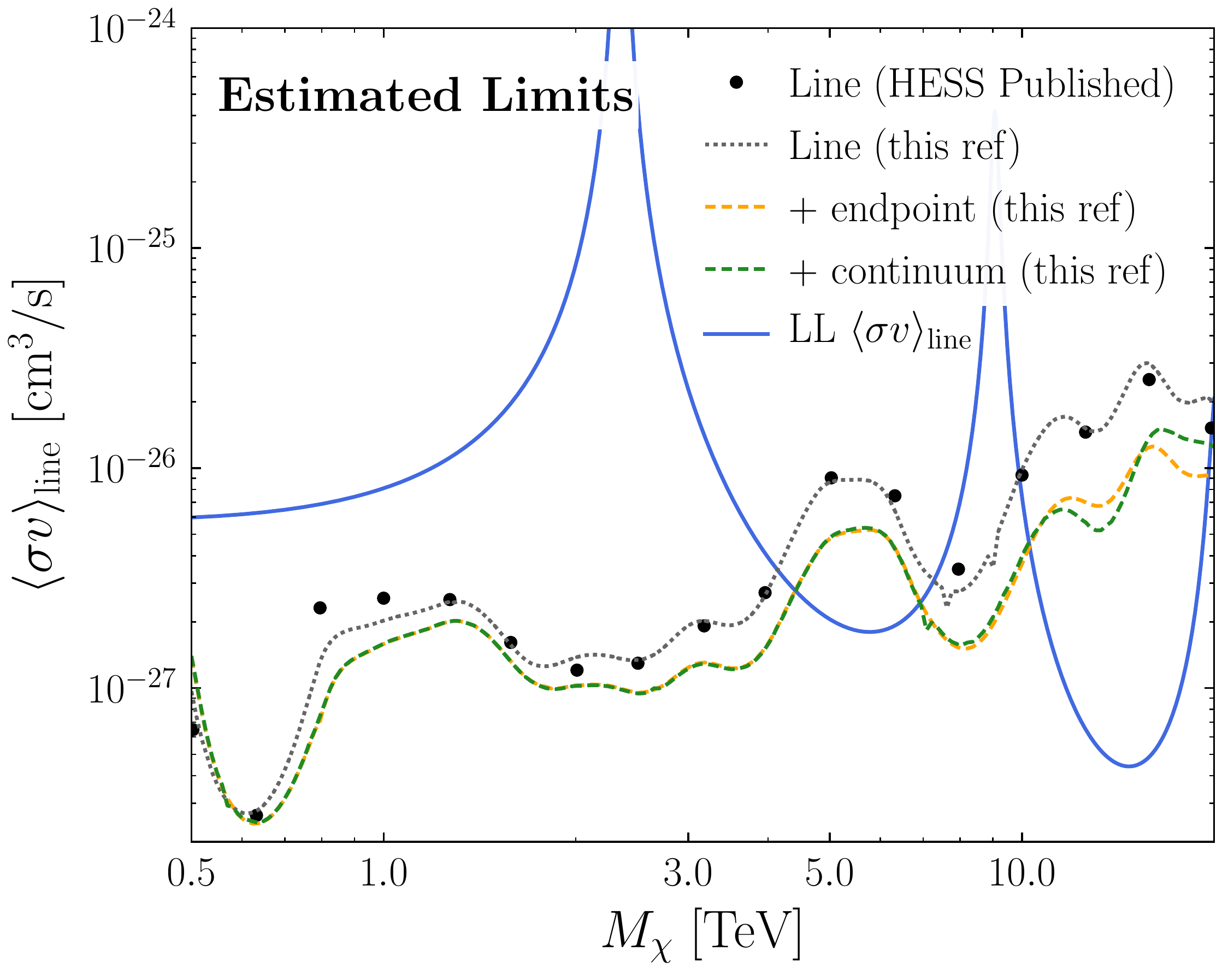}    
}
\vspace{10pt}
\subfloat[]{\label{fig:DM_limits_cont_a}
\includegraphics[width=0.65\columnwidth]{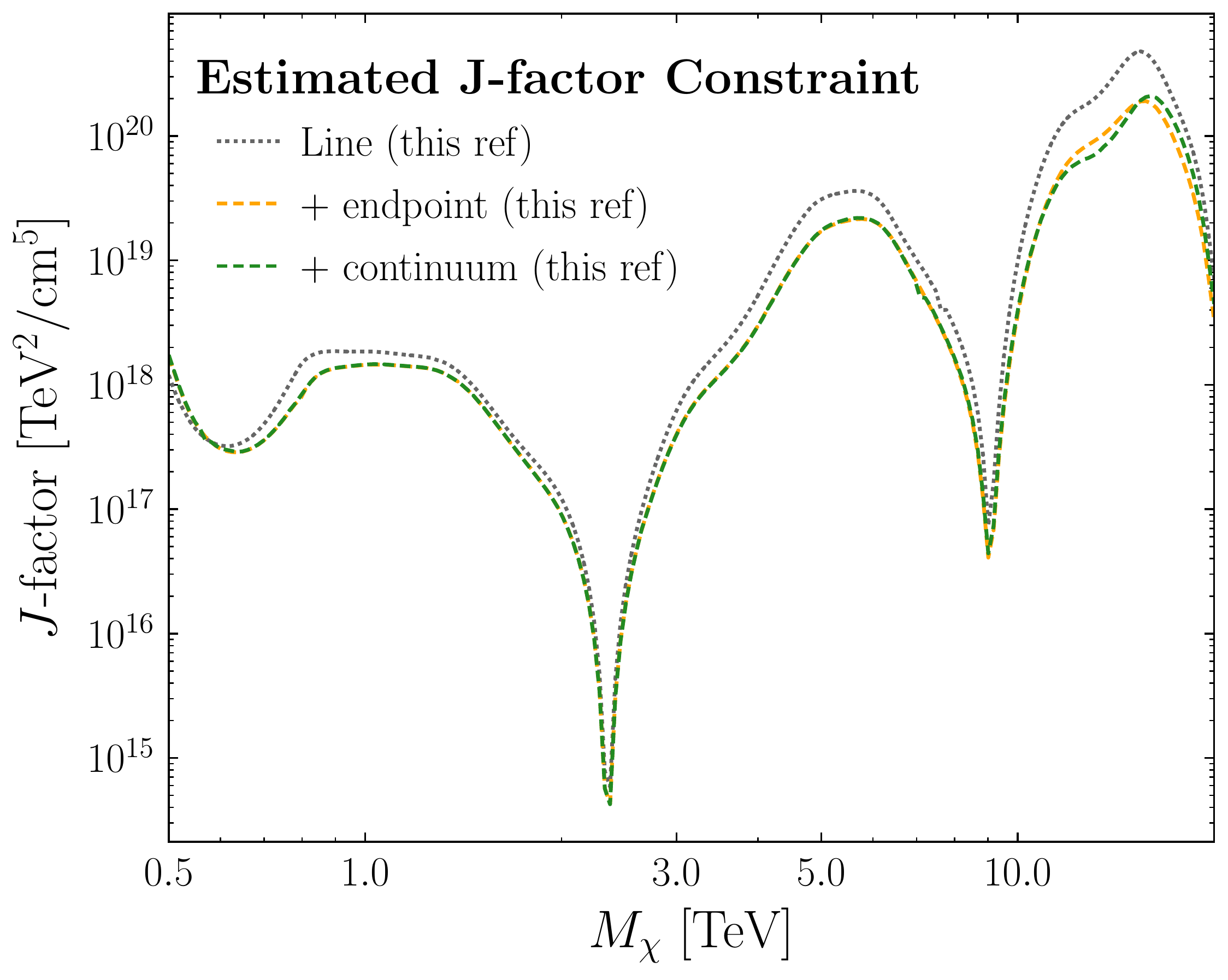}
}
\end{center}
\vspace{-10pt}
\caption{As in Figs.~\ref{fig:DM_limits_b} and~\ref{fig:DM_limits_a}, but showing the impact of adding the continuum contribution from $W$ and $Z$ decays in addition to the endpoint on the constraints. In general these contributions have a much smaller impact than that already resulting from adding in the endpoint spectrum. We caution once more that these are only estimated limits.
}
\label{fig:DM_limits_cont}
\end{figure}

In the main body of this work we presented a calculation of the internal bremsstrahlung (+ initial/final state radiation), or endpoint, contribution to the wino annihilation spectrum. As we mentioned there, another source of photons arises from the final state decay products of the unstable particles that are produced by DM annihilations, such as the $W^\pm$ and $Z$ bosons. In this appendix we estimate the contribution from these additional final states, and show that they have a small impact on the HESS constraints for the thermal wino.  However, they could be interesting for instruments searching for lower energy photons such as \textit{Fermi}.

In order to estimate these contributions, we have added to the line and endpoint spectra the spectrum coming from the decay of $W^\pm$ and $Z$ bosons. The spectrum of photons that arises from their decay is determined using PPPC4DMID~\cite{Cirelli:2010xx} with electroweak corrections turned off,\footnote{The electroweak corrections in PPPC4DMID include a partial accounting of the endpoint corrections that we determined in the main body, which they include following~\cite{Ciafaloni:2010ti}, and so we remove them to avoid double counting. We thank Marco Cirelli for confirming this point. This choice means we are missing the electroweak corrections from the remainder of the continuum spectrum, however we have confirmed these effects are small by directly comparing the spectra to the predictions of \texttt{Pythia} 8.219~\cite{Sjostrand:2006za,Sjostrand:2007gs,Sjostrand:2014zea}, which includes electroweak showering~\cite{Christiansen:2014kba}.} whereas the branching fraction is evaluated differently for the two cases. For annihilation to $W^+\,W^-$, the branching fraction is given by the Sommerfeld-enhanced tree-level cross section for this final state~\cite{Hisano:2004ds,Hisano:2006nn}.\footnote{Note that there is a factor of 2 missing in the off-diagonal terms of $\Gamma_{W^+W^-}$ in Eq.~(28) in Ref.~\cite{Hisano:2004ds}, which is corrected in Ref.~\cite{Hisano:2006nn}.} Radiative corrections to this cross section, which have been shown to be small~\cite{Hryczuk:2011vi}, are not included. To determine the $Z$ production cross section, we use the leading log cross section, which is given by \Eq{eq:LLexclusive} reweighted by $c_{\scriptscriptstyle W}^2/s_{\scriptscriptstyle W}^2$.

In Fig.~\ref{fig:Smoothed_Spectra}, we show the impact on the photon spectrum from DM, after convolving it with the HESS energy resolution, when this continuum contribution is added, for two DM masses. Generically, as we approach $E_\gamma \sim M_\chi$, this continuum emission is a sub-dominant effect.  However, at lower energies it can have substantial impact (note this spectrum is multiplied by $E^{2.7}$ which downweights the flux at lower energies). Nevertheless, such a contribution over many energy bins is hard to distinguish from the 7 parameter background model used by HESS.  These background parameters are profiled over, so that we would not expect this additional emission to make a sizable impact. Indeed, in Fig.~\ref{fig:DM_limits_cont} we demonstrate this point, by repeating the analysis from Sec.~\ref{sec:mockHESS} with the inclusion of the additional continuum photons.  We note the effect of including the continuum becomes more important at higher masses, but is almost always subdominant to the impact of adding in the endpoint emission. Further, the broad nature of the continuum emission can lead to a non-trivial interplay with the background model in fits to the data, and in fact lead to weaker limits for some masses. For example, near $M_{\chi} =20$ TeV in Fig.~\ref{fig:DM_limits_cont}, the additional continuum emission at lower energies drives down the best fit background model, resulting in a reduced background prediction near the dark matter mass where the line and endpoint contributions dominate, and accordingly a weaker limit.

Finally we note in passing that the large contribution from the continuum may be relevant to lower energy instruments such as \textit{Fermi}-LAT. The advantage of such an approach is that we can look at a number of different potential astrophysical sources of DM flux, each associated with partially uncorrelated systematics on their $J$-factors. In this way we can extend the search beyond the Galactic Center and its large uncertainties to look at potentially cleaner environments such as the Milky Way Dwarfs~\cite{Ackermann:2015zua,Fermi-LAT:2016uux} or even galaxy clusters~\cite{Lisanti:2017qoz,Lisanti:2017qlb}. However, note that the effective area of \textit{Fermi}-LAT drops sharply at TeV energies.  This implies that if the DM mass is multi-TeV, the HESS constraints are generally stronger than those from \textit{Fermi}.  Even accounting for the astrophysical uncertainties, the HESS dataset continues to be the best probe of the gamma-rays from thermal wino DM.

\pagebreak

\addcontentsline{toc}{section}{\protect\numberline{}References}%
\bibliography{HDMA}

\providecommand{\href}[2]{#2}\begingroup\raggedright\begin{thebibliography}{100}

\bibitem{Hisano:2002fk}
J.~Hisano, S.~Matsumoto, and M.~M. Nojiri, {\it {Unitarity and higher order
  corrections in neutralino dark matter annihilation into two photons}},  {\em
  Phys. Rev.} {\bf D67} (2003) 075014,
  [\href{http://arxiv.org/abs/hep-ph/0212022}{{\tt hep-ph/0212022}}].

\bibitem{Hisano:2003ec}
J.~Hisano, S.~Matsumoto, and M.~M. Nojiri, {\it {Explosive dark matter
  annihilation}},  {\em Phys. Rev. Lett.} {\bf 92} (2004) 031303,
  [\href{http://arxiv.org/abs/hep-ph/0307216}{{\tt hep-ph/0307216}}].

\bibitem{Hisano:2004ds}
J.~Hisano, S.~Matsumoto, M.~M. Nojiri, and O.~Saito, {\it {Non-perturbative
  effect on dark matter annihilation and gamma ray signature from galactic
  center}},  {\em Phys. Rev.} {\bf D71} (2005) 063528,
  [\href{http://arxiv.org/abs/hep-ph/0412403}{{\tt hep-ph/0412403}}].

\bibitem{Hisano:2006nn}
J.~Hisano, S.~Matsumoto, M.~Nagai, O.~Saito, and M.~Senami, {\it
  {Non-perturbative effect on thermal relic abundance of dark matter}},  {\em
  Phys. Lett.} {\bf B646} (2007) 34--38,
  [\href{http://arxiv.org/abs/hep-ph/0610249}{{\tt hep-ph/0610249}}].

\bibitem{Cirelli:2007xd}
M.~Cirelli, A.~Strumia, and M.~Tamburini, {\it {Cosmology and Astrophysics of
  Minimal Dark Matter}},  {\em Nucl. Phys.} {\bf B787} (2007) 152--175,
  [\href{http://arxiv.org/abs/0706.4071}{{\tt arXiv:0706.4071}}].

\bibitem{ArkaniHamed:2004fb}
N.~Arkani-Hamed and S.~Dimopoulos, {\it {Supersymmetric unification without low
  energy supersymmetry and signatures for fine-tuning at the LHC}},  {\em JHEP}
  {\bf 06} (2005) 073, [\href{http://arxiv.org/abs/hep-th/0405159}{{\tt
  hep-th/0405159}}].

\bibitem{ArkaniHamed:2004yi}
N.~Arkani-Hamed, S.~Dimopoulos, G.~F. Giudice, and A.~Romanino, {\it {Aspects
  of split supersymmetry}},  {\em Nucl. Phys.} {\bf B709} (2005) 3--46,
  [\href{http://arxiv.org/abs/hep-ph/0409232}{{\tt hep-ph/0409232}}].

\bibitem{Giudice:2004tc}
G.~F. Giudice and A.~Romanino, {\it {Split supersymmetry}},  {\em Nucl. Phys.}
  {\bf B699} (2004) 65--89, [\href{http://arxiv.org/abs/hep-ph/0406088}{{\tt
  hep-ph/0406088}}]. [Erratum: Nucl. Phys.B706,487(2005)].

\bibitem{Wells:2004di}
J.~D. Wells, {\it {PeV-scale supersymmetry}},  {\em Phys. Rev.} {\bf D71}
  (2005) 015013, [\href{http://arxiv.org/abs/hep-ph/0411041}{{\tt
  hep-ph/0411041}}].

\bibitem{Pierce:2004mk}
A.~Pierce, {\it {Dark matter in the finely tuned minimal supersymmetric
  standard model}},  {\em Phys. Rev.} {\bf D70} (2004) 075006,
  [\href{http://arxiv.org/abs/hep-ph/0406144}{{\tt hep-ph/0406144}}].

\bibitem{Arvanitaki:2012ps}
A.~Arvanitaki, N.~Craig, S.~Dimopoulos, and G.~Villadoro, {\it {Mini-Split}},
  {\em JHEP} {\bf 02} (2013) 126, [\href{http://arxiv.org/abs/1210.0555}{{\tt
  arXiv:1210.0555}}].

\bibitem{ArkaniHamed:2012gw}
N.~Arkani-Hamed, A.~Gupta, D.~E. Kaplan, N.~Weiner, and T.~Zorawski, {\it
  {Simply Unnatural Supersymmetry}},
  \href{http://arxiv.org/abs/1212.6971}{{\tt arXiv:1212.6971}}.

\bibitem{Hall:2012zp}
L.~J. Hall, Y.~Nomura, and S.~Shirai, {\it {Spread Supersymmetry with Wino LSP:
  Gluino and Dark Matter Signals}},  {\em JHEP} {\bf 01} (2013) 036,
  [\href{http://arxiv.org/abs/1210.2395}{{\tt arXiv:1210.2395}}].

\bibitem{Cirelli:2005uq}
M.~Cirelli, N.~Fornengo, and A.~Strumia, {\it {Minimal dark matter}},  {\em
  Nucl. Phys.} {\bf B753} (2006) 178--194,
  [\href{http://arxiv.org/abs/hep-ph/0512090}{{\tt hep-ph/0512090}}].

\bibitem{Cirelli:2008id}
M.~Cirelli, R.~Franceschini, and A.~Strumia, {\it {Minimal Dark Matter
  predictions for galactic positrons, anti-protons, photons}},  {\em Nucl.
  Phys.} {\bf B800} (2008) 204--220,
  [\href{http://arxiv.org/abs/0802.3378}{{\tt arXiv:0802.3378}}].

\bibitem{Cirelli:2009uv}
M.~Cirelli and A.~Strumia, {\it {Minimal Dark Matter: Model and results}},
  {\em New J. Phys.} {\bf 11} (2009) 105005,
  [\href{http://arxiv.org/abs/0903.3381}{{\tt arXiv:0903.3381}}].

\bibitem{Cirelli:2015bda}
M.~Cirelli, T.~Hambye, P.~Panci, F.~Sala, and M.~Taoso, {\it {Gamma ray tests
  of Minimal Dark Matter}},  {\em JCAP} {\bf 1510} (2015), no.~10 026,
  [\href{http://arxiv.org/abs/1507.05519}{{\tt arXiv:1507.05519}}].

\bibitem{Low:2014cba}
M.~Low and L.-T. Wang, {\it {Neutralino dark matter at 14 TeV and 100 TeV}},
  {\em JHEP} {\bf 08} (2014) 161, [\href{http://arxiv.org/abs/1404.0682}{{\tt
  arXiv:1404.0682}}].

\bibitem{Cirelli:2014dsa}
M.~Cirelli, F.~Sala, and M.~Taoso, {\it {Wino-like Minimal Dark Matter and
  future colliders}},  {\em JHEP} {\bf 10} (2014) 033,
  [\href{http://arxiv.org/abs/1407.7058}{{\tt arXiv:1407.7058}}]. [Erratum:
  JHEP01,041(2015)].

\bibitem{Hill:2011be}
R.~J. Hill and M.~P. Solon, {\it {Universal behavior in the scattering of
  heavy, weakly interacting dark matter on nuclear targets}},  {\em Phys.
  Lett.} {\bf B707} (2012) 539--545,
  [\href{http://arxiv.org/abs/1111.0016}{{\tt arXiv:1111.0016}}].

\bibitem{Hill:2013hoa}
R.~J. Hill and M.~P. Solon, {\it {WIMP-nucleon scattering with heavy WIMP
  effective theory}},  {\em Phys. Rev. Lett.} {\bf 112} (2014) 211602,
  [\href{http://arxiv.org/abs/1309.4092}{{\tt arXiv:1309.4092}}].

\bibitem{Hisano:2015rsa}
J.~Hisano, K.~Ishiwata, and N.~Nagata, {\it {QCD Effects on Direct Detection of
  Wino Dark Matter}},  {\em JHEP} {\bf 06} (2015) 097,
  [\href{http://arxiv.org/abs/1504.00915}{{\tt arXiv:1504.00915}}].

\bibitem{Cohen:2013ama}
T.~Cohen, M.~Lisanti, A.~Pierce, and T.~R. Slatyer, {\it {Wino Dark Matter
  Under Siege}},  {\em JCAP} {\bf 1310} (2013) 061,
  [\href{http://arxiv.org/abs/1307.4082}{{\tt arXiv:1307.4082}}].

\bibitem{Fan:2013faa}
J.~Fan and M.~Reece, {\it {In Wino Veritas? Indirect Searches Shed Light on
  Neutralino Dark Matter}},  {\em JHEP} {\bf 10} (2013) 124,
  [\href{http://arxiv.org/abs/1307.4400}{{\tt arXiv:1307.4400}}].

\bibitem{Cuoco:2017aa}
A.~Cuoco, J.~Heisig, M.~Korsmeier, and M.~Kr{\"a}mer, {\it {Constraining heavy
  dark matter with cosmic-ray antiprotons}},
  \href{http://arxiv.org/abs/1711.05274}{{\tt arXiv:1711.05274}}.

\bibitem{Abramowski:2011hc}
{\bf H.E.S.S.} Collaboration, A.~Abramowski et~al., {\it {Search for a Dark
  Matter annihilation signal from the Galactic Center halo with H.E.S.S}},
  {\em Phys. Rev. Lett.} {\bf 106} (2011) 161301,
  [\href{http://arxiv.org/abs/1103.3266}{{\tt arXiv:1103.3266}}].

\bibitem{Hinton:2004eu}
{\bf HESS} Collaboration, J.~A. Hinton, {\it {The Status of the H.E.S.S.
  project}},  {\em New Astron. Rev.} {\bf 48} (2004) 331--337,
  [\href{http://arxiv.org/abs/astro-ph/0403052}{{\tt astro-ph/0403052}}].

\bibitem{Abramowski:2013ax}
{\bf HESS} Collaboration, A.~Abramowski et~al., {\it {Search for
  Photon-Linelike Signatures from Dark Matter Annihilations with H.E.S.S.}},
  {\em Phys. Rev. Lett.} {\bf 110} (2013) 041301,
  [\href{http://arxiv.org/abs/1301.1173}{{\tt arXiv:1301.1173}}].

\bibitem{Sinnis:2004je}
G.~Sinnis, A.~Smith, and J.~E. McEnery, {\it {HAWC: A Next generation all - sky
  VHE gamma-ray telescope}},  in {\em {On recent developments in theoretical
  and experimental general relativity, gravitation, and relativistic field
  theories. Proceedings, 10th Marcel Grossmann Meeting, MG10, Rio de Janeiro,
  Brazil, July 20-26, 2003. Pt. A-C}}, pp.~1068--1088, 2004.
\newblock \href{http://arxiv.org/abs/astro-ph/0403096}{{\tt astro-ph/0403096}}.

\bibitem{Harding:2015bua}
{\bf HAWC} Collaboration, J.~P. Harding and B.~Dingus, {\it {Dark Matter
  Annihilation and Decay Searches with the High Altitude Water Cherenkov (HAWC)
  Observatory}},  in {\em {Proceedings, 34th International Cosmic Ray
  Conference (ICRC 2015)}}, 2015.
\newblock \href{http://arxiv.org/abs/1508.04352}{{\tt arXiv:1508.04352}}.

\bibitem{Pretz:2015zja}
{\bf HAWC} Collaboration, J.~Pretz, {\it {Highlights from the High Altitude
  Water Cherenkov Observatory}},  in {\em {Proceedings, 34th International
  Cosmic Ray Conference (ICRC 2015)}}, 2015.
\newblock \href{http://arxiv.org/abs/1509.07851}{{\tt arXiv:1509.07851}}.

\bibitem{Consortium:2010bc}
{\bf CTA Consortium} Collaboration, M.~Actis et~al., {\it {Design concepts for
  the Cherenkov Telescope Array CTA: An advanced facility for ground-based
  high-energy gamma-ray astronomy}},  {\em Exper. Astron.} {\bf 32} (2011)
  193--316, [\href{http://arxiv.org/abs/1008.3703}{{\tt arXiv:1008.3703}}].

\bibitem{Weekes:2001pd}
T.~C. Weekes et~al., {\it {VERITAS: The Very energetic radiation imaging
  telescope array system}},  {\em Astropart. Phys.} {\bf 17} (2002) 221--243,
  [\href{http://arxiv.org/abs/astro-ph/0108478}{{\tt astro-ph/0108478}}].

\bibitem{Holder:2006gi}
{\bf VERITAS} Collaboration, J.~Holder et~al., {\it {The first VERITAS
  telescope}},  {\em Astropart. Phys.} {\bf 25} (2006) 391--401,
  [\href{http://arxiv.org/abs/astro-ph/0604119}{{\tt astro-ph/0604119}}].

\bibitem{Geringer-Sameth:2013cxy}
{\bf VERITAS} Collaboration, A.~Geringer-Sameth, {\it {The VERITAS Dark Matter
  Program}},  in {\em {4th International Fermi Symposium Monterey, California,
  USA, October 28-November 2, 2012}}, 2013.
\newblock \href{http://arxiv.org/abs/1303.1406}{{\tt arXiv:1303.1406}}.

\bibitem{FlixMolina:2005hv}
{\bf MAGIC} Collaboration, J.~Flix~Molina, {\it {Planned dark matter searches
  with the MAGIC Telescope}},  in {\em {Proceedings, 40th Rencontres de Moriond
  on Very High Energy Phenomena in the Universe}}, pp.~421--424, 2005.
\newblock \href{http://arxiv.org/abs/astro-ph/0505313}{{\tt astro-ph/0505313}}.

\bibitem{Ahnen:2016qkx}
{\bf Fermi-LAT, MAGIC} Collaboration, M.~L. Ahnen et~al., {\it {Limits to dark
  matter annihilation cross-section from a combined analysis of MAGIC and
  Fermi-LAT observations of dwarf satellite galaxies}},  {\em JCAP} {\bf 1602}
  (2016), no.~02 039, [\href{http://arxiv.org/abs/1601.06590}{{\tt
  arXiv:1601.06590}}].

\bibitem{ArkaniHamed:2008qn}
N.~Arkani-Hamed, D.~P. Finkbeiner, T.~R. Slatyer, and N.~Weiner, {\it {A Theory
  of Dark Matter}},  {\em Phys. Rev.} {\bf D79} (2009) 015014,
  [\href{http://arxiv.org/abs/0810.0713}{{\tt arXiv:0810.0713}}].

\bibitem{Blum:2016nrz}
K.~Blum, R.~Sato, and T.~R. Slatyer, {\it {Self-consistent Calculation of the
  Sommerfeld Enhancement}},  {\em JCAP} {\bf 1606} (2016), no.~06 021,
  [\href{http://arxiv.org/abs/1603.01383}{{\tt arXiv:1603.01383}}].

\bibitem{Hryczuk:2011vi}
A.~Hryczuk and R.~Iengo, {\it {The one-loop and Sommerfeld electroweak
  corrections to the Wino dark matter annihilation}},  {\em JHEP} {\bf 01}
  (2012) 163, [\href{http://arxiv.org/abs/1111.2916}{{\tt arXiv:1111.2916}}].
  [Erratum: JHEP06,137(2012)].

\bibitem{Baumgart:2014vma}
M.~Baumgart, I.~Z. Rothstein, and V.~Vaidya, {\it {Calculating the Annihilation
  Rate of Weakly Interacting Massive Particles}},  {\em Phys. Rev. Lett.} {\bf
  114} (2015) 211301, [\href{http://arxiv.org/abs/1409.4415}{{\tt
  arXiv:1409.4415}}].

\bibitem{Bauer:2014ula}
M.~Bauer, T.~Cohen, R.~J. Hill, and M.~P. Solon, {\it {Soft Collinear Effective
  Theory for Heavy WIMP Annihilation}},  {\em JHEP} {\bf 01} (2015) 099,
  [\href{http://arxiv.org/abs/1409.7392}{{\tt arXiv:1409.7392}}].

\bibitem{Ovanesyan:2014fwa}
G.~Ovanesyan, T.~R. Slatyer, and I.~W. Stewart, {\it {Heavy Dark Matter
  Annihilation from Effective Field Theory}},  {\em Phys. Rev. Lett.} {\bf 114}
  (2015), no.~21 211302, [\href{http://arxiv.org/abs/1409.8294}{{\tt
  arXiv:1409.8294}}].

\bibitem{Baumgart:2014saa}
M.~Baumgart, I.~Z. Rothstein, and V.~Vaidya, {\it {Constraints on Galactic Wino
  Densities from Gamma Ray Lines}},  {\em JHEP} {\bf 04} (2015) 106,
  [\href{http://arxiv.org/abs/1412.8698}{{\tt arXiv:1412.8698}}].

\bibitem{Baumgart:2015bpa}
M.~Baumgart and V.~Vaidya, {\it {Semi-inclusive wino and higgsino annihilation
  to ${\rm LL}^{\prime}$}},  {\em JHEP} {\bf 03} (2016) 213,
  [\href{http://arxiv.org/abs/1510.02470}{{\tt arXiv:1510.02470}}].

\bibitem{Ovanesyan:2016vkk}
G.~Ovanesyan, N.~L. Rodd, T.~R. Slatyer, and I.~W. Stewart, {\it {One-loop
  correction to heavy dark matter annihilation}},  {\em Phys. Rev.} {\bf D95}
  (2017), no.~5 055001, [\href{http://arxiv.org/abs/1612.04814}{{\tt
  arXiv:1612.04814}}].

\bibitem{Beacom:2004pe}
J.~F. Beacom, N.~F. Bell, and G.~Bertone, {\it {Gamma-ray constraint on
  Galactic positron production by MeV dark matter}},  {\em Phys. Rev. Lett.}
  {\bf 94} (2005) 171301, [\href{http://arxiv.org/abs/astro-ph/0409403}{{\tt
  astro-ph/0409403}}].

\bibitem{Birkedal:2005ep}
A.~Birkedal, K.~T. Matchev, M.~Perelstein, and A.~Spray, {\it {Robust gamma ray
  signature of WIMP dark matter}},
  \href{http://arxiv.org/abs/hep-ph/0507194}{{\tt hep-ph/0507194}}.

\bibitem{Bergstrom:2004cy}
L.~Bergstrom, T.~Bringmann, M.~Eriksson, and M.~Gustafsson, {\it {Gamma rays
  from Kaluza-Klein dark matter}},  {\em Phys. Rev. Lett.} {\bf 94} (2005)
  131301, [\href{http://arxiv.org/abs/astro-ph/0410359}{{\tt
  astro-ph/0410359}}].

\bibitem{Bergstrom:2005ss}
L.~Bergstrom, T.~Bringmann, M.~Eriksson, and M.~Gustafsson, {\it {Gamma rays
  from heavy neutralino dark matter}},  {\em Phys. Rev. Lett.} {\bf 95} (2005)
  241301, [\href{http://arxiv.org/abs/hep-ph/0507229}{{\tt hep-ph/0507229}}].

\bibitem{Bringmann:2007nk}
T.~Bringmann, L.~Bergstrom, and J.~Edsjo, {\it {New Gamma-Ray Contributions to
  Supersymmetric Dark Matter Annihilation}},  {\em JHEP} {\bf 01} (2008) 049,
  [\href{http://arxiv.org/abs/0710.3169}{{\tt arXiv:0710.3169}}].

\bibitem{Bauer:2000yr}
C.~W. Bauer, S.~Fleming, D.~Pirjol, and I.~W. Stewart, {\it {An Effective field
  theory for collinear and soft gluons: Heavy to light decays}},  {\em
  Phys.Rev.} {\bf D63} (2001) 114020,
  [\href{http://arxiv.org/abs/hep-ph/0011336}{{\tt hep-ph/0011336}}].

\bibitem{Bauer:2001ct}
C.~W. Bauer and I.~W. Stewart, {\it {Invariant operators in collinear effective
  theory}},  {\em Phys.Lett.} {\bf B516} (2001) 134--142,
  [\href{http://arxiv.org/abs/hep-ph/0107001}{{\tt hep-ph/0107001}}].

\bibitem{Bauer:2001yt}
C.~W. Bauer, D.~Pirjol, and I.~W. Stewart, {\it {Soft collinear factorization
  in effective field theory}},  {\em Phys.Rev.} {\bf D65} (2002) 054022,
  [\href{http://arxiv.org/abs/hep-ph/0109045}{{\tt hep-ph/0109045}}].

\bibitem{Bauer:2011uc}
C.~W. Bauer, F.~J. Tackmann, J.~R. Walsh, and S.~Zuberi, {\it {Factorization
  and Resummation for Dijet Invariant Mass Spectra}},  {\em Phys.Rev.} {\bf
  D85} (2012) 074006, [\href{http://arxiv.org/abs/1106.6047}{{\tt
  arXiv:1106.6047}}].

\bibitem{Larkoski:2014tva}
A.~J. Larkoski, I.~Moult, and D.~Neill, {\it {Toward Multi-Differential Cross
  Sections: Measuring Two Angularities on a Single Jet}},  {\em JHEP} {\bf
  1409} (2014) 046, [\href{http://arxiv.org/abs/1401.4458}{{\tt
  arXiv:1401.4458}}].

\bibitem{Procura:2014cba}
M.~Procura, W.~J. Waalewijn, and L.~Zeune, {\it {Resummation of
  Double-Differential Cross Sections and Fully-Unintegrated Parton Distribution
  Functions}},  {\em JHEP} {\bf 1502} (2015) 117,
  [\href{http://arxiv.org/abs/1410.6483}{{\tt arXiv:1410.6483}}].

\bibitem{Larkoski:2015zka}
A.~J. Larkoski, I.~Moult, and D.~Neill, {\it {Non-Global Logarithms,
  Factorization, and the Soft Substructure of Jets}},
  \href{http://arxiv.org/abs/1501.04596}{{\tt arXiv:1501.04596}}.

\bibitem{Pietrulewicz:2016nwo}
P.~Pietrulewicz, F.~J. Tackmann, and W.~J. Waalewijn, {\it {Factorization and
  Resummation for Generic Hierarchies between Jets}},
  \href{http://arxiv.org/abs/1601.05088}{{\tt arXiv:1601.05088}}.

\bibitem{Neubert:1993um}
M.~Neubert, {\it {Analysis of the photon spectrum in inclusive B ---> X(s)
  gamma decays}},  {\em Phys. Rev.} {\bf D49} (1994) 4623--4633,
  [\href{http://arxiv.org/abs/hep-ph/9312311}{{\tt hep-ph/9312311}}].

\bibitem{Ligeti:1999ea}
Z.~Ligeti, M.~E. Luke, A.~V. Manohar, and M.~B. Wise, {\it {The anti-B --->
  X(s) gamma photon spectrum}},  {\em Phys. Rev.} {\bf D60} (1999) 034019,
  [\href{http://arxiv.org/abs/hep-ph/9903305}{{\tt hep-ph/9903305}}].

\bibitem{Bauer:2000ew}
C.~W. Bauer, S.~Fleming, and M.~E. Luke, {\it {Summing Sudakov logarithms in B
  ---> X(s gamma) in effective field theory}},  {\em Phys. Rev.} {\bf D63}
  (2000) 014006, [\href{http://arxiv.org/abs/hep-ph/0005275}{{\tt
  hep-ph/0005275}}].

\bibitem{Neubert:2004dd}
M.~Neubert, {\it {Renormalization-group improved calculation of the B ---> X(s)
  gamma branching ratio}},  {\em Eur. Phys. J.} {\bf C40} (2005) 165--186,
  [\href{http://arxiv.org/abs/hep-ph/0408179}{{\tt hep-ph/0408179}}].

\bibitem{Becher:2006pu}
T.~Becher and M.~Neubert, {\it {Analysis of Br(anti-B ---> X(s gamma)) at NNLO
  with a cut on photon energy}},  {\em Phys. Rev. Lett.} {\bf 98} (2007)
  022003, [\href{http://arxiv.org/abs/hep-ph/0610067}{{\tt hep-ph/0610067}}].

\bibitem{Bigi:1993ex}
I.~I.~Y. Bigi, M.~A. Shifman, N.~G. Uraltsev, and A.~I. Vainshtein, {\it {On
  the motion of heavy quarks inside hadrons: Universal distributions and
  inclusive decays}},  {\em Int. J. Mod. Phys.} {\bf A9} (1994) 2467--2504,
  [\href{http://arxiv.org/abs/hep-ph/9312359}{{\tt hep-ph/9312359}}].

\bibitem{Mannel:1994pm}
T.~Mannel and M.~Neubert, {\it {Resummation of nonperturbative corrections to
  the lepton spectrum in inclusive B ---> X lepton anti-neutrino decays}},
  {\em Phys. Rev.} {\bf D50} (1994) 2037--2047,
  [\href{http://arxiv.org/abs/hep-ph/9402288}{{\tt hep-ph/9402288}}].

\bibitem{Wilson:1969zs}
K.~G. Wilson, {\it {Nonlagrangian models of current algebra}},  {\em Phys.
  Rev.} {\bf 179} (1969) 1499--1512.

\bibitem{Korchemsky:1994jb}
G.~P. Korchemsky and G.~F. Sterman, {\it {Infrared factorization in inclusive B
  meson decays}},  {\em Phys. Lett.} {\bf B340} (1994) 96--108,
  [\href{http://arxiv.org/abs/hep-ph/9407344}{{\tt hep-ph/9407344}}].

\bibitem{Salam:2001bd}
G.~P. Salam and D.~Wicke, {\it {Hadron masses and power corrections to event
  shapes}},  {\em JHEP} {\bf 05} (2001) 061,
  [\href{http://arxiv.org/abs/hep-ph/0102343}{{\tt hep-ph/0102343}}].

\bibitem{Mateu:2012nk}
V.~Mateu, I.~W. Stewart, and J.~Thaler, {\it {Power Corrections to Event Shapes
  with Mass-Dependent Operators}},  {\em Phys.Rev.} {\bf D87} (2013), no.~1
  014025, [\href{http://arxiv.org/abs/1209.3781}{{\tt arXiv:1209.3781}}].

\bibitem{Collins:1989bt}
J.~C. Collins, {\it {Sudakov form-factors}},  {\em Adv. Ser. Direct. High
  Energy Phys.} {\bf 5} (1989) 573--614,
  [\href{http://arxiv.org/abs/hep-ph/0312336}{{\tt hep-ph/0312336}}].

\bibitem{Chiu:2007dg}
J.-y. Chiu, F.~Golf, R.~Kelley, and A.~V. Manohar, {\it {Electroweak
  Corrections in High Energy Processes using Effective Field Theory}},  {\em
  Phys. Rev.} {\bf D77} (2008) 053004,
  [\href{http://arxiv.org/abs/0712.0396}{{\tt arXiv:0712.0396}}].

\bibitem{Chiu:2007yn}
J.-y. Chiu, F.~Golf, R.~Kelley, and A.~V. Manohar, {\it {Electroweak Sudakov
  corrections using effective field theory}},  {\em Phys. Rev. Lett.} {\bf 100}
  (2008) 021802, [\href{http://arxiv.org/abs/0709.2377}{{\tt
  arXiv:0709.2377}}].

\bibitem{Chiu:2008vv}
J.-y. Chiu, R.~Kelley, and A.~V. Manohar, {\it {Electroweak Corrections using
  Effective Field Theory: Applications to the LHC}},  {\em Phys. Rev.} {\bf
  D78} (2008) 073006, [\href{http://arxiv.org/abs/0806.1240}{{\tt
  arXiv:0806.1240}}].

\bibitem{Chiu:2009mg}
J.-y. Chiu, A.~Fuhrer, R.~Kelley, and A.~V. Manohar, {\it {Factorization
  Structure of Gauge Theory Amplitudes and Application to Hard Scattering
  Processes at the LHC}},  {\em Phys. Rev.} {\bf D80} (2009) 094013,
  [\href{http://arxiv.org/abs/0909.0012}{{\tt arXiv:0909.0012}}].

\bibitem{Chiu:2009ft}
J.-y. Chiu, A.~Fuhrer, R.~Kelley, and A.~V. Manohar, {\it {Soft and Collinear
  Functions for the Standard Model}},  {\em Phys. Rev.} {\bf D81} (2010)
  014023, [\href{http://arxiv.org/abs/0909.0947}{{\tt arXiv:0909.0947}}].

\bibitem{Fuhrer:2010eu}
A.~Fuhrer, A.~V. Manohar, J.-y. Chiu, and R.~Kelley, {\it {Radiative
  Corrections to Longitudinal and Transverse Gauge Boson and Higgs
  Production}},  {\em Phys. Rev.} {\bf D81} (2010) 093005,
  [\href{http://arxiv.org/abs/1003.0025}{{\tt arXiv:1003.0025}}].

\bibitem{Larkoski:2015kga}
A.~J. Larkoski, I.~Moult, and D.~Neill, {\it {Analytic Boosted Boson
  Discrimination}},  {\em JHEP} {\bf 05} (2016) 117,
  [\href{http://arxiv.org/abs/1507.03018}{{\tt arXiv:1507.03018}}].

\bibitem{Larkoski:2017iuy}
A.~J. Larkoski, I.~Moult, and D.~Neill, {\it {Analytic Boosted Boson
  Discrimination at the Large Hadron Collider}},
  \href{http://arxiv.org/abs/1708.06760}{{\tt arXiv:1708.06760}}.

\bibitem{Larkoski:2017cqq}
A.~J. Larkoski, I.~Moult, and D.~Neill, {\it {Factorization and Resummation for
  Groomed Multi-Prong Jet Shapes}},
  \href{http://arxiv.org/abs/1710.00014}{{\tt arXiv:1710.00014}}.

\bibitem{Caswell:1985ui}
W.~E. Caswell and G.~P. Lepage, {\it {Effective Lagrangians for Bound State
  Problems in QED, QCD, and Other Field Theories}},  {\em Phys. Lett.} {\bf
  167B} (1986) 437--442.

\bibitem{Bodwin:1994jh}
G.~T. Bodwin, E.~Braaten, and G.~P. Lepage, {\it {Rigorous QCD analysis of
  inclusive annihilation and production of heavy quarkonium}},  {\em Phys.
  Rev.} {\bf D51} (1995) 1125--1171,
  [\href{http://arxiv.org/abs/hep-ph/9407339}{{\tt hep-ph/9407339}}]. [Erratum:
  Phys. Rev.D55,5853(1997)].

\bibitem{Luke:1999kz}
M.~E. Luke, A.~V. Manohar, and I.~Z. Rothstein, {\it {Renormalization group
  scaling in nonrelativistic QCD}},  {\em Phys. Rev.} {\bf D61} (2000) 074025,
  [\href{http://arxiv.org/abs/hep-ph/9910209}{{\tt hep-ph/9910209}}].

\bibitem{Rothstein:1999vz}
I.~Z. Rothstein, {\it {NRQCD: A Critical review}},
  \href{http://arxiv.org/abs/hep-ph/9911276}{{\tt hep-ph/9911276}}.

\bibitem{Rothstein:2003mp}
I.~Z. Rothstein, {\it {TASI lectures on effective field theories}},
  \href{http://arxiv.org/abs/hep-ph/0308266}{{\tt hep-ph/0308266}}.

\bibitem{Hoang:2002ae}
A.~H. Hoang, {\it {Heavy quarkonium dynamics}},
  \href{http://arxiv.org/abs/hep-ph/0204299}{{\tt hep-ph/0204299}}.

\bibitem{Fan:2010gt}
J.~Fan, M.~Reece, and L.-T. Wang, {\it {Non-relativistic effective theory of
  dark matter direct detection}},  {\em JCAP} {\bf 1011} (2010) 042,
  [\href{http://arxiv.org/abs/1008.1591}{{\tt arXiv:1008.1591}}].

\bibitem{Fitzpatrick:2012ix}
A.~L. Fitzpatrick, W.~Haxton, E.~Katz, N.~Lubbers, and Y.~Xu, {\it {The
  Effective Field Theory of Dark Matter Direct Detection}},  {\em JCAP} {\bf
  1302} (2013) 004, [\href{http://arxiv.org/abs/1203.3542}{{\tt
  arXiv:1203.3542}}].

\bibitem{Neubert:1993mb}
M.~Neubert, {\it {Heavy quark symmetry}},  {\em Phys. Rept.} {\bf 245} (1994)
  259--396, [\href{http://arxiv.org/abs/hep-ph/9306320}{{\tt hep-ph/9306320}}].

\bibitem{Manohar:2000dt}
A.~V. Manohar and M.~B. Wise, {\it {Heavy quark physics}},  {\em Camb. Monogr.
  Part. Phys. Nucl. Phys. Cosmol.} {\bf 10} (2000) 1--191.

\bibitem{Luke:1992cs}
M.~E. Luke and A.~V. Manohar, {\it {Reparametrization invariance constraints on
  heavy particle effective field theories}},  {\em Phys. Lett.} {\bf B286}
  (1992) 348--354, [\href{http://arxiv.org/abs/hep-ph/9205228}{{\tt
  hep-ph/9205228}}].

\bibitem{Heinonen:2012km}
J.~Heinonen, R.~J. Hill, and M.~P. Solon, {\it {Lorentz invariance in heavy
  particle effective theories}},  {\em Phys. Rev.} {\bf D86} (2012) 094020,
  [\href{http://arxiv.org/abs/1208.0601}{{\tt arXiv:1208.0601}}].

\bibitem{Manohar:1999xd}
A.~V. Manohar and I.~W. Stewart, {\it {Renormalization group analysis of the
  QCD quark potential to order v**2}},  {\em Phys. Rev.} {\bf D62} (2000)
  014033, [\href{http://arxiv.org/abs/hep-ph/9912226}{{\tt hep-ph/9912226}}].

\bibitem{Manohar:2000hj}
A.~V. Manohar and I.~W. Stewart, {\it {The QCD heavy quark potential to order
  v**2: One loop matching conditions}},  {\em Phys. Rev.} {\bf D62} (2000)
  074015, [\href{http://arxiv.org/abs/hep-ph/0003032}{{\tt hep-ph/0003032}}].

\bibitem{Ibe:2012sx}
M.~Ibe, S.~Matsumoto, and R.~Sato, {\it {Mass Splitting between Charged and
  Neutral Winos at Two-Loop Level}},  {\em Phys. Lett.} {\bf B721} (2013)
  252--260, [\href{http://arxiv.org/abs/1212.5989}{{\tt arXiv:1212.5989}}].

\bibitem{Beneke:2012tg}
M.~Beneke, C.~Hellmann, and P.~Ruiz-Femenia, {\it {Non-relativistic pair
  annihilation of nearly mass degenerate neutralinos and charginos I. General
  framework and S-wave annihilation}},  {\em JHEP} {\bf 03} (2013) 148,
  [\href{http://arxiv.org/abs/1210.7928}{{\tt arXiv:1210.7928}}]. [Erratum:
  JHEP10,224(2013)].

\bibitem{Beneke:2014gja}
M.~Beneke, C.~Hellmann, and P.~Ruiz-Femenia, {\it {Non-relativistic pair
  annihilation of nearly mass degenerate neutralinos and charginos III.
  Computation of the Sommerfeld enhancements}},  {\em JHEP} {\bf 05} (2015)
  115, [\href{http://arxiv.org/abs/1411.6924}{{\tt arXiv:1411.6924}}].

\bibitem{Beneke:2016ync}
M.~Beneke, A.~Bharucha, F.~Dighera, C.~Hellmann, A.~Hryczuk, S.~Recksiegel, and
  P.~Ruiz-Femenia, {\it {Relic density of wino-like dark matter in the MSSM}},
  {\em JHEP} {\bf 03} (2016) 119, [\href{http://arxiv.org/abs/1601.04718}{{\tt
  arXiv:1601.04718}}].

\bibitem{Braaten:2017kci}
E.~Braaten, E.~Johnson, and H.~Zhang, {\it {Zero-Range Effective Field Theory
  for Resonant Wino Dark Matter II. Coulomb Resummation}},
  \href{http://arxiv.org/abs/1708.07155}{{\tt arXiv:1708.07155}}.

\bibitem{Braaten:2017gpq}
E.~Braaten, E.~Johnson, and H.~Zhang, {\it {Zero-Range Effective Field Theory
  for Resonant Wino Dark Matter I. Framework}},
  \href{http://arxiv.org/abs/1706.02253}{{\tt arXiv:1706.02253}}.

\bibitem{Manohar:2002fd}
A.~V. Manohar, T.~Mehen, D.~Pirjol, and I.~W. Stewart, {\it {Reparameterization
  invariance for collinear operators}},  {\em Phys. Lett.} {\bf B539} (2002)
  59--66, [\href{http://arxiv.org/abs/hep-ph/0204229}{{\tt hep-ph/0204229}}].

\bibitem{Chay:2002vy}
J.~Chay and C.~Kim, {\it {Collinear effective theory at subleading order and
  its application to heavy - light currents}},  {\em Phys. Rev.} {\bf D65}
  (2002) 114016, [\href{http://arxiv.org/abs/hep-ph/0201197}{{\tt
  hep-ph/0201197}}].

\bibitem{Bauer:2002aj}
C.~W. Bauer, D.~Pirjol, and I.~W. Stewart, {\it {Factorization and endpoint
  singularities in heavy to light decays}},  {\em Phys. Rev.} {\bf D67} (2003)
  071502, [\href{http://arxiv.org/abs/hep-ph/0211069}{{\tt hep-ph/0211069}}].

\bibitem{Rothstein:2016bsq}
I.~Z. Rothstein and I.~W. Stewart, {\it {An Effective Field Theory for Forward
  Scattering and Factorization Violation}},  {\em JHEP} {\bf 08} (2016) 025,
  [\href{http://arxiv.org/abs/1601.04695}{{\tt arXiv:1601.04695}}].

\bibitem{Moult:2017xpp}
I.~Moult, M.~P. Solon, I.~W. Stewart, and G.~Vita, {\it {Fermionic Glauber
  Operators and Quark Reggeization}},
  \href{http://arxiv.org/abs/1709.09174}{{\tt arXiv:1709.09174}}.

\bibitem{Manohar:2006nz}
A.~V. Manohar and I.~W. Stewart, {\it {The Zero-Bin and Mode Factorization in
  Quantum Field Theory}},  {\em Phys.Rev.} {\bf D76} (2007) 074002,
  [\href{http://arxiv.org/abs/hep-ph/0605001}{{\tt hep-ph/0605001}}].

\bibitem{Beneke:2002ni}
M.~Beneke and T.~Feldmann, {\it {Multipole expanded soft collinear effective
  theory with nonAbelian gauge symmetry}},  {\em Phys. Lett.} {\bf B553} (2003)
  267--276, [\href{http://arxiv.org/abs/hep-ph/0211358}{{\tt hep-ph/0211358}}].

\bibitem{Beneke:2002ph}
M.~Beneke, A.~P. Chapovsky, M.~Diehl, and T.~Feldmann, {\it {Soft collinear
  effective theory and heavy to light currents beyond leading power}},  {\em
  Nucl. Phys.} {\bf B643} (2002) 431--476,
  [\href{http://arxiv.org/abs/hep-ph/0206152}{{\tt hep-ph/0206152}}].

\bibitem{Chay:2004zn}
J.~Chay, C.~Kim, Y.~G. Kim, and J.-P. Lee, {\it {Soft Wilson lines in
  soft-collinear effective theory}},  {\em Phys. Rev.} {\bf D71} (2005) 056001,
  [\href{http://arxiv.org/abs/hep-ph/0412110}{{\tt hep-ph/0412110}}].

\bibitem{Arnesen:2005nk}
C.~M. Arnesen, J.~Kundu, and I.~W. Stewart, {\it {Constraint equations for
  heavy-to-light currents in SCET}},  {\em Phys. Rev.} {\bf D72} (2005) 114002,
  [\href{http://arxiv.org/abs/hep-ph/0508214}{{\tt hep-ph/0508214}}].

\bibitem{Bauer:2002nz}
C.~W. Bauer, S.~Fleming, D.~Pirjol, I.~Z. Rothstein, and I.~W. Stewart, {\it
  {Hard scattering factorization from effective field theory}},  {\em
  Phys.Rev.} {\bf D66} (2002) 014017,
  [\href{http://arxiv.org/abs/hep-ph/0202088}{{\tt hep-ph/0202088}}].

\bibitem{Becher:2011dz}
T.~Becher and G.~Bell, {\it {Analytic Regularization in Soft-Collinear
  Effective Theory}},  {\em Phys. Lett.} {\bf B713} (2012) 41--46,
  [\href{http://arxiv.org/abs/1112.3907}{{\tt arXiv:1112.3907}}].

\bibitem{Chiu:2011qc}
J.-y. Chiu, A.~Jain, D.~Neill, and I.~Z. Rothstein, {\it {The Rapidity
  Renormalization Group}},  {\em Phys. Rev. Lett.} {\bf 108} (2012) 151601,
  [\href{http://arxiv.org/abs/1104.0881}{{\tt arXiv:1104.0881}}].

\bibitem{Chiu:2012ir}
J.-Y. Chiu, A.~Jain, D.~Neill, and I.~Z. Rothstein, {\it {A Formalism for the
  Systematic Treatment of Rapidity Logarithms in Quantum Field Theory}},  {\em
  JHEP} {\bf 05} (2012) 084, [\href{http://arxiv.org/abs/1202.0814}{{\tt
  arXiv:1202.0814}}].

\bibitem{Mitridate:2017oky}
A.~Mitridate, M.~Redi, J.~Smirnov, and A.~Strumia, {\it {Dark Matter as a
  weakly coupled Dark Baryon}},  \href{http://arxiv.org/abs/1707.05380}{{\tt
  arXiv:1707.05380}}.

\bibitem{Bauer:2016kkv}
C.~W. Bauer and N.~Ferland, {\it {Resummation of electroweak Sudakov logarithms
  for real radiation}},  {\em JHEP} {\bf 09} (2016) 025,
  [\href{http://arxiv.org/abs/1601.07190}{{\tt arXiv:1601.07190}}].

\bibitem{Chen:2016wkt}
J.~Chen, T.~Han, and B.~Tweedie, {\it {Electroweak Splitting Functions and High
  Energy Showering}},  \href{http://arxiv.org/abs/1611.00788}{{\tt
  arXiv:1611.00788}}.

\bibitem{Sveshnikov:1995vi}
N.~Sveshnikov and F.~Tkachov, {\it {Jets and quantum field theory}},  {\em
  Phys.Lett.} {\bf B382} (1996) 403--408,
  [\href{http://arxiv.org/abs/hep-ph/9512370}{{\tt hep-ph/9512370}}].

\bibitem{Korchemsky:1997sy}
G.~P. Korchemsky, G.~Oderda, and G.~F. Sterman, {\it {Power corrections and
  nonlocal operators}},  \href{http://arxiv.org/abs/hep-ph/9708346}{{\tt
  hep-ph/9708346}}. [AIP Conf. Proc.407,988(1997)].

\bibitem{Lee:2006nr}
C.~Lee and G.~F. Sterman, {\it {Momentum Flow Correlations from Event Shapes:
  Factorized Soft Gluons and Soft-Collinear Effective Theory}},  {\em
  Phys.Rev.} {\bf D75} (2007) 014022,
  [\href{http://arxiv.org/abs/hep-ph/0611061}{{\tt hep-ph/0611061}}].

\bibitem{iain_notes}
I.~W. Stewart and C.~W. Bauer, ``Lectures on the soft-collinear effective
  theory.''
  \url{http://ocw.mit.edu/courses/physics/8-851-effective-field-theory-spring-2013/lecture-notes/MIT8_851S13_scetnotes.pdf}.

\bibitem{Bloch:1937pw}
F.~Bloch and A.~Nordsieck, {\it {Note on the Radiation Field of the electron}},
   {\em Phys. Rev.} {\bf 52} (1937) 54--59.

\bibitem{Kinoshita:1962ur}
T.~Kinoshita, {\it {Mass singularities of Feynman amplitudes}},  {\em J. Math.
  Phys.} {\bf 3} (1962) 650--677.

\bibitem{Lee:1964is}
T.~D. Lee and M.~Nauenberg, {\it {Degenerate Systems and Mass Singularities}},
  {\em Phys. Rev.} {\bf 133} (1964) B1549--B1562. [,25(1964)].

\bibitem{Ciafaloni:1998xg}
P.~Ciafaloni and D.~Comelli, {\it {Sudakov enhancement of electroweak
  corrections}},  {\em Phys. Lett.} {\bf B446} (1999) 278--284,
  [\href{http://arxiv.org/abs/hep-ph/9809321}{{\tt hep-ph/9809321}}].

\bibitem{Ciafaloni:1999ub}
P.~Ciafaloni and D.~Comelli, {\it {Electroweak Sudakov form-factors and
  nonfactorizable soft QED effects at NLC energies}},  {\em Phys. Lett.} {\bf
  B476} (2000) 49--57, [\href{http://arxiv.org/abs/hep-ph/9910278}{{\tt
  hep-ph/9910278}}].

\bibitem{Ciafaloni:2000df}
M.~Ciafaloni, P.~Ciafaloni, and D.~Comelli, {\it {Bloch-Nordsieck violating
  electroweak corrections to inclusive TeV scale hard processes}},  {\em Phys.
  Rev. Lett.} {\bf 84} (2000) 4810--4813,
  [\href{http://arxiv.org/abs/hep-ph/0001142}{{\tt hep-ph/0001142}}].

\bibitem{Farhi:1977sg}
E.~Farhi, {\it {A QCD Test for Jets}},  {\em Phys.Rev.Lett.} {\bf 39} (1977)
  1587--1588.

\bibitem{Dasgupta:2001sh}
M.~Dasgupta and G.~Salam, {\it {Resummation of nonglobal QCD observables}},
  {\em Phys.Lett.} {\bf B512} (2001) 323--330,
  [\href{http://arxiv.org/abs/hep-ph/0104277}{{\tt hep-ph/0104277}}].

\bibitem{Banfi:2002hw}
A.~Banfi, G.~Marchesini, and G.~Smye, {\it {Away from jet energy flow}},  {\em
  JHEP} {\bf 0208} (2002) 006, [\href{http://arxiv.org/abs/hep-ph/0206076}{{\tt
  hep-ph/0206076}}].

\bibitem{Caron-Huot:2015bja}
S.~Caron-Huot, {\it {Resummation of non-global logarithms and the BFKL
  equation}},  \href{http://arxiv.org/abs/1501.03754}{{\tt arXiv:1501.03754}}.

\bibitem{Becher:2015hka}
T.~Becher, M.~Neubert, L.~Rothen, and D.~Y. Shao, {\it {Effective Field Theory
  for Jet Processes}},  {\em Phys. Rev. Lett.} {\bf 116} (2016), no.~19 192001,
  [\href{http://arxiv.org/abs/1508.06645}{{\tt arXiv:1508.06645}}].

\bibitem{Becher:2016mmh}
T.~Becher, M.~Neubert, L.~Rothen, and D.~Y. Shao, {\it {Factorization and
  Resummation for Jet Processes}},  \href{http://arxiv.org/abs/1605.02737}{{\tt
  arXiv:1605.02737}}.

\bibitem{Larkoski:2016zzc}
A.~J. Larkoski, I.~Moult, and D.~Neill, {\it {The Analytic Structure of
  Non-Global Logarithms: Convergence of the Dressed Gluon Expansion}},  {\em
  JHEP} {\bf 11} (2016) 089, [\href{http://arxiv.org/abs/1609.04011}{{\tt
  arXiv:1609.04011}}].

\bibitem{Abbate:2010xh}
R.~Abbate, M.~Fickinger, A.~H. Hoang, V.~Mateu, and I.~W. Stewart, {\it {Thrust
  at $N^3LL$ with Power Corrections and a Precision Global Fit for
  alphas(mZ)}},  {\em Phys.Rev.} {\bf D83} (2011) 074021,
  [\href{http://arxiv.org/abs/1006.3080}{{\tt arXiv:1006.3080}}].

\bibitem{Korchemsky:1987wg}
G.~P. Korchemsky and A.~V. Radyushkin, {\it {Renormalization of the Wilson
  Loops Beyond the Leading Order}},  {\em Nucl. Phys. B} {\bf 283} (1987)
  342--364.

\bibitem{Bell:2010gi}
G.~Bell, J.~H. Kuhn, and J.~Rittinger, {\it {Electroweak Sudakov Logarithms and
  Real Gauge-Boson Radiation in the TeV Region}},  {\em Eur. Phys. J.} {\bf
  C70} (2010) 659--671, [\href{http://arxiv.org/abs/1004.4117}{{\tt
  arXiv:1004.4117}}].

\bibitem{Manohar:2014vxa}
A.~Manohar, B.~Shotwell, C.~Bauer, and S.~Turczyk, {\it {Non-cancellation of
  electroweak logarithms in high-energy scattering}},  {\em Phys. Lett.} {\bf
  B740} (2015) 179--187, [\href{http://arxiv.org/abs/1409.1918}{{\tt
  arXiv:1409.1918}}].

\bibitem{Rinchiuso:2017kfn}
{\bf H. E. S. S.} Collaboration, L.~Rinchiuso, E.~Moulin, A.~Viana,
  C.~Van~Eldik, and J.~Veh, {\it {Dark matter gamma-ray line searches toward
  the Galactic Center halo with H.E.S.S. I}},  {\em PoS} {\bf ICRC2017} (2017)
  893, [\href{http://arxiv.org/abs/1708.08358}{{\tt arXiv:1708.08358}}].

\bibitem{Rinchiuso:2017pcx}
{\bf for the H. E. S. S.} Collaboration, L.~Rinchiuso and E.~Moulin, {\it {Dark
  matter searches toward the Galactic Centre halo with H.E.S.S}},  2017.
\newblock \href{http://arxiv.org/abs/1711.08634}{{\tt arXiv:1711.08634}}.

\bibitem{Lefranc:2016fgn}
V.~Lefranc, E.~Moulin, P.~Panci, F.~Sala, and J.~Silk, {\it {Dark Matter in
  $\gamma$ lines: Galactic Center vs dwarf galaxies}},  {\em JCAP} {\bf 1609}
  (2016), no.~09 043, [\href{http://arxiv.org/abs/1608.00786}{{\tt
  arXiv:1608.00786}}].

\bibitem{Lisanti:2017qoz}
M.~Lisanti, S.~Mishra-Sharma, N.~L. Rodd, B.~R. Safdi, and R.~H. Wechsler, {\it
  {Mapping Extragalactic Dark Matter Annihilation with Galaxy Surveys: A
  Systematic Study of Stacked Group Searches}},
  \href{http://arxiv.org/abs/1709.00416}{{\tt arXiv:1709.00416}}.

\bibitem{Pato:2015tja}
M.~Pato and F.~Iocco, {\it {The Dark Matter Profile of the Milky Way: a
  Non-parametric Reconstruction}},  {\em Astrophys. J.} {\bf 803} (2015), no.~1
  L3, [\href{http://arxiv.org/abs/1504.03317}{{\tt arXiv:1504.03317}}].

\bibitem{Pieri:2009je}
L.~Pieri, J.~Lavalle, G.~Bertone, and E.~Branchini, {\it {Implications of
  High-Resolution Simulations on Indirect Dark Matter Searches}},  {\em Phys.
  Rev.} {\bf D83} (2011) 023518, [\href{http://arxiv.org/abs/0908.0195}{{\tt
  arXiv:0908.0195}}].

\bibitem{Navarro:1996gj}
J.~F. Navarro, C.~S. Frenk, and S.~D.~M. White, {\it {A Universal density
  profile from hierarchical clustering}},  {\em Astrophys. J.} {\bf 490} (1997)
  493--508, [\href{http://arxiv.org/abs/astro-ph/9611107}{{\tt
  astro-ph/9611107}}].

\bibitem{Rolke:2004mj}
W.~A. Rolke, A.~M. Lopez, and J.~Conrad, {\it {Limits and confidence intervals
  in the presence of nuisance parameters}},  {\em Nucl. Instrum. Meth.} {\bf
  A551} (2005) 493--503, [\href{http://arxiv.org/abs/physics/0403059}{{\tt
  physics/0403059}}].

\bibitem{Cowan:2010js}
G.~Cowan, K.~Cranmer, E.~Gross, and O.~Vitells, {\it {Asymptotic formulae for
  likelihood-based tests of new physics}},  {\em Eur. Phys. J.} {\bf C71}
  (2011) 1554, [\href{http://arxiv.org/abs/1007.1727}{{\tt arXiv:1007.1727}}].
  [Erratum: Eur. Phys. J.C73,2501(2013)].

\bibitem{Chan:2015tna}
T.~K. Chan, D.~Kere{\v s}, J.~O{\~n}orbe, P.~F. Hopkins, A.~L. Muratov, C.~A.
  Faucher-Gigu{\`e}re, and E.~Quataert, {\it {The impact of baryonic physics on
  the structure of dark matter haloes: the view from the FIRE cosmological
  simulations}},  {\em Mon. Not. Roy. Astron. Soc.} {\bf 454} (2015), no.~3
  2981--3001, [\href{http://arxiv.org/abs/1507.02282}{{\tt arXiv:1507.02282}}].

\bibitem{2015MNRAS.448..713P}
M.~{Portail}, C.~{Wegg}, O.~{Gerhard}, and I.~{Martinez-Valpuesta}, {\it
  {Made-to-measure models of the Galactic box/peanut bulge: stellar and total
  mass in the bulge region}},  {\em \mnras} {\bf 448} (Mar., 2015) 713--731,
  [\href{http://arxiv.org/abs/1502.00633}{{\tt arXiv:1502.00633}}].

\bibitem{Hooper:2016ggc}
D.~Hooper, {\it {The Density of Dark Matter in the Galactic Bulge and
  Implications for Indirect Detection}},  {\em Phys. Dark Univ.} {\bf 15}
  (2017) 53--56, [\href{http://arxiv.org/abs/1608.00003}{{\tt
  arXiv:1608.00003}}].

\bibitem{Beneke:2016jpw}
M.~Beneke, A.~Bharucha, A.~Hryczuk, S.~Recksiegel, and P.~Ruiz-Femenia, {\it
  {The last refuge of mixed wino-Higgsino dark matter}},  {\em JHEP} {\bf 01}
  (2017) 002, [\href{http://arxiv.org/abs/1611.00804}{{\tt arXiv:1611.00804}}].

\bibitem{Krall:2017xij}
R.~Krall and M.~Reece, {\it {Last Electroweak WIMP Standing: Pseudo-Dirac
  Higgsino Status and Compact Stars as Future Probes}},
  \href{http://arxiv.org/abs/1705.04843}{{\tt arXiv:1705.04843}}.

\bibitem{Bauer:2003pi}
C.~W. Bauer and A.~V. Manohar, {\it {Shape function effects in B ---> X(s)
  gamma and B ---> X(u) l anti-nu decays}},  {\em Phys. Rev.} {\bf D70} (2004)
  034024, [\href{http://arxiv.org/abs/hep-ph/0312109}{{\tt hep-ph/0312109}}].

\bibitem{Bosch:2004th}
S.~W. Bosch, B.~O. Lange, M.~Neubert, and G.~Paz, {\it {Factorization and shape
  function effects in inclusive B meson decays}},  {\em Nucl. Phys.} {\bf B699}
  (2004) 335--386, [\href{http://arxiv.org/abs/hep-ph/0402094}{{\tt
  hep-ph/0402094}}].

\bibitem{Becher:2006qw}
T.~Becher and M.~Neubert, {\it {Toward a NNLO calculation of the anti-B --->
  X(s) gamma decay rate with a cut on photon energy. II. Two-loop result for
  the jet function}},  {\em Phys. Lett.} {\bf B637} (2006) 251--259,
  [\href{http://arxiv.org/abs/hep-ph/0603140}{{\tt hep-ph/0603140}}].

\bibitem{Ligeti:2008ac}
Z.~Ligeti, I.~W. Stewart, and F.~J. Tackmann, {\it {Treating the b quark
  distribution function with reliable uncertainties}},  {\em Phys.Rev.} {\bf
  D78} (2008) 114014, [\href{http://arxiv.org/abs/0807.1926}{{\tt
  arXiv:0807.1926}}].

\bibitem{Baumgart:2018yed}
M.~Baumgart, T.~Cohen, E.~Moulin, I.~Moult, L.~Rinchiuso, N.~L. Rodd, T.~R.
  Slatyer, I.~W. Stewart, and V.~Vaidya, {\it {Precision Photon Spectra for
  Wino Annihilation}},  {\em JHEP} {\bf 01} (2019) 036,
  [\href{http://arxiv.org/abs/1808.08956}{{\tt arXiv:1808.08956}}].

\bibitem{Beneke:2019vhz}
M.~Beneke, A.~Broggio, C.~Hasner, K.~Urban, and M.~Vollmann, {\it {Resummed
  photon spectrum from dark matter annihilation for intermediate and narrow
  energy resolution}},  \href{http://arxiv.org/abs/1903.08702}{{\tt
  arXiv:1903.08702}}.

\bibitem{Chiu:2009yx}
J.-y. Chiu, A.~Fuhrer, A.~H. Hoang, R.~Kelley, and A.~V. Manohar, {\it
  {Soft-Collinear Factorization and Zero-Bin Subtractions}},  {\em Phys. Rev.}
  {\bf D79} (2009) 053007, [\href{http://arxiv.org/abs/0901.1332}{{\tt
  arXiv:0901.1332}}].

\bibitem{Cirelli:2010xx}
M.~Cirelli, G.~Corcella, A.~Hektor, G.~Hutsi, M.~Kadastik, P.~Panci, M.~Raidal,
  F.~Sala, and A.~Strumia, {\it {PPPC 4 DM ID: A Poor Particle Physicist
  Cookbook for Dark Matter Indirect Detection}},  {\em JCAP} {\bf 1103} (2011)
  051, [\href{http://arxiv.org/abs/1012.4515}{{\tt arXiv:1012.4515}}].
  [Erratum: JCAP1210,E01(2012)].

\bibitem{Ciafaloni:2010ti}
P.~Ciafaloni, D.~Comelli, A.~Riotto, F.~Sala, A.~Strumia, and A.~Urbano, {\it
  {Weak Corrections are Relevant for Dark Matter Indirect Detection}},  {\em
  JCAP} {\bf 1103} (2011) 019, [\href{http://arxiv.org/abs/1009.0224}{{\tt
  arXiv:1009.0224}}].

\bibitem{Sjostrand:2006za}
T.~Sjostrand, S.~Mrenna, and P.~Z. Skands, {\it {PYTHIA 6.4 Physics and
  Manual}},  {\em JHEP} {\bf 05} (2006) 026,
  [\href{http://arxiv.org/abs/hep-ph/0603175}{{\tt hep-ph/0603175}}].

\bibitem{Sjostrand:2007gs}
T.~Sjostrand, S.~Mrenna, and P.~Z. Skands, {\it {A Brief Introduction to PYTHIA
  8.1}},  {\em Comput. Phys. Commun.} {\bf 178} (2008) 852--867,
  [\href{http://arxiv.org/abs/0710.3820}{{\tt arXiv:0710.3820}}].

\bibitem{Sjostrand:2014zea}
{\it {An Introduction to PYTHIA 8.2}},  {\em Comput. Phys. Commun.} {\bf 191}
  (2015) 159--177, [\href{http://arxiv.org/abs/1410.3012}{{\tt
  arXiv:1410.3012}}].

\bibitem{Christiansen:2014kba}
{\it {Weak Gauge Boson Radiation in Parton Showers}},  {\em JHEP} {\bf 04}
  (2014) 115, [\href{http://arxiv.org/abs/1401.5238}{{\tt arXiv:1401.5238}}].

\bibitem{Ackermann:2015zua}
{\bf Fermi-LAT} Collaboration, M.~Ackermann et~al., {\it {Searching for Dark
  Matter Annihilation from Milky Way Dwarf Spheroidal Galaxies with Six Years
  of Fermi Large Area Telescope Data}},  {\em Phys. Rev. Lett.} {\bf 115}
  (2015), no.~23 231301, [\href{http://arxiv.org/abs/1503.02641}{{\tt
  arXiv:1503.02641}}].

\bibitem{Fermi-LAT:2016uux}
{\bf DES, Fermi-LAT} Collaboration, A.~Albert et~al., {\it {Searching for Dark
  Matter Annihilation in Recently Discovered Milky Way Satellites with
  Fermi-LAT}},  {\em Astrophys. J.} {\bf 834} (2017), no.~2 110,
  [\href{http://arxiv.org/abs/1611.03184}{{\tt arXiv:1611.03184}}].

\bibitem{Lisanti:2017qlb}
M.~Lisanti, S.~Mishra-Sharma, N.~L. Rodd, and B.~R. Safdi, {\it {A Search for
  Dark Matter Annihilation in Galaxy Groups}},
  \href{http://arxiv.org/abs/1708.09385}{{\tt arXiv:1708.09385}}.

\end{thebibliography}\endgroup
\bibliographystyle{JHEP}

\end{document}